\documentclass{aa}  

\usepackage{natbib,twoopt}
\usepackage[colorlinks =true, urlcolor =blue, linkcolor =blue, citecolor =blue]{hyperref} 
\bibpunct{(}{)}{;}{a}{}{,}            
\makeatletter
  \newcommandtwoopt{\citeads}[3][][]{\href{http://adsabs.harvard.edu/abs/#3}%
    {\def\hyper@linkstart##1##2{}%
     \let\hyper@linkend\@empty\citealp[#1][#2]{#3}}}
  \newcommandtwoopt{\citepads}[3][][]{\href{http://adsabs.harvard.edu/abs/#3}%
    {\def\hyper@linkstart##1##2{}%
     \let\hyper@linkend\@empty\citep[#1][#2]{#3}}}
  \newcommandtwoopt{\citetads}[3][][]{\href{http://adsabs.harvard.edu/abs/#3}%
    {\def\hyper@linkstart##1##2{}%
     \let\hyper@linkend\@empty\citet[#1][#2]{#3}}}
  \newcommandtwoopt{\citeyearads}[3][][]%
    {\href{http://adsabs.harvard.edu/abs/#3}
    {\def\hyper@linkstart##1##2{}%
     \let\hyper@linkend\@empty\citeyear[#1][#2]{#3}}}
\makeatother

\usepackage{graphicx}
\usepackage{caption}
\usepackage{upgreek}
\usepackage{multirow}
\usepackage{placeins}
\usepackage[normalem]{ulem}
\usepackage{txfonts}

\usepackage{tikz,hyperref}

\definecolor{lime}{HTML}{A6CE39}
\DeclareRobustCommand{\orcidicon}{
        \begin{tikzpicture}
        \draw[lime, fill =lime] (0,0) 
        circle [radius =0.16] 
        node[white] {{\fontfamily{qag}\selectfont \tiny ID}};
        \draw[white, fill =white] (-0.0625,0.095) 
        circle [radius =0.007];
        \end{tikzpicture}
        \hspace{-2mm}
}

\foreach \x in {A, ..., Z}{\expandafter\xdef\csname orcid\x\endcsname{\noexpand\href{https://orcid.org/\csname orcidauthor\x\endcsname}
                        {\noexpand\orcidicon}}
}

\begin{document}

     \title{Nickel- and iron-rich clumps in planetary nebulae: New discoveries and emission-line diagnostics}

   \author{K. Bouvis\inst{1}\fnmsep\inst{2}{\orcidA{}}
          \and
          S. Akras\inst{1}{\orcidB{}}
          \and
          H. Monteiro\inst{3,4}{\orcidC{}}
          \and
          L. Konstantinou\inst{1}\fnmsep\inst{2}{\orcidD{}}
          \and 
          P. Boumis\inst{1}{\orcidE{}}
          \and
          J. Garcia-Rojas\inst{5,6}{\orcidF{}}
          \and
          D.R. Gon\c{c}alves\inst{7}{\orcidK{}}
          \and
          I. Aleman\inst{8}{\orcidH{}}
          \and
          A. Monreal-Ibero\inst{9}{\orcidI{}}
          \and
          J. Cami\inst{10,11,12}{\orcidG{}}
          }

   \institute{Institute for Astronomy, Astrophysics, Space Applications and Remote Sensing, National Observatory of Athens,
              GR 15236 Penteli, Greece\\
              \email{kbouvis@noa.gr, stavrosakras@noa.gr}
         \and
            Department of Physics, University of Patras, Patras, 26504 Rio, Greece
         \and 
            School of Physics and Astronomy, Cardiff University, Queen's Buildings, The Parade, Cardiff CF24 3AA, UK
        \and 
            Instituto de F\'{i}sica e Qu\'{i}mica, Universidade Federal de Itajuba, Av. BPS 1303-Pinheirinho, 37500-903, \'{I}tajuba, Brazil
         \and 
             Instituto de Astrof\'isica de Canarias, E-38205 La Laguna, Tenerife, Spain
         \and
             Departamento de Astrof\'isica, Universidad de La Laguna, E-38206 La Laguna, Tenerife, Spain
         \and
             Observat\'orio do Valongo, Universidade Federal do Rio de Janeiro, Ladeira Pedro Antonio 43, 20080-090, Rio de Janeiro, Brazil
        \and
            Laborat\'{o}rio Nacional de Astrof\'{i}sica, Rua dos Estados Unidos, 154, Bairro das Na\c{c}\~{o}es, Itajub\'{a}, MG, 37504-365, Brazil     
        \and
            Leiden Observatory, Leiden University, P.O. Box 9513, 2300 RA Leiden, The Netherlands
        \and
             Department of Physics and Astronomy, University of Western Ontario, London, ON N6A 3K7, Canada
        \and 
            Institute for Earth and Space Exploration, University of Western Ontario, London, ON N6A 3K7, Canada
         \and 
            SETI Institute, 189 Bernardo Ave, Suite 100, Mountain View, CA 94043, USA
         }
   \date{Received XXXXX, XXXX; accepted XXXXXX, XXXX}
 
  \abstract
   {Integral field spectroscopy (IFS) offers a distinct advantage for studying extended sources by enabling spatially resolved emission maps for several emission lines without the need for specific filters.} 
   {This study aims to conduct a detailed analysis of iron and nickel emission lines in 12 planetary nebulae (PNe) using integral field unit (IFU) data from MUSE to provide valuable insights into their formation and evolution mechanisms.}
   {New diagnostic line ratios, combined with machine-learning algorithms, were used to distinguish excitation mechanisms such as shock and photoionization. Electron densities and elemental abundances were estimated for different atomic data using the {\sc PyNeb} package. The contribution of fluorescent excitation of nickel lines was also examined.}
   {A total of 16 iron- and nickel-rich clumps are detected in seven out of 12 PNe. New clumps are discovered in NGC~3132 and IC~4406. The most prominent lines are [Fe~{\sc ii}] 8617~$\AA$ and [Ni~{\sc ii}] 7378~$\AA$. Both emission lines are observed emanating directly from the low-ionization structures (LIS) of NGC~3242, NGC~7009, and NGC~6153, as well as from clumps in NGC~6369 and Tc~1. Their abundances are found to be below the solar values, indicating that a fraction of Fe and Ni remains depleted in dust grains. The depletion factors exhibit a strong correlation over a wide range of values. A machine-learning approach allows us to classify ten out of 16 clumps as shock-excited and to establish a new shock/photoionization selection criterion: log([Ni~{\sc ii}] 7378~$\AA$/H$\upalpha$) $\&$ log([Fe~{\sc ii}] 8617~$\AA$/H$\upalpha$) > $-$2.20.}
   {}


   \maketitle

\section{Introduction}
\label{intro}
The powerful technique of integral field spectroscopy (IFS) has revolutionized the way we study extended astronomical objects. Integral field units (IFUs) were initially employed in the late 1980s with TIGER/CFHT \citep{TIGER}, and since then numerous instruments have been developed. Most IFUs included in planetary nebula (PN) studies are FLAMES/VLT \citep{FLAMES}, VIMOS/VLT \citep{VIMOS}, SINFONI/VLT \citep{SINFONI}, MUSE/VLT \citep{bacon2010}, MEGARA/GTC \citep{megara}, NIRSPEC/JWST \citep{NIRSPEC}, and MIRI/JWST \citep{MIRI}. The wavelength coverage of most IFUs spans from near ultraviolet (NUV) to infrared (IR) wavelengths, while the field of view varies from a few arcseconds to nearly one arcminute. 

Unlike previous imaging studies limited by the available filters, IFUs offer an unprecedented ability to spatially resolve emission lines from a wide range of species with different ionization states. Before the advent of IFS, knowledge of the emission region was mostly constrained by the slit position in spectroscopic studies. 

A wide variety of astronomical objects have been studied using IFUs, including galaxies, active galactic nuclei (AGNs), supernova remnants (SNRs), H~{\sc ii} regions, young stellar objects (YSOs), Herbig-Haro (HH) objects, and PNe, among others. Regarding PNe, which are the main focus of this work, IFU observations have provided unprecedented findings and novel insights into their physical and chemical properties, revealing details about ionization structures, abundance variations, and dynamical features. Several studies have demonstrated the significant advantage of imaging spectroscopy over traditional long-slit spectroscopy in the study of extended PNe \citep[e.g.,][]{matsuura2007,tsamis2008,ali2015,ali2017,dopita2017,walsh2018,ibero2020,rechy-garcia2020,akras2022}.
However, the majority of the studies are focused on the brightest and more common emission lines. Notable exceptions are the studies by \citet{jorge2022}, \citet{akras24C}, \citet{gomez_llanos_2024}, and \citet{lydia2025}, which have investigated a broader range of emission lines in PNe, including the first spatial distribution of the near-IR emission line of atomic carbon [C~{\sc i}]~8727~$\AA$ in low-ionization structures \citep[LISs;][]{denise2001,akras2016,mari2023}. This motivated us to look into the available MUSE data of PNe for less common and generally faint emission lines such as [Ni~{\sc ii}] 7378~$\AA$ and [Fe~{\sc ii}] 8617~$\AA$. 

Planetary nebulae are formed by the interaction of the slow asymptotic giant branch (AGB) wind and the fast post-AGB wind \citep[e.g.,][]{kwok1978,balick1987,Icke1988}. The nebular gas is ionized by the ultraviolet (UV) photons of the hot central star, and its spectrum is characterized by discrete emission lines (recombination or collisionally excited). Although PNe constitute only a small fraction of the lifetime of low-mass stars ($\sim$$10^4$ yr) between AGB and white-dwarf phases, they play a key role in the chemical enhancement of the interstellar medium. The most common elements observed in PNe are H, He, N, S, O, Cl, Ar, C, and Ne; and in rarer cases Fe, Ni, Ca, P, Mg, K, Cr, Mn, Se, Kr, and Xe \citep[e.g.,][]{jorge2015, sterling2017}. Exploring and analyzing a wide variety of elements in PNe provides further insights into the stellar evolution of low-mass stars, both single and in binary systems.

Iron and nickel lines have been detected in a wide variety of gaseous nebulae, indicating a close correlation between them. More specifically, [Ni~{\sc ii}] 7412~$\AA$, 7378~$\AA,$ and 11910~$\AA;$ and [Fe~{\sc ii}] 5158~$\AA$, 5262~$\AA$, 7155~$\AA$, 8617~$\AA$, 12570~$\AA,$ and 16440~$\AA$ emission lines appear strong in SNRs \citep{dennefeld1986,jerkstrand2015A,crab_jwst}, HH objects \citep{brugel1981,mesa2009,reiter2019}, and Seyfert galaxies \citep{halpern1986,henry1988}, but they are fainter in H~{\sc ii} regions \citep{osterbrock1990,inglada2016}, circumstellar nebulae of luminous blue variables \citep{barlow1994,meaburn2000}, and PNe \citep{jorge2013,delgado_inglada_2014}. Table \ref{Fe_Ni_prev_stud} lists the PNe where optical forbidden emission lines of singly ionized nickel and/or iron have been detected from previous long-slit or IFU observations. 

In SNRs and some HH objects, nickel and iron emission lines are usually detected in regions influenced by shocks. In particular, high [Fe~{\sc ii}]~1.257 $\upmu$m/P$\upalpha$ and [Fe~{\sc ii}]~1.644 $\upmu$m/Br$\upgamma$ line ratios are considered reliable shock tracers \citep{graham1987,crab_jwst}, whereas low values are attributed to UV radiation \citep[e.g.,][and references therein]{akras24Fe}.

Neutral and singly ionized Fe and Ni have comparable ionization potentials (7.9 eV for Fe$^{0}$, 7.6 eV for Ni$^{0}$, and 16.2 eV for Fe$^{+}$, 18.2 eV for Ni$^{+}$), indicating that they coexist in partially ionized zones (PIZs). Fe$^{+}$ and Ni$^{+}$ are sustained in these regions because their ionization potentials are higher than H~{\sc i} (13.6~eV), which acts as a shield. The most prominent and brightest [Ni~{\sc ii}] and [Fe~{\sc ii}] emission lines are centered at 7378~$\AA$ and 8617~$\AA$, respectively. [Ni~{\sc ii}] 7378~$\AA$ originates from $^{2}$D$_{5/2}$-$^{2}$F$_{7/2}$ transition, while [Fe~{\sc ii}] 8617~$\AA$ from the $^{4}$F$_{9/2}$-$^{4}$P$_{5/2}$ transition. The [Fe~{\sc ii}] and [Ni~{\sc ii}] lines have critical electron densities above $10^6$ cm$^{-3}$ and their excitation energies are very close: $\Delta$E([Fe~{\sc ii}] 8617~$\AA$)= 1.67~eV and $\Delta$E([Ni~{\sc ii}] 7378~$\AA$)=1.68 eV \citep{bautista1996}.

This study focuses on the first detection of nickel and iron emissions from clumps embedded in PNe. Sect.~\ref{section2} describes the PNe sample and their properties, while Sect.~\ref{section3} provides details of the conducted spectroscopic analysis. Sect.~\ref{section4} presents the main findings of our study, and Sect.~\ref{section5} outlines our machine-learning approach to interpreting the results and estimating additional physical parameters (e.g., density, abundances, depletion factors). Finally, Sect.~\ref{section6} discusses the origin of Fe and Ni in PNe, and Sect.~\ref{section7} summarizes the conclusions of our study. 

\begin{table}[]
\centering
\caption{PNe where optical forbidden emissions of [Ni~{\sc ii}] and/or [Fe~{\sc ii}] were detected.}
\begin{tabular}{|c|c|}
\hline
Object    & Reference                     \\ \hline
IC~2165$^{\dag}$ & \citet{hyung1994}              \\ \hline
M~1-91    & \citet{rodriguez2001}              \\ \hline
Mz~3      & \citet{zhang2003}                  \\ \hline
IC~418    & \citet{williams2003,dopita2017}               \\ \hline
NGC~5315    & \citet{peimpert2004}               \\ \hline
NGC~40$^{\dag}$    & \multirow{4}{*}{\citet{liu2004}}   \\ \cline{1-1}
NGC~6720$^{\dag}$    &                               \\ \cline{1-1}
NGC~6741$^{\dag}$    &                               \\ \cline{1-1}
NGC~6884$^{\dag}$    &                               \\ \hline
NGC~7027  & \citet{zhang2005}                  \\ \hline
H~1-42$^{\dag}$    & \multirow{4}{*}{\citet{wang2007}}   \\ \cline{1-1}
M~3-7$^{\dag}$  &                               \\ \cline{1-1}
NGC~6565$^{\dag}$    &                               \\ \cline{1-1}
NGC~6620$^{\dag}$    &                               \\ \hline
NGC~7009$^{\dag\dag}$  & \citet{fang2011}                   \\ \hline
Cn~1-5    & \multirow{8}{*}{\citet{jorge2013}}   \\ \cline{1-1}
He~2-86   &                               \\ \cline{1-1}
M~1-25    &                               \\ \cline{1-1}
M~1-30$^{\dag}$    &                               \\ \cline{1-1}
M~1-32    &                               \\ \cline{1-1}
M~1-61    &                               \\ \cline{1-1}
PC~14$^{\dag}$     &                               \\ \cline{1-1}
Pe~1-1    &                               \\ \hline
H~1-40    & \multirow{7}{*}{\citet{jorge2018}}   \\ \cline{1-1}
Hen~2-73$^{\dag}$ &                               \\ \cline{1-1}
Hen~2-96  &                               \\ \cline{1-1}
Hen~2-158$^{\dag}$ &                               \\ \cline{1-1}
M~1-31    &                               \\ \cline{1-1}
M~1-33    &                               \\ \cline{1-1}
M~1-60    &                               \\ \hline
Hen~2-459$^{\dag}$ & \multirow{5}{*}{\citet{manea2022}} \\ \cline{1-1}
K~3-17$^{\dag}$    &                               \\ \cline{1-1}
K~3-60$^{\dag}$    &                               \\ \cline{1-1}
M~2-43$^{\dag}$    &                               \\ \cline{1-1}
M~3-35$^{\dag}$    &                               \\ \hline
IC~4663$^{\dag\dag}$   & \citet{mohery2023}                \\ \hline
IRAS 22568+6141 & {\citet{roldan2024}} \\ \hline
\end{tabular}
\label{Fe_Ni_prev_stud}
\tablefoot{$^{\dag}$ Only [Fe~{\sc ii}] lines were detected. $^{\dag\dag}$ [Ni~{\sc ii}] lines were detected, but not at 7378~$\AA$.}
\end{table}

\section{ESO archival data and sample description}
\label{section2}

We gathered most of the available and already reduced MUSE data of PNe from the ESO archive, where the standard data-reduction process was carried out using ESO pipelines\footnote{No post-pipeline treatment was applied to the data.} \citep[a summary of reduction steps can be found in Sect. 2 of][]{jorge2022}. Table \ref{data_pne} presents the list of PNe in our study with information on their observations. 

Multi Unit Spectroscopic Explorer \citep[MUSE;][]{bacon2010} is an IFU mounted on the UT4 of the Very Large Telescope (VLT) in Cerro Paranal, Chile. The wide field mode (WFM) of MUSE provides a field of view of 1$^{\prime}$ $\times$ 1$^{\prime}$ and spatial resolution of 0.2\arcsec~per pixel, while the wavelength range spans from 4800~$\AA$ to 9300~$\AA$ in the nominal spectral mode and from 4600~$\AA$ to 9300~$\AA$ in the extended spectral mode. The resolving power is 1770 at the blue part of the spectrum and 3590 at the red part. The outcome from an IFU, such as MUSE, is a data cube with two spatial dimensions and a spectral one. 

\begin{table}[h!]
\caption{PNe with available MUSE data in the ESO archive used in this work.}
\centering
\begin{tabular}{|c|c|l|c|} 
 \hline
 Target & Exp.(s) & Program ID & P.I. \\ [0.5ex] 
 \hline
 NGC~3242 & 900 & 097.D-0241(A) & Corradi R.L.M.\\
 \hline
 NGC~6153 & 2320 & 097.D-0241(A) & Corradi R.L.M.\\
 \hline
 Hf~2-2 & 9000 & 097.D-0241(A) & Corradi R.L.M.\\
 \hline
 M~1-42 & 3150 & 097.D-0241(A) & Corradi R.L.M.\\
 \hline
 NGC~6778 & 2250 & 097.D-0241(A) & Corradi R.L.M.\\
 \hline
 \multirow{2}{*}{NGC~7009} & \multirow{2}{*}{3300} & 097.D-0241(A)  & Corradi R.L.M.\\ \cline{3-4}
                           &                       & 60.A-9347(A) & Walsh J. \\ 
 \hline
 NGC~3132 & 600 & 60.A-9100(A) & MUSE team \\
 \hline
 IC~418 & 600 & 60.A-9100(A) & MUSE team \\
 \hline
 NGC~6369 & 480 & 60.A-9100(H) & MUSE team \\
 \hline
 NGC~6563 & 480 & 60.A-9100(H) & MUSE team \\
 \hline
 IC~4406 & 180 & 60.A-9100(G) & MUSE team \\
 \hline
 Tc~1$^{\dag}$ & 595 & 105.20R7.001 & Cami J. \\
 \hline
\end{tabular}
\label{data_pne}
\tablefoot{The total integration time, the program ID, and the P.I. are also provided. $^{\dag}$For full details on the reduction process, see Walsh et al. (in prep.).}
\end{table}

NGC~3242 is a multiple-shell PN with a bright rim and a pair of symmetric LISs. It is ionized by an O-type star with a spectral type O(H) \citep[][ and references therein]{weidmann2020}, $T_{\rm eff}$=80\,000 K, and log(L/L$_{\odot}$) = 2.86 \citep{pottasch2008}. A triple central system has also been proposed for this nebula \citep{soker_1992}. Moreover, NGC~3242 has been detected in X-ray emission with XMM-Newton \citep[][]{ruiz2011}.

NGC~6153 is a 3.2 kyr \citep{gonzalez2021} elliptical nebula with two bright regions at the end of its minor axis and several smaller clumps. A weak-emission-line star (wels) is present at the center with $T_{\rm eff}$=110\,000~K and log(L/L$_{\odot}$) = 3.76 \citep{gonzalez2021}. There is probably an unseen stellar companion, as recent Gaia measurements indicate \citep{chorney2021}. Also, the abundance discrepancy factor (ADF) of the nebula was found to be high \citep[][]{liu2000, gomez_llanos_2024}, which is probably due to the presence of a binary system (see Sect.~\ref{sec:corr}).

Hf~2-2 is a spherical nebula with a bright and fragmented ring seen in [N~{\sc ii}] images. A short-orbital-period ($\sim$0.4 days) post-common-envelope binary system has been found at the center of this nebula \citep{lutz1998,bond2000}. Furthermore, it is characterized by high ADF \citep[][]{liu2006,jorge2022}, and its central star has an O(H)3 spectral type \citep{weidmann2020}. 

M~1-42 is an elliptical nebula with symmetrical clumps alongside its major axis. In addition, a high ADF has been derived for this nebula \citep[][]{liu2001,jorge2022}, probably indicating the presence of a binary system.

NGC~6778 is a 4.4 kyr \citep{tocknell2014} irregular nebula with jet-like structures. NGC~6778 hosts a close binary central star that has undergone a common-envelope phase with short orbital period \citep[$\sim$0.15 days,][]{miszalski2011}, which is consistent with the high ADF measured in this nebula \citep[][]{jones2016,jorge2022}.

NGC~7009 is a 1.9 kyr \citep{gonzalez2021} elliptical nebula with two pairs of LISs located along its major axis. H$_2$ emission was recently detected in the outer pair of LISs \citep{akras2020_H2}. The central star of the PN (CSPN) is an O(H) spectral-type star with $T_{\rm eff}$=82\,000 K and log(L/L$_{\odot}$) = 3.97 \citep{mendez1992}.

NGC~3132 is a young, elliptical, and molecule-rich PN \citep{kastner2024}, likely with a multi-stellar system \citep{demarco2022}. The CS has $T_{\rm eff}$=100\,000~K, log(L/L$_{\odot}$)= 2.19 \citep{frew2008}, and an A2V spectral type \citep{weidmann2020}.

IC~418 is a 1.6 kyr \citep[][and references therein]{weidmann2020} spherical nebula with a CSPN of $T_{\rm eff}$=38\,000 K, log(L/L$_{\odot}$)~=~3.77, and the spectral type O7fp. Moreover, emission from polycyclic aromatic hydrocarbons (PAHs) at 13.2 $\upmu$m and fullerenes at 17.4 $\upmu$m have been detected in this nebula \citep{diaz2018}.

NGC~6369 has a complex morphology with a bright and fragmented rim. Its central star is an oxygen-rich Wolf-Rayet star with a spectral type of [WO~3] \citep[][and references therein]{weidmann2020}, $T_{\rm eff}$=70\,000 K, and log(L/L$_{\odot}$) = 3.38 \citep{pottasch2008}. The presence of a stellar companion has also been proposed based on the Gaia measurements \citep{chorney2021}. Molecular hydrogen and PAHs have also been identified in this nebula \citep{ramos2012}.

NGC~6563 is a 6.35 kyr elliptical nebula. Its central star is characterized by $T_{\rm eff}$=123\,000~K and log(L/L$_{\odot}$) = 1.84, and its progenitor mass was estimated $\sim$2.9 M$_{\odot}$ \citep{gonzalez2021}.

IC~4406 has a prolate spheroid shape, and its CSPN has been classified as a Wolf-Rayet star [WR] \citep[][and references therein]{weidmann2020}. Recent Gaia measurements suggest the existence of a binary system at the center of the nebula \citep{chorney2021}. Moreover, CO and H$_2$ emission have been found alongside the nebula's major axis \citep{sahai1991,ramoslarios2022}.

Tc~1 is a young, slightly elongated spheroid and fullerene-rich nebula \citep{Cami2010} with a spectacular round halo. Its central star has a spectral type of O(H)5-9f \citep[][and references therein]{weidmann2020} with 30\,000 K<$T_{\rm eff}$<34\,000 K and 3.3$<$log(L/L$_{\odot}$)$<$3.6 \citep{otsuka2014, aleman2019}. In addition, \citet{Ali2023} identified the CS of Tc~1 as a variable star, based on GAIA DR3 data, linking it to a possible binary system.

\section{Spectroscopic analysis}
\label{section3}
\subsection{Emission-line extraction}
\label{3.1}
\begin{figure}[h!]     
    \centering{\includegraphics[width=0.5\textwidth]{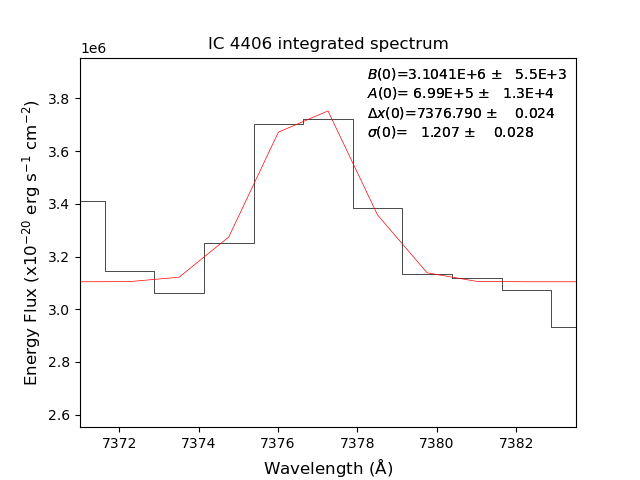}}
    \caption{Fitting procedure of [Ni~{\sc ii}] 7378~$\AA$ in IC~4406 data cube. The black line represents the observation and the red line shows the Gaussian fitting of the emission line.}
    \label{extraction}
\end{figure}

Each MUSE data cube underwent a careful inspection for the detection of Fe and Ni along with C and Ca faint emission lines. We searched the available literature to identify the strongest emission lines within the MUSE wavelength range from 4650~$\AA$ to 9300~$\AA$ for these elements, as indicated in Table \ref{PN_EMIS_LIST}. Emission lines were extracted from the data cubes by fitting a Gaussian profile and subtracting the continuum emission using a Python code. The code is capable of simultaneously fitting multiple distributions and gives both the emission-line maps and the corresponding error maps as output. Fig.~\ref{extraction} displays, as an illustrative example, the fitting of the [Ni~{\sc ii}] 7378~$\AA$ emission line in the data cube of IC~4406.

\begin{table*}[h!]
\caption{Emission lines detected from the MUSE data in our sample of PNe.}
\label{PN_EMIS_LIST}
\centering 
\resizebox{2\columnwidth}{!}{%
\begin{tabular}{|c c c c c c c c c c c c c|}
 \hline
 Emission lines & NGC~3242 & NGC~6153 & Hf~2-2 & M~1-42 & NGC~6778 & NGC~7009 & NGC~3132 & IC~418 & NGC~6369 & NGC~6563 & IC~4406 & Tc~1 \\ [0.5ex] 
 \hline
[Fe~{\sc iii}] 4658 \AA & X & - & - & \checkmark & - & \checkmark & X & X & X & X & X & X \\
\hline
[Fe~{\sc iii}] 4701 \AA & X & - & - & - & - & \checkmark & X & X & X & X & X & X \\
\hline
[Fe~{\sc iii}] 4881 \AA & \checkmark & \checkmark & - & - & - & \checkmark & \checkmark & \checkmark & - & - & - & \checkmark \\
\hline
[Fe~{\sc iii}] 5270 \AA & \checkmark & \checkmark & - & - & - & \checkmark & \checkmark & \checkmark & - & - & - & \checkmark \\
\hline
[Fe~{\sc ii}] 5158 \AA & - & - & - & - & - & \checkmark & \checkmark & \checkmark & - & - & - & \checkmark \\
\hline
[Fe~{\sc ii}] 5262 \AA & - & - & - & - & - & - & \checkmark & \checkmark & - & - & - & - \\
\hline
[Fe~{\sc ii}] 7155 \AA & \checkmark & \checkmark & - & - & - & \checkmark & \checkmark & \checkmark & \checkmark & - & \checkmark & \checkmark \\
\hline
[Fe~{\sc ii}] 8617 \AA & \checkmark & \checkmark & - & - & - & \checkmark & \checkmark & \checkmark & \checkmark & - & \checkmark & \checkmark \\
\hline
[Ni~{\sc ii}] 7378 \AA & \checkmark & \checkmark & - & - & - & \checkmark & \checkmark & \checkmark & \checkmark & - & \checkmark & \checkmark \\
\hline
[Ca~{\sc ii}] 7291 \AA & - & - & - & - & - & - & ? & \checkmark & - & - & - & \checkmark \\
\hline
[C~{\sc i}] 8727 \AA & \checkmark & \checkmark & - & \checkmark & \checkmark & \checkmark & \checkmark & \checkmark & \checkmark & \checkmark & \checkmark & \checkmark \\
\hline
\end{tabular}
}
\tablefoot{Dash ("-") indicates non-detection, while "X" denotes lines that fall outside the MUSE wavelength range because the corresponding objects were not observed in extended spectral mode.}
\end{table*}

Given that the optical iron and nickel lines are nearly two or three orders of magnitude fainter than H$\upbeta$, we employed three steps to assure their detection. Emission lines were extracted from both the original data cube from binned versions (2 $\times$ 2 or 3 $\times$ 3), where adjacent pixel values are combined to enhance sensitivity at the expense of reducing the spatial resolution. While binning increases signal-to-noise ratio, it can introduce uncertainties when applied to regions with spatially varying nebular backgrounds. This is why we used binned data solely to verify the detection in the original cube; all flux measurements presented in this study were derived exclusively from the original data cubes. Furthermore, we cross-checked the literature for potential emission lines or skylines close to the wavelengths of interest. In most cases,\footnote{Except [Fe~{\sc iii}] 4658~$\AA$ and [Fe~{\sc iii}] 5270~$\AA,$ which can be affected by C~{\sc iv} 4657, 4658~$\AA,$ and [Fe~{\sc ii}] 5269~$\AA$, respectively.} no strong emission lines were found nearby, allowing us to rule out the contribution of other elements and possible misidentification. Moreover, we compared the resulting emission-line maps with their corresponding error maps, which were generated during the fitting procedure. These errors include uncertainties from the instruments, the observational errors, and possible systematics from blending with nearby lines and sky substraction residuals. To ensure the reliability of our findings, we established a criterion requiring the flux-to-error ratio > 3 for each pixel in the extracted emission-line maps. Lastly, spectra in the wavelength range of 7350-7400~$\AA$ and 8550-8750~$\AA$ were extracted for several regions in order to visually support our detections. The spectra are presented in Appendix \ref{int_spec}, where [Fe~{\sc ii}]~8617~$\AA$ and [Ni~{\sc ii}]~7378~$\AA$ are indicated with vertical orange and blue lines, respectively.

\subsection{Flux measurements}
Besides the extraction of the spectra, we computed the fluxes of the H$\upalpha$, H$\upbeta$, Paschen 10 (P10: 9015~$\AA$), [O~{\sc i}] 6300~$\AA$, [Fe~{\sc ii}] 8617~$\AA,$ and [Ni~{\sc ii}]~7378~$\AA$ emission lines. The first three were used to compute the extinction coefficient (c(H$\upbeta$))\footnote{The extinction coefficient was calculated assuming H$\upalpha$/H$\upbeta$=2.86 and P10/H$\upbeta$=0.0185, which are the expected values for standard plasma conditions $T_{\rm e}$=10\,000~K \citep{morisset2023} and density-bounded nebulae \citep[Case B approximation,][]{osterbrock2006}.} using the {\sc PyNeb} 1.1.19 package \citep{luridiana2015, morisset2020} and correct the fluxes for interstellar extinction (see Table \ref{flux_table}). As a proof of concept, the extinction coefficient in NGC~3132 was estimated from both H$\upalpha$/H$\upbeta$ and P10/H$\upbeta$ ratios, with differences of only $\sim$5$\%$ when using the integrated values over the whole nebula.

The selected regions, where line fluxes were measured, are constrained by the [Ni~{\sc ii}]~7378~$\AA$ line, since it is the weakest. The regions are displayed on the [O~{\sc i}] 6300~$\AA$ emission map of each PN in Appendix \ref{supplementary_maps}. Our H$\upbeta$ and [O~{\sc i}] fluxes as well the interstellar extinction are, within errors, in agreement with previous studies (NGC~3242, c(H$\upbeta$)=0.14 \citet{lydia2025}; NGC~6153, c(H$\upbeta$)=1.2 \citet{gomez_llanos_2024}; NGC~7009, c(H$\upbeta$)=0.12 \citet{walsh2018, akras2022}; NGC~3132, c(H$\upbeta$)=0.14 \citet{ibero2020}; NGC~6369, c(H$\upbeta$)=2.12 \citet{pottasch2008}; IC~4409, c(H$\upbeta$)=0.27 \citet{corradi1997}; Tc~1, c(H$\upbeta$) ranges from 0.28-0.44 \citet{frew2013,aleman2019}). This assures that our process for calculating the emission-line fluxes is reliable.

\section{Results}
\label{section4}

\begin{table*}[]
\caption{Emission-line intensities normalized to H$\upbeta$ = 100 for the clump regions in our sample of PNe.}
\label{flux_table}
\centering
\resizebox{\textwidth}{!}{%
\begin{tabular}{|c|c|c|c|c|c|c|c|c|c|c|}
\hline
Object                    & Clump & Area (pix$^2$) & F(H$\upbeta$) & c(H$\upbeta$) & [O~{\sc i}] 6300~$\AA$ & [Fe~{\sc ii}] 8617~$\AA$ & [Ni~{\sc ii}] 7378~$\AA$ & log([Ni~{\sc ii}]/H$\upalpha$) & log([Fe~{\sc ii}]/H$\upalpha$) & Excitation \\ \hline
\multirow{2}{*}{NGC~3242} & N     & 1.8& 12.27 (1.5 $\%$) & 0.14$^{+0.02}_{-0.02}$ & 4.09 (1.3 $\%$) & 0.04 (35 $\%$) & 0.02 (55 $\%$) & $-$4.17$^{+0.19}_{-0.34}$ & $-$3.85$^{+0.13}_{-0.19}$ & UV \\ \cline{2-11} 
                          & S     & 2.9& 17.69 (1.8 $\%$)& 0.16$^{+0.02}_{-0.02}$ & 7.41 (0.9 $\%$)  & 0.04 (32 $\%$) & 0.02 (48 $\%$) & $-$4.09$^{+0.17}_{-0.28}$ & $-$3.85$^{+0.12}_{-0.17}$ & UV \\ \hline
\multirow{2}{*}{NGC~6153} & E     & 13.8& 12.42 (0.6 $\%$) & 1.15$^{+0.01}_{-0.01}$ $^{\dag}$ & 14.57 (0.8 $\%$) & 0.05 (33 $\%$) & 0.03 (55 $\%$) & $-$4.02$^{+0.19}_{-0.35}$ & $-$3.78$^{+0.12}_{-0.17}$ & SS \\ \cline{2-11} 
                          & W     & 8.2& 7.59 (0.5 $\%$) & 1.18$^{+0.01}_{-0.01}$ $^{\dag}$ & 10.37 (1 $\%$) & 0.06 (26 $\%$) & 0.03 (45 $\%$) & $-$3.93$^{+0.16}_{-0.25}$ & $-$3.67$^{+0.10}_{-0.13}$ & SS \\ \hline
\multirow{2}{*}{NGC~7009} & E     & 15.2& 21.6 (0.5 $\%$) & 0.14$^{+0.01}_{-0.01}$ $^{\dag}$ & 23.49 (0.7 $\%$) & 0.1 (51 $\%$) & 0.06 (51 $\%$) & $-$3.67$^{+0.18}_{-0.30}$ & $-$3.45$^{+0.18}_{-0.31}$ & SS \\ \cline{2-11} 
                          & W     & 36.3& 31.9 (0.6 $\%$) & 0.15$^{+0.01}_{-0.01}$ $^{\dag}$ & 24.99 (1 $\%$) & 0.13 (53 $\%$) & 0.09 (51 $\%$) & $-$3.50$^{+0.18}_{-0.31}$ & $-$3.34$^{+0.18}_{-0.33}$ & SS \\ \hline    
\multirow{3}{*}{NGC~3132} & NW    & 12.8& 18.3 (0.7 $\%$) & 0.13$^{+0.01}_{-0.01}$ & 8.63 (2.3 $\%$) & 0.55 (14 $\%$) & 0.42 (17 $\%$) & $-$2.82$^{+0.07}_{-0.08}$ & $-$2.72$^{+0.06}_{-0.06}$ & FS \\ \cline{2-11} 
                          & NE    & 7.8& 6.03 (0.9 $\%$) & 0.17$^{+0.01}_{-0.01}$ & 18.9 (1.8 $\%$) & 0.58 (21 $\%$) & 0.49 (22 $\%$) & $-$2.76$^{+0.09}_{-0.11}$ & $-$2.69$^{+0.08}_{-0.10}$ & FS \\ \cline{2-11} 
                          & SE    & 9.4& 24.8 (0.4 $\%$) & 0.11$^{+0.01}_{-0.01}$ & 6.97 (1.8 $\%$) & 0.13 (30 $\%$) & 0.10 (32 $\%$) & $-$3.43$^{+0.12}_{-0.17}$ & $-$3.35$^{+0.11}_{-0.15}$ & UV \\ \hline
\multirow{2}{*}{NGC~6369} & NE    & 22.1& 0.275 (13 $\%$) & 2.13$^{+0.20}_{-0.18}$ & 15.53 (19 $\%$) & 0.96 (45 $\%$) & 1.08 (56 $\%$) & $-$2.42$^{+0.17}_{-0.29}$ & $-$2.47$^{+0.16}_{-0.25}$ & FS \\ \cline{2-11} 
                          & SW    & 8.0& 0.426 (5.2 $\%$) & 2.28$^{+0.08}_{-0.08}$ & 12.37 (6.8 $\%$) & 0.23 (40 $\%$) & 0.21 (59 $\%$) & $-$3.13$^{+0.20}_{-0.39}$ & $-$3.1$^{+0.15}_{-0.22}$ & UV \\ \hline
\multirow{3}{*}{IC~4406}  & NE    & 1.5& 0.09 (18.5 $\%$) & 0.36$^{+0.26}_{-0.22}$ & 70.29 (17.3 $\%$) & 10.74 (59.1 $\%$) & 7.43 (48.1 $\%$) & $-$1.55$^{+0.17}_{-0.28}$ & $-$1.39$^{+0.20}_{-0.39}$ & FS \\ \cline{2-11}
                          & SE    & 6.1& 0.52 (11.1 $\%$) & 0.26$^{+0.15}_{-0.14}$ & 68.21 (9.7 $\%$) & 7.95 (46.4 $\%$) & 4.98 (45.0 $\%$) & $-$1.75$^{+0.16}_{-0.26}$ & $-$1.55$^{+0.17}_{-0.27}$ & FS \\ \cline{2-11}
                          & W     & 20.3& 2.61 (9.1 $\%$) & 0.26$^{+0.12}_{-0.11}$ & 36.28 (12.7 $\%$) & 5.38 (50.4 $\%$) & 4.03 (50.4 $\%$) & $-$1.85$^{+0.18}_{-0.31}$ & $-$1.72$^{+0.18}_{-0.30}$ & FS \\ \hline                 
\multirow{2}{*}{Tc~1}     & NW    & 10.2& 224 (0.2 $\%$)& 0.32$^{+0.003}_{-0.003}$ $^{\dag}$ & 0.29 (1.6 $\%$) & 0.01 (16 $\%$) & 0.004 (39 $\%$) & $-$4.85$^{+0.14}_{-0.21}$ & $-$4.46$^{+0.06}_{-0.07}$ & UV \\ \cline{2-11} 
                          & SE    &  3.1& 52.5 (0.1 $\%$) & 0.27$^{+0.001}_{-0.001}$ $^{\dag}$ & 0.42 (1.5 $\%$) & 0.015 (9.3 $\%$) & 0.008 (22 $\%$) & $-$4.55$^{+0.08}_{-0.10}$ & $-$4.28$^{+0.04}_{-0.04}$ & UV \\ \hline
\end{tabular}
}
\tablefoot{F(H$\upbeta$) corresponds to the de-reddened flux in unit $\times$10$^{-14}$ erg cm$^{-2}$ s$^{-1}$ corrected for interstellar extinction (F(H$\upbeta$)$_{\rm dered}$=F(H$\upbeta$)$_{\rm red}$$\cdot$e$^{\rm c(H\upbeta)}$). The values noted with $^{\dag}$ were corrected using the extinction from the P10/H$\upbeta$ ratio, since H$\upalpha$ was saturated. Values in parentheses represent the uncertainty. The last column lists the excitation mechanism of each clump, UV (photoionization), SS (slow shocks), and FS (fast shocks); these were predicted using the KNN algorithm (see Sect.~\ref{sec:ML}).}
\end{table*}

\subsection{NGC~3242}
\label{sec_3242}
\begin{figure*}[h!]     
    \centering{\includegraphics[width=1\textwidth]{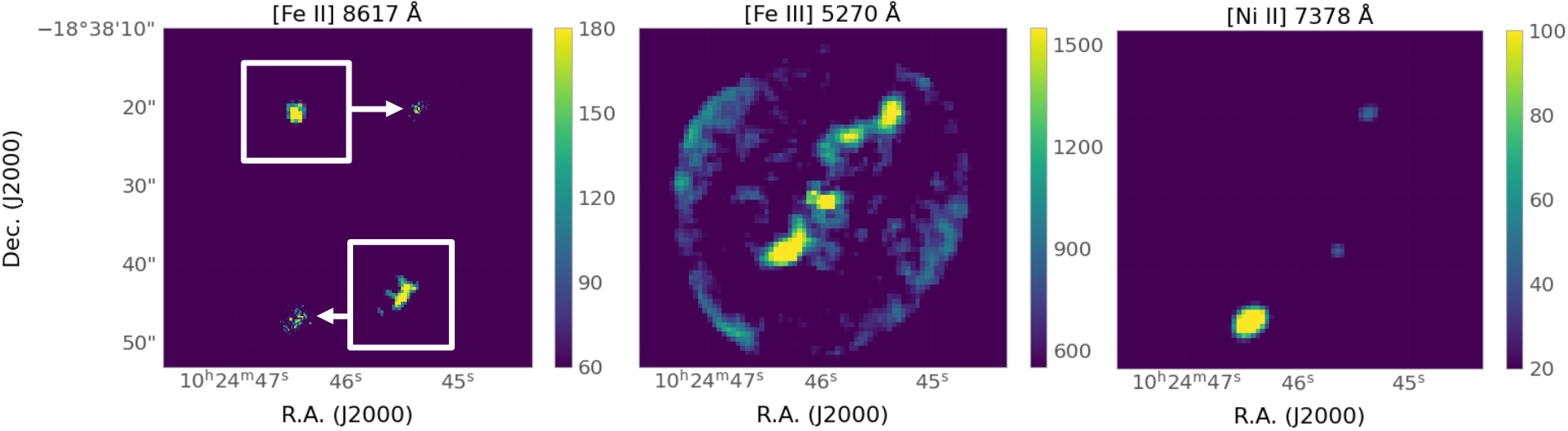}}
    \caption{NGC~3242 emission-line maps of [Fe~{\sc ii}] 8617~$\AA$, [Fe~{\sc iii}] 5270~$\AA$, and [Ni~{\sc ii}] 7378~$\AA$. White boxes represent zoomed-in views of smoothed and higher contrast versions of the selected regions. [Ni~{\sc ii}] and [Fe~{\sc iii}] emission-line maps are smoothed and scaled down 3 times from the original. Color-bar values are in units of 10$^{-20}$ erg s$^{-1}$ cm$^{-2}$.}
    \label{ngc3242_text}
\end{figure*}

The NW and SE LISs of NGC~3242 are well known and studied \citep[e.g.,][]{pottasch2008, monteiro2013}. However, these studies focused primarily on the most prominent emission lines, except \citet{lydia2025}, which was the first to pay attention to the faint [C~{\sc i}] emission. In this work, we focused on the detection of the emission lines from heavier element lines, such as [Fe~{\sc iii}] 4881~$\AA$ and 5270~$\AA$, [Fe~{\sc ii}] 7155~$\AA,$ and 8617~$\AA$ and [Ni~{\sc ii}] 7378~$\AA$ (Figs. \ref{ngc3242_text} and \ref{ngc3242_append}) \citep[some were already mentioned in][]{lydia2025}. The [Fe~{\sc ii}] and [Ni~{\sc ii}] emission lines clearly emanate from the LISs of NGC~3242 \citep[k1 and k4 features in][]{munoz2015}.

\begin{figure}[h!]     
    \centering{\includegraphics[width=0.39\textwidth]{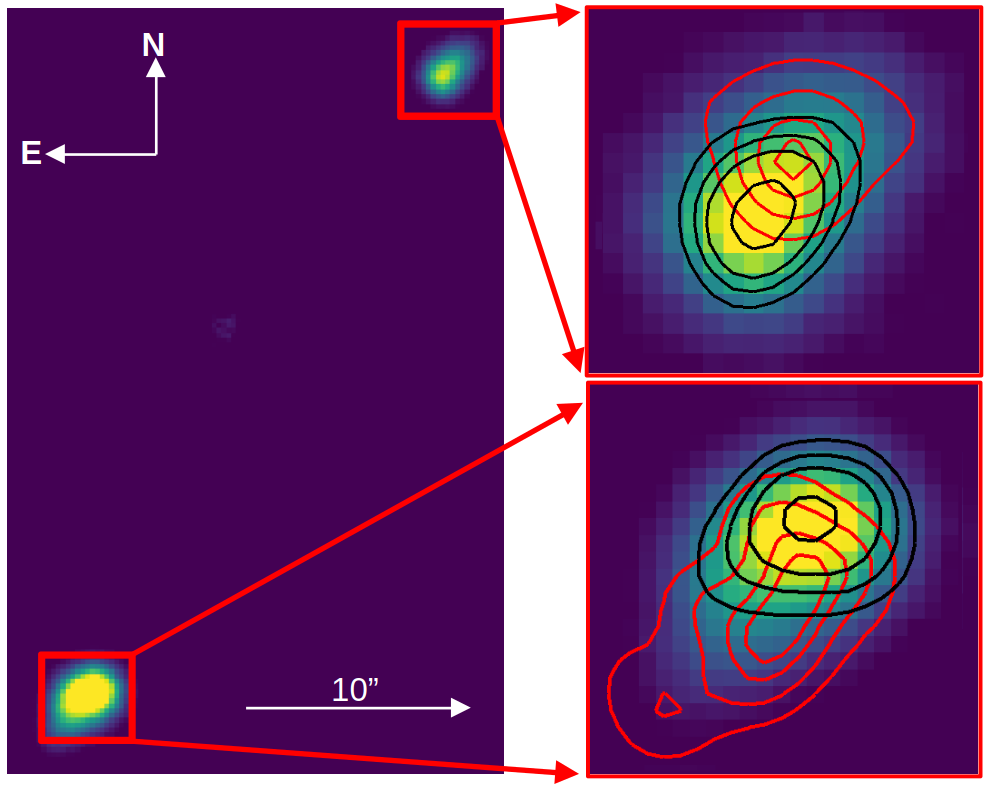}}
    \caption{NGC~3242 [O~{\sc i}] 6300~$\AA$ emission map with [C~{\sc i}] 8727~$\AA$ (black) and [Fe~{\sc ii}] 8617~$\AA$ (red) line contours overlaid.}
    \label{ngc3242_contours}
\end{figure}

Further analysis has revealed an intriguing spatial offset between [Fe~{\sc ii}] 8617~$\AA$ and [C~{\sc i}] 8727~$\AA$, which is of $\sim$0.5\arcsec~for the NW LIS and $\sim$1\arcsec~for the SE LIS (Fig.~\ref{ngc3242_contours}). More specifically, [Fe~{\sc ii}] emission peaks at larger radial distances from the central star compared to the [C~{\sc i}] line. A similar behavior is observed for the [Ni~{\sc ii}] 7378~$\AA$ line, but with higher uncertainties. This offset could likely reflect the ionization gradient, different shock fronts, or even possible differences in dust absorption. On the other hand, [Fe~{\sc iii}] 5270~$\AA$ emission shows a jet-like structure, which is probably associated with the LISs and was also noted by \citet{lydia2025}. The excitation mechanisms of the [Fe~{\sc ii}], [Fe~{\sc iii}], and [Ni~{\sc ii}] lines observed in this nebula are discussed in Sect.~\ref{section4}.

\subsection{NGC~6153}
\begin{figure*}[h!]     
    \centering{\includegraphics[width=1\textwidth]{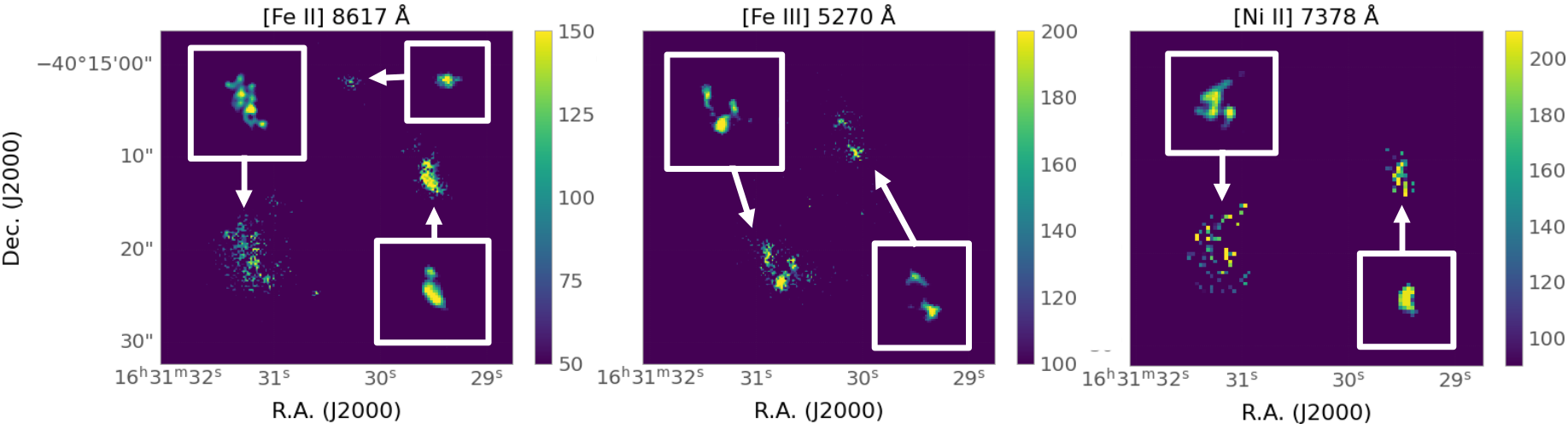}}
    \caption{NGC~6153 emission-line maps of [Fe~{\sc ii}] 8617~$\AA$, [Fe~{\sc iii}] 5270~$\AA,$ and [Ni~{\sc ii}] 7378~$\AA$. White boxes represent zoomed-in views of smoothed and higher contrast versions of the selected regions. [Ni~{\sc ii}] 7378~$\AA$ emission-line map is scaled down by a factor of 2 from the original. Color-bar values are in units of 10$^{-20}$ erg s$^{-1}$ cm$^{-2}$.}
    \label{ngc6153_text}
\end{figure*}

NGC~6153 displays two prominent LISs at the eastern and western parts of the nebula, as illustrated by the [O~{\sc i}] 6300~$\AA$ line map (Fig.~\ref{ngc6153_append}). We verified the detection of the emission lines [Fe~{\sc iii}]~4881~$\AA$ and 5270~$\AA$, [Fe~{\sc ii}] 7155~$\AA$ and 8617~$\AA$, and [Ni~{\sc ii}] 7378~$\AA$ and [C~{\sc i}] 8727~$\AA$ (Fig.~\ref{ngc6153_text}). [Fe~{\sc ii}] 8617~$\AA$ and [Ni~{\sc ii}] 7378~$\AA$ lines are found to emanate from the same regions, while [Fe~{\sc iii}] 5270~$\AA$ comes from two district regions with different orientation.

\subsection{Hf~2-2}
Hf~2–2 shows a spherical shell that is highly fragmented into knots, which are prominent in low-ionization lines (e.g., [O~{\sc i}] 6300~$\AA$). Neither Fe nor Ni emission lines were detected in this PN (Fig.~\ref{hf2-2_append}). 

\subsection{M~1-42}
M~1-42 is characterized by an ellipsoidal shell and five LISs along the major axis of the nebula, which are likely associated with a jet-like 
structure \citep{guerrero2013}. The LISs were only detected in the typical low-ionization lines, such as [N~{\sc ii}] 6548~$\AA$ and 6584~$\AA$, [S~{\sc ii}] 6716~$\AA,$ and 6731~$\AA$ and [O~{\sc i}] 6300~$\AA$ (numbers 1 to 5 in Fig.~\ref{m1-42_append}). The [C~{\sc i}] 8727~$\AA$ emission line was successfully identified at the western and eastern edges of the nebula. No Fe or Ni lines were detected in this nebula either.

\subsection{NGC~6778}
NGC~6778 was recently examined by \citet{akras2022} and \citet{jorge2022}. The [C~{\sc i}]~8727~$\AA$  originates from the inner filamentary structure, alongside neutral oxygen 6300~$\AA$ (see Fig.~\ref{ngc6778_append}). There is an indication (very low signal-to-noise  ratio) of [Ni~{\sc ii}]~7378~$\AA$ emission; thus, further investigation is required to confirm its possible detection.

\subsection{NGC~7009}
\begin{figure}[h!]     
    \centering{\includegraphics[width=0.5\textwidth]{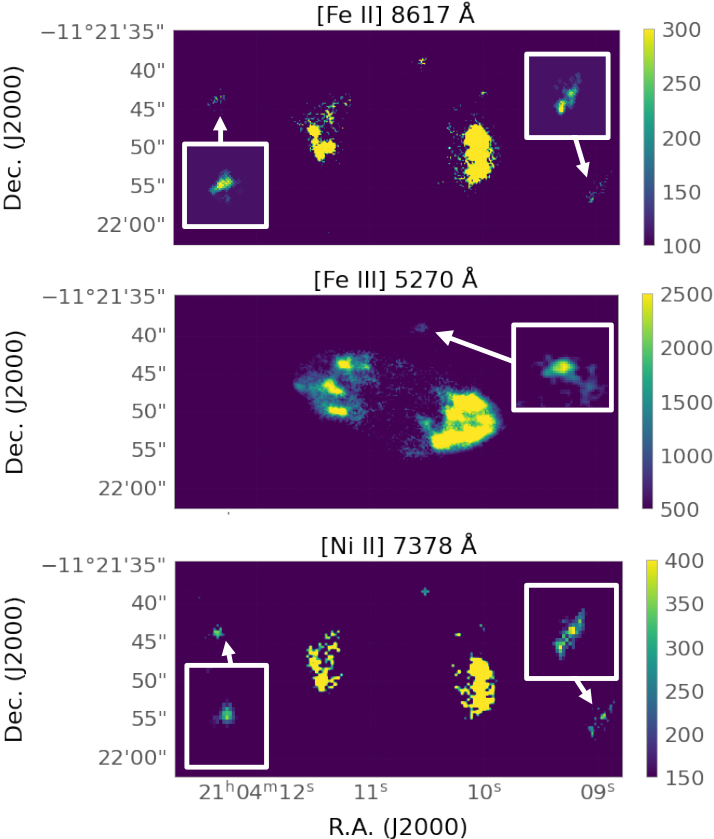}}
    \caption{NGC~7009 emission-line maps of [Fe~{\sc ii}] 8617~$\AA$, [Fe~{\sc iii}] 5270~$\AA,$ and [Ni~{\sc ii}] 7378~$\AA$. White squares represent zoomed-in views of smoothed and higher contrast versions of the selected regions. [Ni~{\sc ii}] 7378~$\AA$ emission-line map is scaled down by a factor of 2 from the original. Color-bar values are in units of 10$^{-20}$ erg s$^{-1}$ cm$^{-2}$.}
    \label{ngc7009_text}
\end{figure}

The Saturn nebula is one of the most studied PNe in the literature \citep[e.g.,][]{goncalves2003,walsh2018, akras2022, akras24C}. It is characterized by an inner ellipsoidal structure and two pairs of LISs. Recently, molecular hydrogen emission was detected in the outer LISs \citep{akras2020_H2}, alongside the [C~{\sc i}] 8727~$\AA$ and [Fe~{\sc ii}] 1.644 $\upmu$m emission lines \citep{akras24Fe,akras24C}. By scrutinizing the MUSE data cube, we can also report the detection of doubly ionized iron lines ([Fe~{\sc iii}] 4658~$\AA$, 4701~$\AA$, 4881~$\AA$, and 5270~$\AA$) and singly ionized iron and nickel lines ([Fe~{\sc ii}] 5158~$\AA$, 7155~$\AA$, 8617~$\AA$, and [Ni~{\sc ii}] 7378~$\AA$) (Figs \ref{ngc7009_text} and \ref{ngc7009_append}). Some of these lines were previously detected in the deep, long-slit spectra from \citet{fang2011}; however, the slit did not cover the outer pair of LISs. The MUSE IFU data have revealed that the [Fe~{\sc ii}] and [Ni~{\sc ii}] lines emanate from both the inner and outer pair of LISs \citep[K1 to K4,][]{goncalves2003}, while [Fe~{\sc iii}] is detected only from the inner LISs (K2 and K3) (see Fig.~\ref{ngc7009_text}). 

\begin{figure}[h!]     
    \centering{\includegraphics[width=0.5\textwidth]{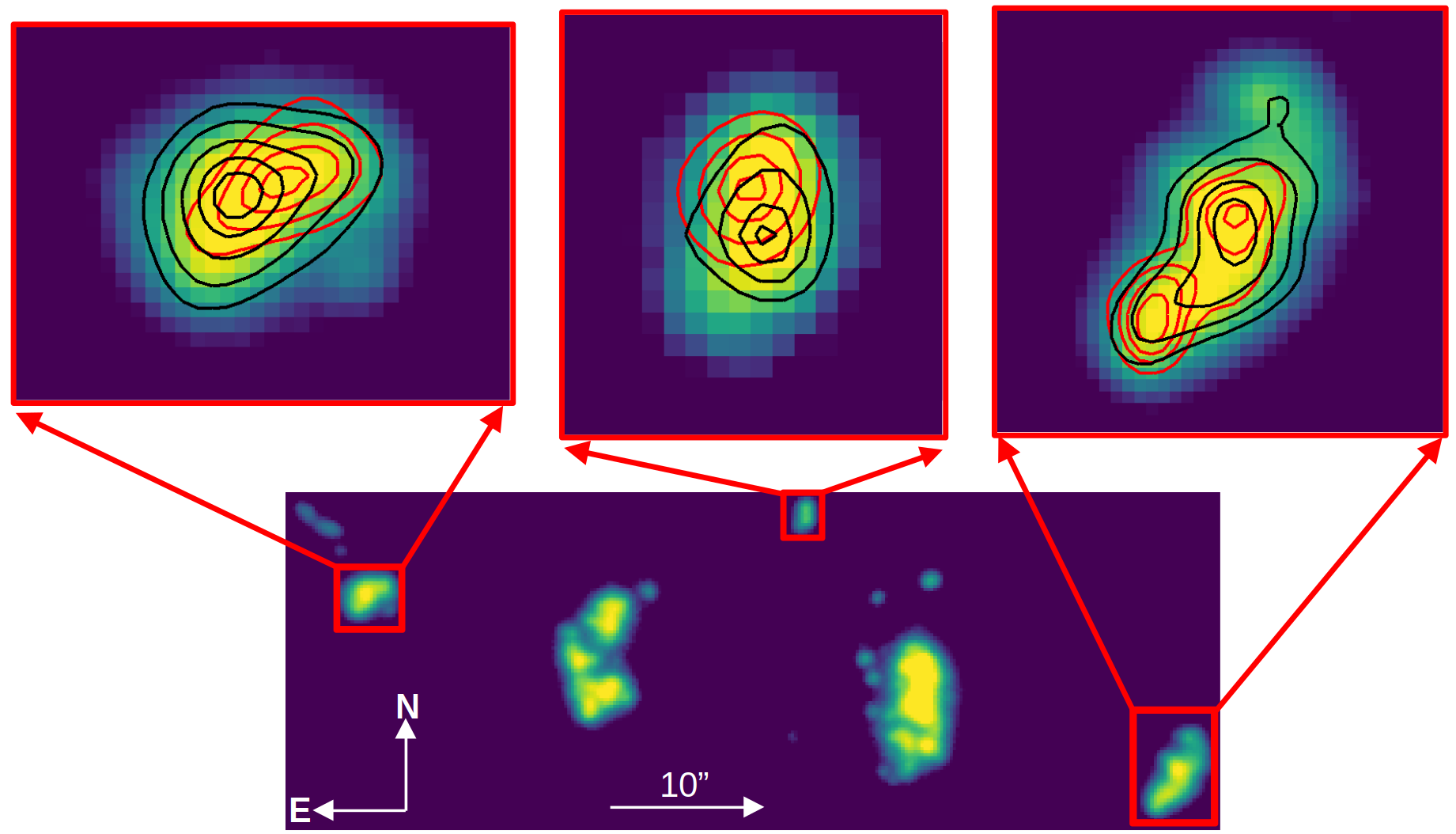}}
    \caption{NGC~7009 [O~{\sc i}] 6300~$\AA$ emission map with [C~{\sc i}] 8727~$\AA$ (black) and [Fe~{\sc ii}] 8617~$\AA$ (red) contours displayed on top of it.}
    \label{ngc7009_contours}
\end{figure}

Similarly to the case of NGC~3242, a spatial offset between the [Fe~{\sc ii}] 8617~$\AA$ and [C~{\sc i}] 8727~$\AA$ lines have been found in the outer pair of LIS as well as the northern clump. In particular, the peak of [Fe~{\sc ii}] 8617~$\AA$ is located 0.4\arcsec~closer to the central star in the E LIS (K1), and 0.5\arcsec~further away in the northern clump (see Fig.~\ref{ngc7009_contours}). As mentioned in Sect. \ref{sec_3242}, additional observations are needed in order to clarify whether ionization stratification or dynamical effects are responsible for this offset.

\subsection{NGC~3132}
\begin{figure*}[h!]     
    \centering{\includegraphics[width=1\textwidth]{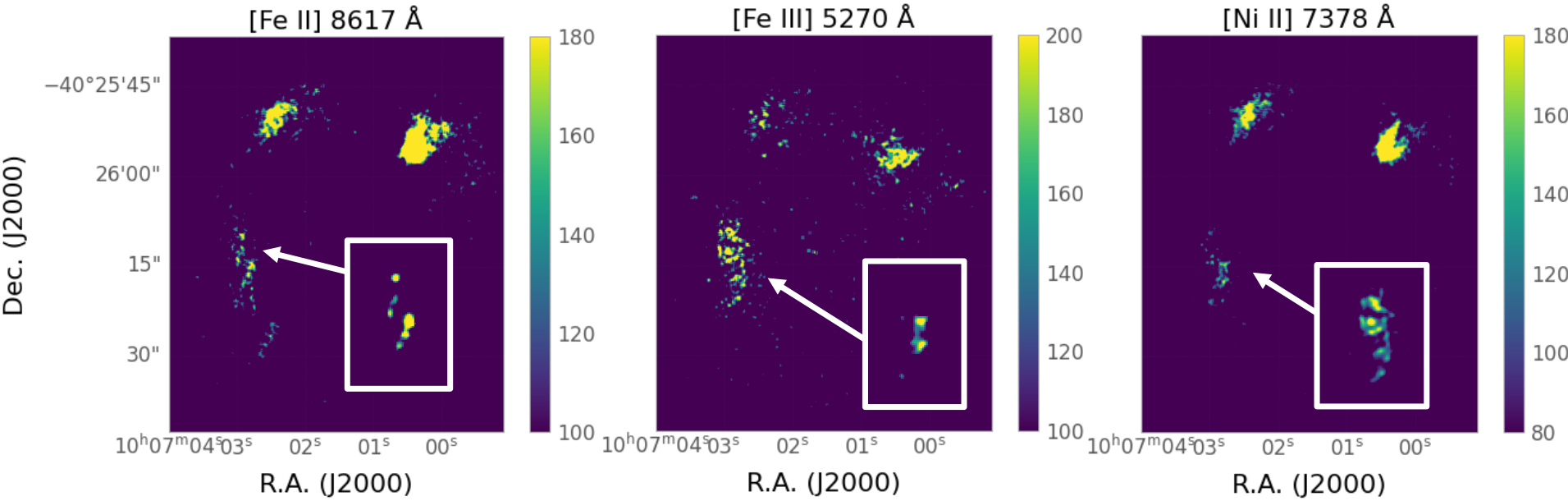}}
    \caption{NGC~3132 smoothed emission-line maps of [Fe~{\sc ii}] 8617~$\AA$, [Fe~{\sc iii}] 5270~$\AA,$ and [Ni~{\sc ii}] 7378~$\AA$. White boxes represent zoomed-in views of smoothed and higher contrast versions of the new detected regions. Color-bar values are in units of 10$^{-20}$ erg s$^{-1}$ cm$^{-2}$.}
    \label{ngc3132_text}
\end{figure*}

Emission lines from iron and nickel have been detected in three new nebular structures in NGC~3132 \citep[Fig.~\ref{ngc3132_text}; see also][]{bouvis2025}. The NW clump, the brightest one, also shows signs of weak [Ca~{\sc ii}] 7291~$\AA$ emission, with a very low signal-to-noise ratio of $\sim$ 1.3. These emission lines probably indicate shock excitation in these regions, but photoionization cannot be ruled out yet. The [C~{\sc i}]~8727~$\AA$ line is only observed at the rim of the nebula (Fig. \ref{ngc3132_append}), but not in the [Fe~{\sc ii}]- and [Ni~{\sc ii}]-emitting regions.

\subsection{IC~418}
\begin{figure*}[h!]     
    \centering{\includegraphics[width=1\textwidth]{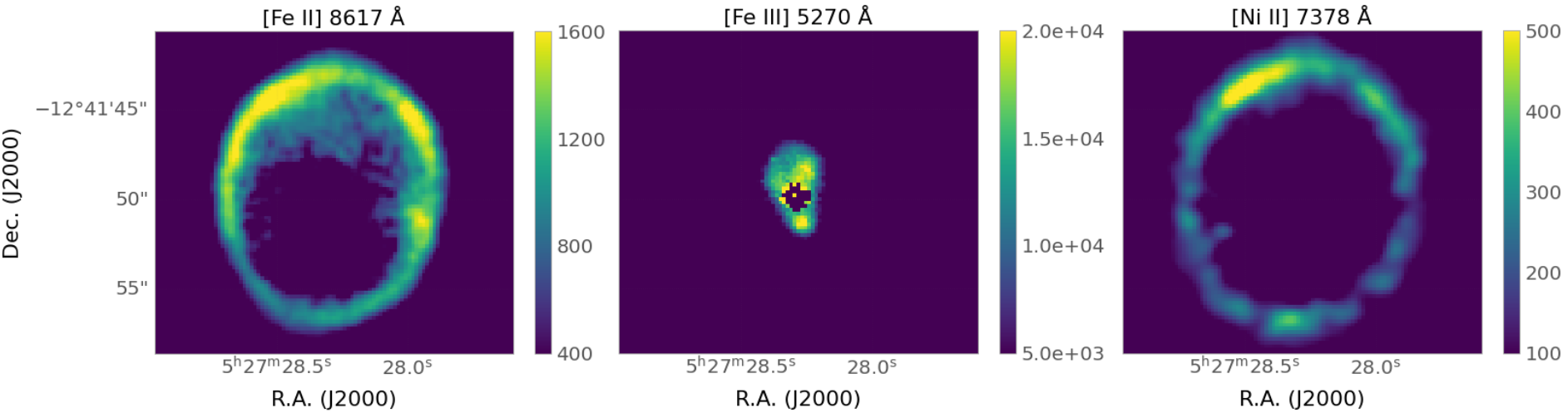}}
    \caption{IC~418 emission-line maps of [Fe~{\sc ii}] 8617~$\AA$, [Fe~{\sc iii}] 5270~$\AA,$ and [Ni~{\sc ii}] 7378~$\AA$. [Ni~{\sc ii}] and [Fe~{\sc ii}] emission-line maps are smoothed and scaled down 2 times from the original. Color-bar values are in units of 10$^{-20}$ erg s$^{-1}$ cm$^{-2}$.}
    \label{ic418_text}
\end{figure*}

IC~418 is characterized by a simple spherical morphology and it has been extensively studied by \citet{ibero2022}. At the edge of the spherical shell some iron, nickel, calcium, and carbon emission lines are detected (Figs \ref{ic418_text}, \ref{ic418_append}). Higher ionization lines, such as doubly ionized iron, are clearly observed closer to the central star.

\subsection{NGC~6369}
\begin{figure*}[h!]     
    \centering{\includegraphics[width=0.75\textwidth]{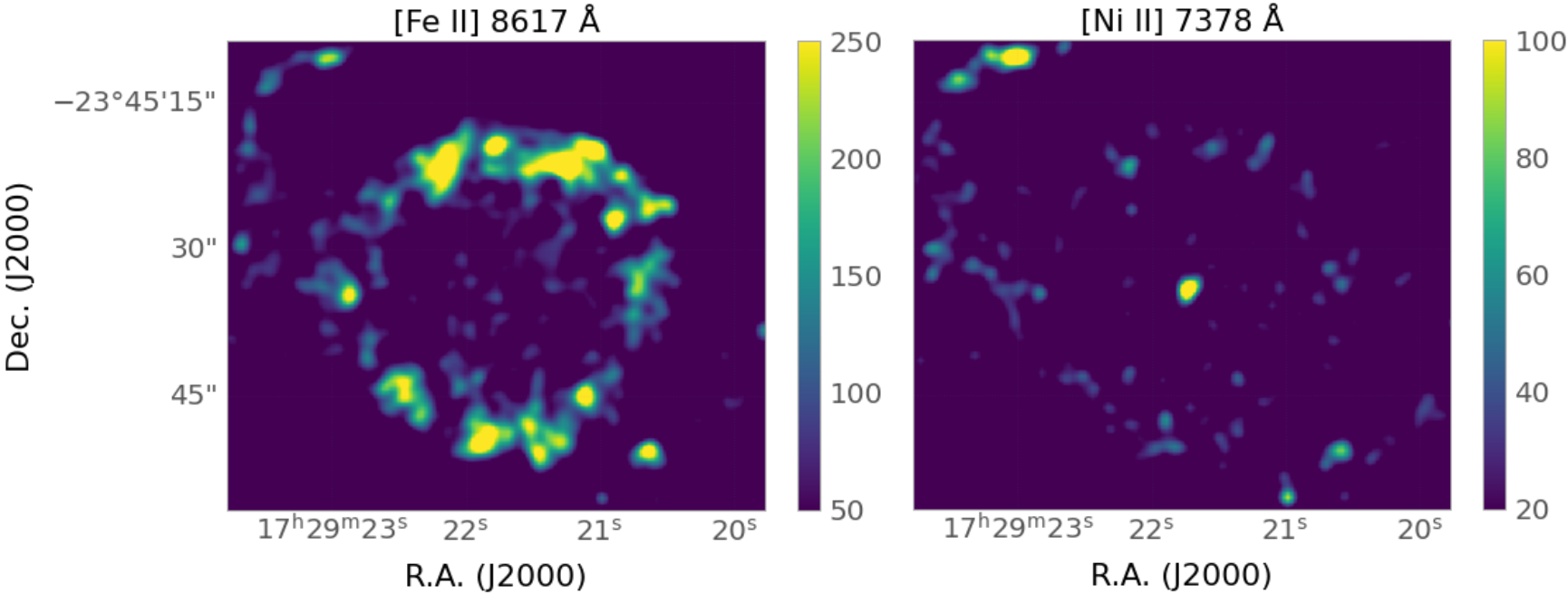}}
    \caption{NGC~6369 emission-line maps of [Fe~{\sc ii}] 8617~$\AA$ and [Ni~{\sc ii}] 7378~$\AA$. The emission-line maps are smoothed and scaled down 2 times from the original. Color-bar values are in units of 10$^{-20}$ erg s$^{-1}$ cm$^{-2}$.}
    \label{ngc6369_text}
\end{figure*}

NGC~6369 exhibits a complex structure characterized by a fragmented bright rim around the central star surrounded by some smaller filamentary structures. [Fe~{\sc ii}] 7155~$\AA$ and 8617~$\AA$, [Ni~{\sc ii}]~7378~$\AA,$ and [C~{\sc i}] 8727~$\AA$ emission lines were detected in this nebula (Figures \ref{ngc6369_text} and \ref{ngc6369_append}). It is worth noting that singly ionized iron and nickel seem to co-exist with atomic carbon and oxygen. 

\subsection{NGC~6563}
Beyond the typical emission lines of PNe, [C~{\sc i}] 8727~$\AA$ was also detected in NGC~6563 (Fig.~\ref{ngc6563_append}). Most of the emission lines originate from the rim of the nebula.

\subsection{IC~4406}
\begin{figure}[h!]     
   \centering{\includegraphics[width=0.5\textwidth]{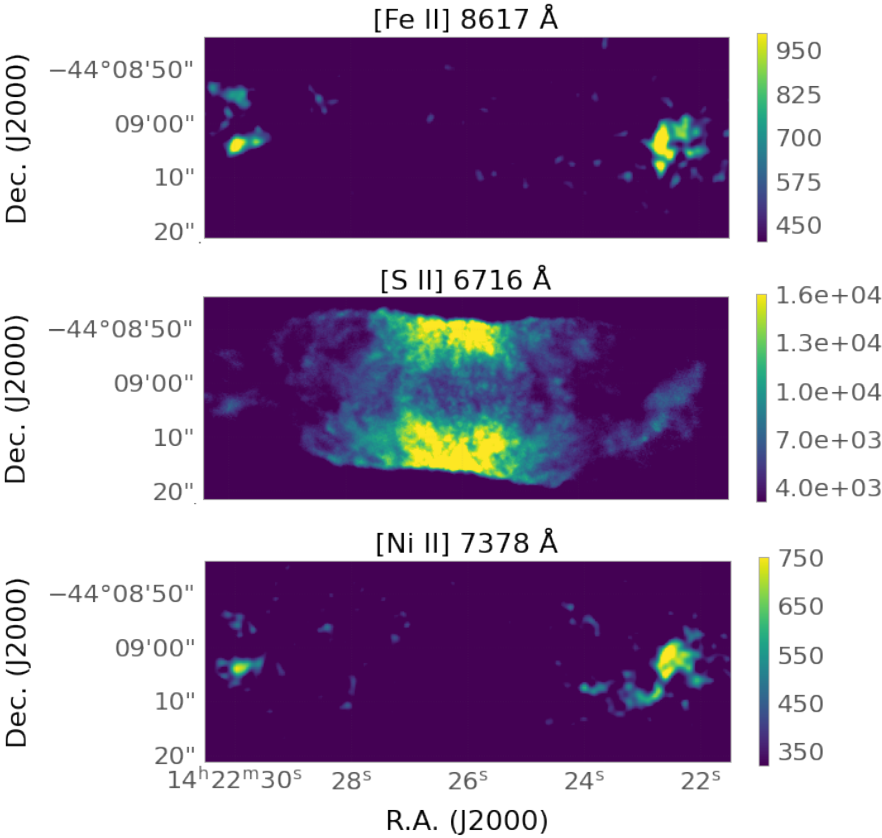}}
   \caption{IC~4406 emission-line maps of [Fe~{\sc ii}] 8617~$\AA$, [S~{\sc ii}] 6716~$\AA,$ and [Ni~{\sc ii}] 7378~$\AA$. [Fe~{\sc ii}] and [Ni~{\sc ii}] emission-line maps are smoothed and scaled down 2 times from the original. Color-bar values are in units of 10$^{-20}$ erg s$^{-1}$ cm$^{-2}$.}
   \label{ic4406_text}
\end{figure}

Three new clumps were unveiled at the eastern and western edges of the ellipsoidal PN IC~4406. [Fe~{\sc ii}] 7155~$\AA$, 8617~$\AA,$ and [Ni~{\sc ii}] 7378~$\AA$ emission lines emanate directly from the clumps (Fig.~\ref{ic4406_text}), while [C~{\sc i}] 8727~$\AA$ and [O~{\sc i}] 6300~$\AA$ emissions are detected at the central part of the nebula (Fig.~\ref{ic4406_append}). The clumps are also bright in the typical low-ionization lines, such as [N~{\sc ii}] and [S~{\sc ii}].

\subsection{Tc~1}
\begin{figure*}[h!]     
    \centering{\includegraphics[width=1\textwidth]{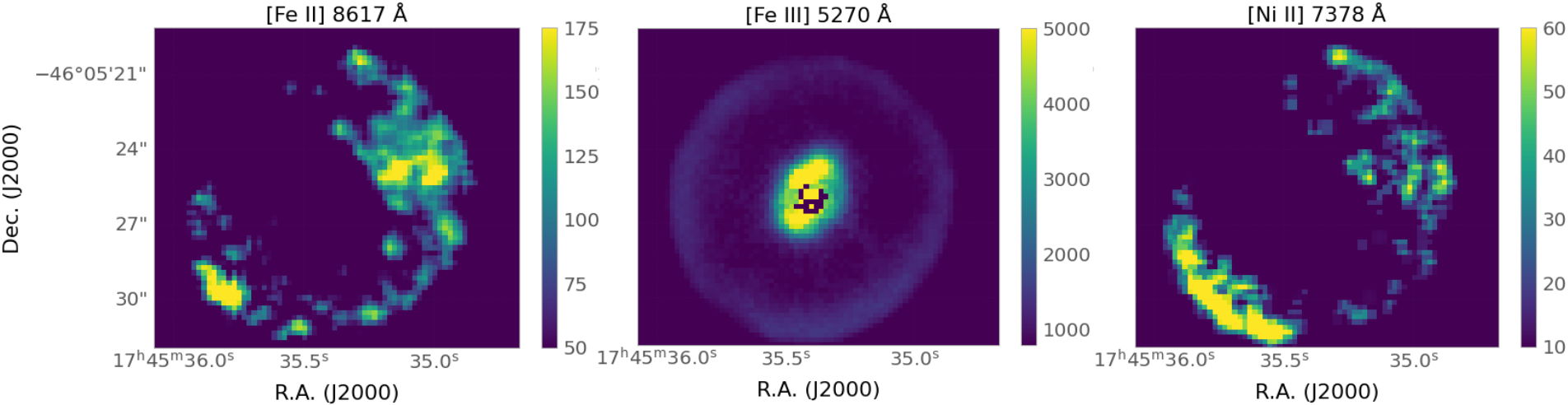}}
    \caption{Tc~1 emission-line maps of [Fe~{\sc ii}] 8617~$\AA$, [Fe~{\sc iii}] 5270~$\AA,$ and [Ni~{\sc ii}] 7378~$\AA$. [Ni~{\sc ii}] and [Fe~{\sc ii}] emission-line maps are smoothed. Color-bar values are in units of 10$^{-20}$ erg s$^{-1}$ cm$^{-2}$.}
    \label{tc1_text}
\end{figure*}

Our analysis of this nebula shows a large knotty and filamentary halo, surrounding the main nebula (Fig.~\ref{tc1_append}). Emission lines from heavy elements were not detected in the halo, instead they are concentrated in the main spherical shell \citep{otsuka2014,aleman2019}, emanating from small knots. [Fe~{\sc iii}] 4881~$\AA$ and 5270~$\AA$ lines emanate from the very central region of the nebula, while [Fe~{\sc ii}] 5158~$\AA$, 7155~$\AA$ and 8617~$\AA$, [Ni~{\sc ii}] 7378~$\AA,$ and [Ca~{\sc ii}]~7291~$\AA$ originate from the outer knotty shell (Fig.~\ref{tc1_text}). None of these emission lines have been previously reported for this nebula.

\section{Discussion}
\label{section5}
\subsection{Correlation of Ni and Fe clumps with PN types}
\label{sec:corr}
Our analysis of the available MUSE data from a sample of 12 PNe has revealed emission of heavy elements such as Fe and Ni in several clumps. The use of IFU data offers us a unique opportunity to explore the spatial distribution of the [Fe~{\sc ii}] 5158~$\AA$, 5262~$\AA$, 7155~$\AA,$ and 8617~$\AA$ and [Ni~{\sc ii}] 7378~$\AA$ emission lines for the first time in PNe. [Ni~{\sc ii}] and [Fe~{\sc ii}] lines were detected in eight out of 12 PNe, and in every case they were found to be co-spatial, verifying the close correlation between these ions \citep{bautista1996}. 

In Table \ref{statistic_table}, we mention the detection or non-detection of the [Ni~{\sc ii}] and [Fe~{\sc ii}] lines in different types of PNe. In this table, a summary of our results is also provided, where various PNe types are analyzed. In our study, the term “LIS” refers to discrete, low-ionization regions that are prominent in [N~{\sc ii}] and [S~{\sc ii}] emission lines, while “Molecule-Rich” PNe denote nebulae where fullerenes, CO, or H$_2$ emissions have been detected. PNe marked with an asterisk (*) are identified as binary candidates based on GAIA measurements \citep{chorney2021, Ali2023}. Additionally, high-ADF PNe (>5) are likely associated with the presence of a binary system \citep{wesson2018}. Notably, some studies suggest that LISs are byproducts of common-envelope evolution in binary PNe \citep{miszalski2009, Jones2017}, a process that can produce soft X-ray emission \citep{kastner2012, freeman2014}.

According to our analysis, Fe- and Ni-rich clumps are found in molecule-rich PNe, X-ray emitting PNe, and, in some cases, PNe with LISs. Moreover, PNe that host a binary system at their center appear to produce such clumps, but more data are needed in order to reach a solid statistical conclusion. Lastly, two PNe of our sample (NGC~6369 and IC~4406), where [Fe~{\sc ii}] and [Ni~{\sc ii}] lines were detected, host a Wolf-Rayet-type central star.  

\begin{table}[]
\centering
\caption{Detection of [Ni~{\sc ii}] and [Fe~{\sc ii}] in different PN types.}
\begin{tabular}{|c|l|l|}
\hline
Category                                 & \multicolumn{1}{c|}{PNe}& \multicolumn{1}{c|}{{[}Ni~{\sc ii}{]} \& {[}Fe~{\sc ii}{]}}\\ \hline
\multirow{4}{*}{PNe with LISs}           & NGC~3242                & \multicolumn{1}{c|}{\checkmark}                          \\ \cline{2-3} 
                                         & NGC~6153                & \multicolumn{1}{c|}{\checkmark}                          \\ \cline{2-3} 
                                         & NGC~7009                & \multicolumn{1}{c|}{\checkmark}                          \\ \cline{2-3} 
                                         & M~1-42                   & \multicolumn{1}{c|}{X}                                   \\ \hline
\multirow{5}{*}{Molecule-Rich PNe}       & NGC~3132                & \multicolumn{1}{c|}{\checkmark}                          \\ \cline{2-3} 
                                         & IC~418                  & \multicolumn{1}{c|}{\checkmark}                          \\ \cline{2-3} 
                                         & NGC~6369                & \multicolumn{1}{c|}{\checkmark}                          \\ \cline{2-3} 
                                         & IC~4406                 & \multicolumn{1}{c|}{\checkmark}                          \\ \cline{2-3} 
                                         & Tc~1                    & \multicolumn{1}{c|}{\checkmark}                          \\ \hline
\multirow{5}{*}{X-ray emission PNe}      & NGC~3242                & \multicolumn{1}{c|}{\checkmark}                          \\ \cline{2-3} 
                                         & NGC~6153                & \multicolumn{1}{c|}{\checkmark}                          \\ \cline{2-3} 
                                         & NGC~7009                & \multicolumn{1}{c|}{\checkmark}                          \\ \cline{2-3} 
                                         & NGC~6369                & \multicolumn{1}{c|}{\checkmark}                          \\ \cline{2-3} 
                                         & IC~418                  & \multicolumn{1}{c|}{\checkmark}                          \\ \hline
\multirow{6}{*}{Binary CSPNe}            & NGC~6153*               & \multicolumn{1}{c|}{\checkmark}                          \\ \cline{2-3} 
                                         & Hf~2-2                  & \multicolumn{1}{c|}{X}                                   \\ \cline{2-3} 
                                         & NGC~6778                & \multicolumn{1}{c|}{X}                                   \\ \cline{2-3} 
                                         & NGC~3132                & \multicolumn{1}{c|}{\checkmark}                          \\ \cline{2-3} 
                                         & NGC~6369*               & \multicolumn{1}{c|}{\checkmark}                          \\ \cline{2-3} 
                                         & IC~4406*                & \multicolumn{1}{c|}{\checkmark}
                                                   \\ \cline{2-3} 
                                         & Tc~1*                   & \multicolumn{1}{c|}{\checkmark}     
                                         \\ \hline
\end{tabular}
\label{statistic_table}
\tablefoot{PNe with asterisks indicate binary candidates based on Gaia measurements.}
\end{table}

\subsection{Doubly versus singly ionized iron}
\label{fe2_fe3}

The unique capability of IFU data to extract emission-line maps makes it possible to examine the spatial distribution of [Fe~{\sc iii}] and [Fe~{\sc ii}] emissions. A filamentary structure that extended along the SE-NW direction was unveiled in [Fe~{\sc iii}] 5270~$\AA$, probably associated with a hidden jet-like feature in NGC~3242 (Fig.~\ref{ngc3242_text}). We note that the SE and NW clumps coincide with the k2 and k4 clumps from \cite{munoz2015} and are likely linked to the interaction between the jet-like structure and the rim. The [Fe~{\sc iii}] 5270~$\AA$ structure  in NGC~6153 is orientated at a higher position angle (150$\degr$) than the [Fe~{\sc ii}] 8617~$\AA$ emitting regions (115$\degr$) (Fig.~\ref{ngc6153_text}). In the case of NGC~3132, the [Fe~{\sc iii}] 5270~$\AA$ line peaks closer to the nebula's center compared to [Fe~{\sc ii}] lines, which is expected given its higher ionization potential. Notably, the SE clump appears brighter and more extended in the doubly ionized iron emission map than in the singly ionized one (Fig.~\ref{ngc3132_text}). As for NGC~7009, the K2 and K3 clumps (the inner pair of LISs) appear noticeably different in the [Fe~{\sc iii}] and [Fe~{\sc ii}] lines (Fig.~\ref{ngc7009_text}). Once again, the former lines show a peak closer to the central star and cover a wider area than the latter. The spatial distribution of [Fe~{\sc ii}] in Tc~1 (Fig.~\ref{tc1_text}) matches the knots at the edge of the spherical shell detected in [O~{\sc i}] (Fig.~\ref{tc1_append}), while [Fe~{\sc iii}] emission is concentrated in the inner region around the central star. Lastly, a similar [Fe~{\sc ii}] and [Fe~{\sc iii}] line distribution is found in IC~418 (Fig.~\ref{ic418_text}). 

The observed spatial offset between the singly and doubly ionized iron lines provides clues about their different origins and nature. In addition, emission from singly ionized iron alone is sufficient to compute the total iron abundance in the clumps, as higher-ionization ions are absent (see Sect. \ref{subsec:ngc7009}). If both [Fe~{\sc ii}] and [Fe~{\sc iii}] are excited by shocks, the presence of [Fe~{\sc iii}] emission indicates a faster shock, provided that the mechanical energy is sufficient both to release iron from dust and to ionize it twice. 

\subsection{Ionic, atomic, and molecular stratification of the clump}
\label{subsec:clumps_ion}

\begin{figure}[h!]     
    \centering{\includegraphics[width=1\columnwidth]{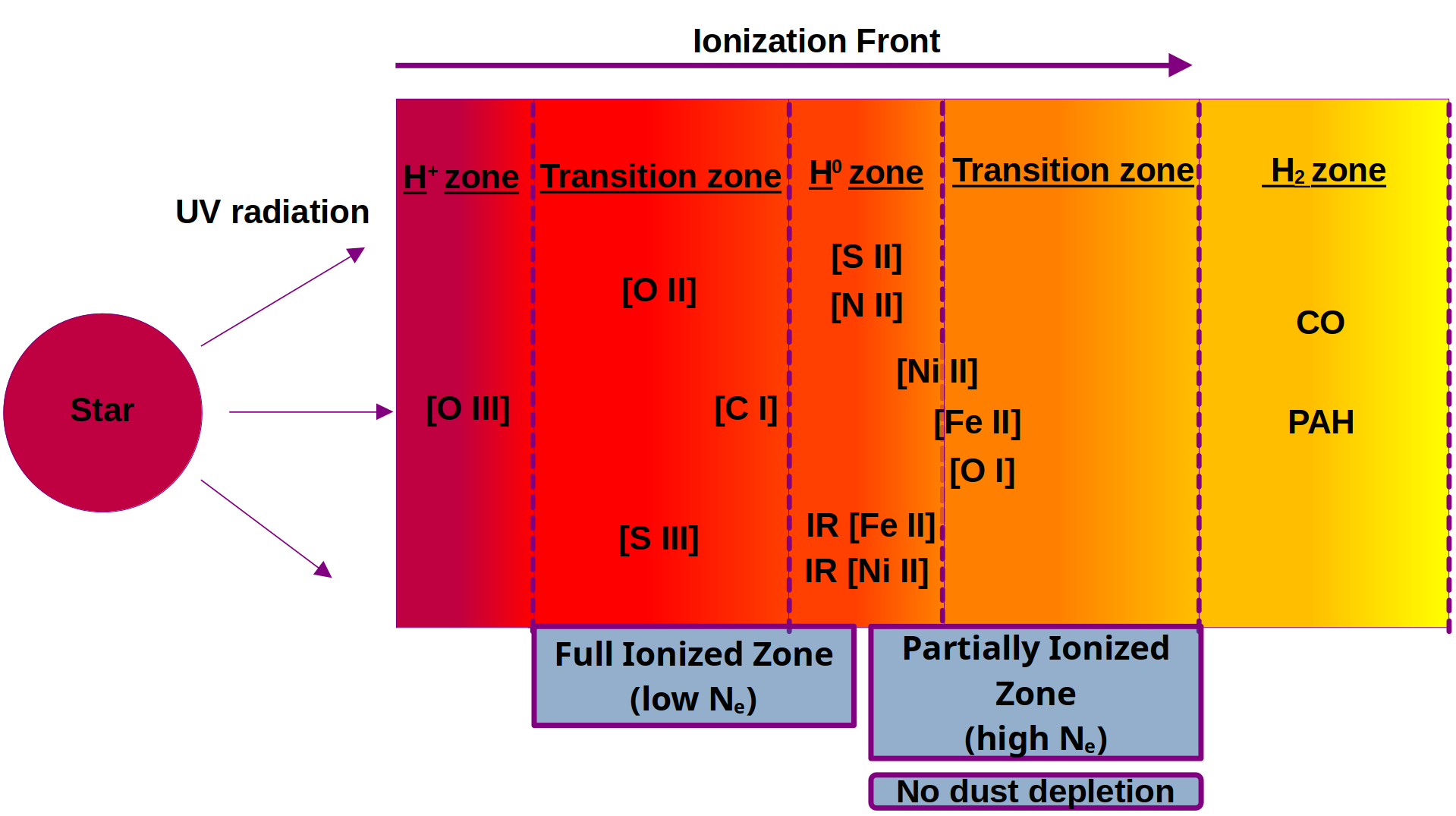}}
    \caption{Ionization structure of dense clump illuminated by UV radiation. The zone sizes are not to scale.}
    \label{sketch}
\end{figure}

The expected ionic stratification of clumps illuminated by the UV radiation from the central stars of PNe is illustrated in Fig.~\ref{sketch}. High-ionization emission lines, such as [O~{\sc iii}], emanate from the H$^{+}$ zone, followed by the [O~{\sc ii}] and [S~{\sc iii}] lines, which arise near and/or within the H$^{+}\rightarrow$~H$^{0}$ transition zone. \citet{macalpine2007} suggested that [C~{\sc i}] emission also arises from this transition zone. Deeper into the clump, [S~{\sc ii}] and [N~{\sc ii}] lines predominantly originate from the H$^{0}$ region, while [Ni~{\sc ii}], [Fe~{\sc ii}], and [O~{\sc i}] arise from the H$^{0}$ zone and the H$^{0} \rightarrow$ H$_2$ transition zone \citep{hudgins1990}. Finally, CO and possibly PAHs are present in the H$_2$ zone \citep{bosman2015}.

The low-density fully ionized zone (FIZ) extends from the H$^{0}$ zone to H$^{+}\rightarrow~$H$^{0}$ transition zone, where strong [S~{\sc ii}] and [N~{\sc ii}] lines originate \citep{bautista1996}. [N~{\sc ii}] and [O~{\sc i]} are emitted by different regions due to the fact that N$^{+}$ requires slightly higher energy photons ($\sim$14.5 eV) than the O$^{+}$ ($\sim$13.6~eV). Infrared lines of [Ni~{\sc ii}] (e.g., 1.191 $\upmu$m) and [Fe~{\sc ii}] (e.g., 1.257~$\upmu$m and 1.644 $\upmu$m), which have lower critical densities compared to the optical ones, originate from the H$^{0}$ zone and lower density gas (FIZs) \citep{bautista1998}.

In contrast, the high-density, partially ionized zones \citep[$\sim$10$^6$ cm$^{-3}$; 10 times denser than FIZs;][]{bautista1998} extend from H$^{0}$ region to H$^{0} \rightarrow$ H$_2$ transition zone, where [Ni~{\sc ii}], [Fe~{\sc ii}], and [O~{\sc i}] lines originate \citep{bautista1996}. The [Ni~{\sc ii}] emission can also extend further into the FIZ than the [Fe~{\sc ii}] emission due to its slightly higher ionization potential \citep[18.2 eV vs. 16.2 eV;][]{lucy1995}. Previous studies on PIZs indicate that the abundance ratio Fe$^+$/O$^0$ is comparable with the Fe/O ratio, suggesting the dominance of Fe$^+$ and O$^0$ ions in PIZs \citep{bautista1996}. This implies that iron depletion in PIZs is likely negligible compared to other ions \citep{bautista1998}.

The spatial offset between the [Fe~{\sc ii}] and [C~{\sc i}] emissions found in NGC~3242 and NGC~7009 LISs supports the aforementioned ionization structure of UV-illuminated clumps (Figs.~\ref{ngc3242_contours} and \ref{ngc7009_contours}, respectively). Although the presence of iron in the LISs may suggest a shock interaction, line ratios ([N~{\sc ii}]/H$\upalpha$, [S~{\sc ii}]/H$\upalpha$) from \citet{lydia2025} and \citet{akras2022} indicate a photo-dominated gas for these cases or only a minor contribution from weak shockwaves. For the remaining PNe in our sample, no spatial offset was measured due to the low spatial resolution of the data or the orientation of the nebulae. 


\subsection{[Ni~{\sc ii}] and [Fe~{\sc ii}] emission-line ratios}
\label{subsec:Ni_Fe_ratios}

Given the fact that nickel and iron are co-spatial in gaseous nebulae and possess similar ionization potentials, we explore a possible correlation between these lines' fluxes. \citet{bautista1996} explored the correlation between the log([Ni~{\sc ii}]~7378~$\AA$/H$\upalpha$) and log([Fe~{\sc ii}]~8617~$\AA$/H$\upalpha$) line ratios for a sample of Seyfert galaxies, SNRs, HH objects, and Orion. Their analysis unveiled a one-to-one relationship, including only one photo-dominated source, Orion. Hence, we decided to reconstruct and reexamine the aforementioned ratios, incorporating our PNe findings as well as recent data from the literature \citep{brugel1981,dennefeld1983,dennefeld1986,russel1990,osterbrock1992,rudy1994,rodriguez2001,williams2003,fang2011,jorge2013,giannini2015,dopita2018}. In total, data from 21 PNe (12 from our study), six HH objects, Orion, and 11 SNRs were used.

\begin{figure}[h!]     
    \centering{\includegraphics[width=0.45\textwidth]{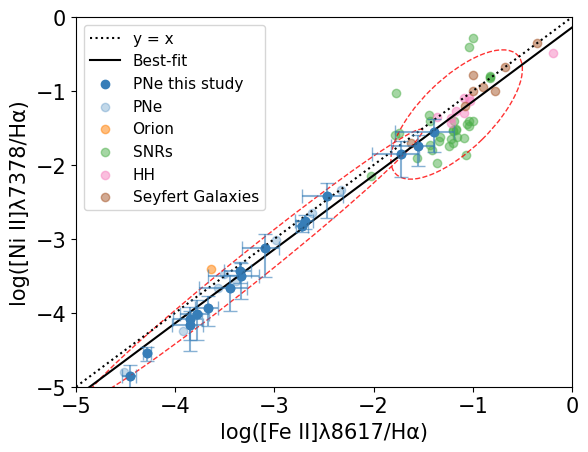}}
    \caption{log([Ni~{\sc ii}] 7378~$\AA$/H$\upalpha$) versus log([Fe~{\sc ii}] 8617~$\AA$/H$\upalpha$) for SNRs (green), HH (pink), Orion (orange), Seyfert galaxies (brown), PNe (light blue), and PNe from this study (blue). Confidence ellipses (dashed red ellipses) include 86.47$\%$ (2$\sigma$) of PNe-Orion data points and SNRs-HH-Seyfert galaxy data points.}
    \label{Ni_Fe_ratio}
\end{figure}

The linear correlation between [Ni~{\sc ii}] and [Fe~{\sc ii}] is demonstrated in Fig.~\ref{Ni_Fe_ratio}. Using a least-squares fitting method, the best-fit line is $y = x - 0.14$, with a goodness of fit for the linear regression of $R^2=0.95$. Our best-fit line is slightly different from the 1:1 relation reported by \citet{bautista1996}. Furthermore, the correlation coefficient (i.e., Pearson's r) for the entire sample is $r=0.97$, while for the PNe-Orion subset it is $0.99$, and for the SNRs-HH-Seyfert galaxies' subset it is $0.70$. This suggests that the former subset exhibits a higher dispersion than the latter.

\begin{figure}[h!]     
    \centering{\includegraphics[width=0.46\textwidth]{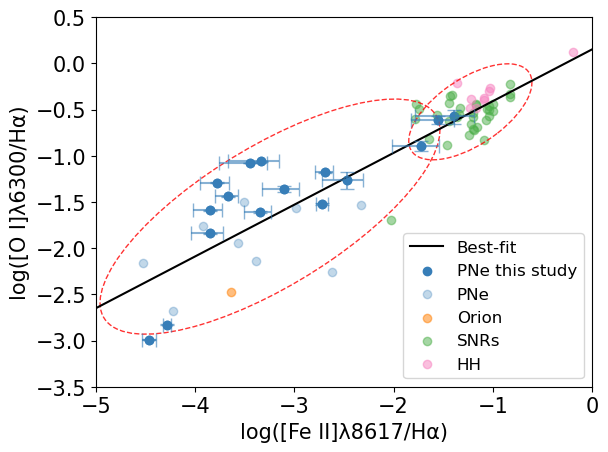}}
    \caption{log([O~{\sc i}] 6300~$\AA$/H$\upalpha$) versus log([Fe~{\sc ii}] 8617~$\AA$/H$\upalpha$) for SNRs (green), HH (pink), Orion (orange), PNe (light blue), and PNe from this study (blue). Confidence ellipses (dashed red ellipses) include 86.47$\%$ (2$\sigma$) of PNe-Orion data points and SNR-HH data points.}
    \label{O_Fe_ratio}
\end{figure}

Figure~\ref{O_Fe_ratio} illustrates the log([O~{\sc i}] 6300~$\AA$/H$\upalpha$) versus log([Fe~{\sc ii}] 8617~$\AA$/H$\upalpha$) diagram, where a strong correlation between [O~{\sc i}] and [Fe~{\sc ii}] is also visible. Once again, we obtained a best fit of $y = 0.56 x + 0.15$, with a goodness of fit for the linear regression of $R^2=0.81$. The correlation coefficient of the entire sample is $r=0.90$, while for the PNe-Orion and SNRs-HH subsets, it is significantly lower: $r=0.73$ and $r=0.61$, respectively. These results reflect the higher dispersion of both subsets in the log([O~{\sc i}] $\lambda$6300/H$\alpha$) line ratio. We attribute this increased dispersion to poor sky background subtraction, which is a known issue of MUSE data \citep{weil2020}.

The error bars in both diagrams represent the fitting procedure error (Sect.~\ref{3.1}) and should be considered lower limits. Additionally, 2$\upsigma$ confidence ellipses (86.47$\%$) for the emission lines of the two subsets are also shown in both diagrams by red dashed lines (Figs.~\ref{Ni_Fe_ratio} and \ref{O_Fe_ratio}).  

Overall, it is evident that the two subsets ---SNRs-Seyfert galaxies-HH objects and PNe-Orion--- occupy distinct regions. We argue that these diagrams are very useful for distinguishing shock-dominated and photo-dominated regions.

\subsection{Machine-learning clustering and classification}
\label{sec:ML}
\begin{figure}[h!]     
    \centering{\includegraphics[width=0.5\textwidth]{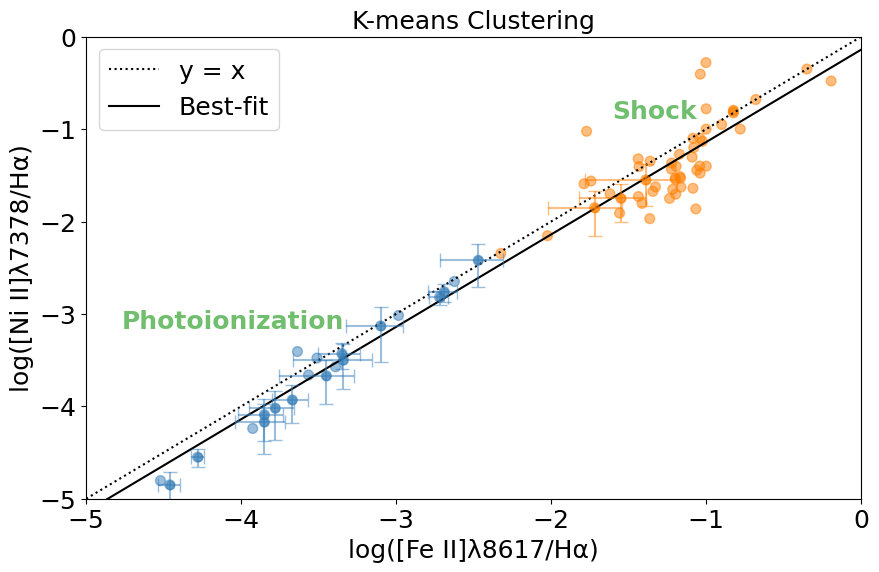}}
    \caption{K-means clustering for log([Ni~{\sc ii}] 7378~$\AA$/H$\upalpha$) versus log([Fe~{\sc ii}] 8617~$\AA$/H$\upalpha$) diagram. Sources classified as shock-dominated are shown in orange, while photoionized sources are shown in blue. The least-squares best-fit line for all data is represented by a solid black line, and the one-to-one correlation is indicated by a dashed black line.}
    \label{ni_fe_k-means}
\end{figure}

To verify our previous conclusions on the dominated excitation mechanisms in the SNRs-Seyfert galaxies-HH objects and PNe-Orion subsets, the unsupervised machine-learning algorithm K-Means \citep{kmeans} was used for clustering our sample. The K-Means algorithm is a type of clustering method that divides data into a predefined number of clusters (in this case two) based on feature similarity. The algorithm works by assigning each data point to the nearest cluster centroid and iteratively updating the centroids until convergence. The resulting shock cluster contains all the HH, Seyfert galaxies, SNRs, and the clumps from IC~4406, while the photoionization cluster includes the Orion, PNe, and the remaining PNe clumps (Fig.~\ref{ni_fe_k-means}). To assess the performance of the clustering, we manually split it into three folds, and for each fold we estimated a confusion matrix and all the following metrics. Finally, we present the average confusion matrix (Fig.~\ref{kmeans_conf_matrix}) and metrics of each fold:
\begin{itemize}
    \item The silhouette score ($\sim$0.74) ranges from $-$1 to 1, where a value closer to 1 indicates well-separated and well-formed clusters.
    \item The Davies-Bouldin index ($\sim$0.35) measures the average similarity ratio of each cluster to its most similar one, with lower values indicating better separation.
    \item The Calinsky-Harabasz index ($\sim$119) evaluates the ratio of the sum of between-cluster dispersion to within-cluster dispersion, with higher values indicating better clustering.
\end{itemize}
Excluding IC~4406 values (see Sect.~\ref{5.7}), we conclude that log([Ni~{\sc ii}]~7378~$\AA$/H$\upalpha$) and log([Fe~{\sc ii}]~8617~$\AA$/H$\upalpha$) $<$ $-$2.20 is the upper limit for photo-dominated regions and the lower limit for shock-dominated regions. 

We also examined the theoretical log([O~{\sc i}] 6300~$\AA$/H$\upalpha$) versus log([Fe~{\sc ii}] 8617~$\AA$/H$\upalpha$) diagram by employing the predictions from UV and shock models. Data from the Mexican Million Models database \citep[3Mdb\footnote{3Mdb (photoionization part) is publicly available at \href{https://sites.google.com/site/mexicanmillionmodels/}{https://sites.google.com/site/mexicanmillionmodels/}.},][]{morisset2015} were employed. We used photoionization models from project PNe$^*$, which were generated using {\sc cloudy} 17.01 \citep{cloudy2017}. $T_{\rm eff}$ and L for the central source were adopted from \citet{delgado_inglada_2014}, considering a blackbody energy distribution, the same solar abundances (log(O/H)=-3.36), and a constant density law (see Table \ref{3mdb_uv_params}).

The grid of shock models \citep{3mdb_shock2019}\footnote{3Mdb (shock part) is publicly available at \href{http://3mdb.astro.unam.mx:3686/}{http://3mdb.astro.unam.mx:3686/}.} was created using {\sc mappings} \citep{mappings2008, mappings2013, mappings2017}. We selected a grid of complete and incomplete shock models, which are further categorized as fast and slow shocks \citep[for slow shocks, see][]{alarie2019} with the following properties: (i) solar abundances for fast shocks \citep{allen2008} and PNe abundances for the slow shocks \citep[][]{delgado_inglada_2014}; (ii) a magnetic-field strength <10 $\upmu$G; (iii) pre-shock densities of 10-10$^4$ cm$^{-3}$ for slow shocks and 10-10$^3$ cm$^{-3}$ for fast shocks; (iv) pre-shock temperatures < 15\,000 K; (v) cut-off temperatures < 15\,000 K (for incomplete shocks only); and (vi) fast- and slow-shock velocities of 100-1\,000 km/s and 10-100 km/s, respectively (see Table \ref{3mdb_shock_params}). Similar parameters were employed for PNe analysis in \citet{mari2023b}.

We recall that iron and nickel are primarily locked in dust grains in the interstellar medium. Fast shocks are expected to destroy dust much more efficiently, resulting in highly depleted heavy elements in the gas phase. Since shock models do not account for dust destruction \citep{allen2008}, the gas-phase abundance of heavy elements, rather than the total (dust + gas) abundance, should be used as input in the shock models. Hence, solar abundances were selected for the fast-shock models, as the subsolar abundances (similar to those observed in PNe) failed to reproduce iron emission for HH objects and SNRs. In HH objects and SNRs environments, where high-velocity shocks are prevalent, significant amounts of iron are released from dust grains into the gas phase \citep{DS96}. Consequently, the solar value \citep[log(Fe/H)=$-$4.63,][]{DS96, allen2008}\footnote{According to \citet{DS96}, the log(Fe/H)=-4.63 includes a 0.2 dex depletion factor.} is more representative than the subsolar PNe abundance \citep[log(Fe/H)=$-$6.55,][]{allen2008,dopita2005}. Overall, the solar abundance reproduces the observed iron emission in HH objects and SNRs. 

\begin{figure}[h!]     
    \centering{\includegraphics[width=1.1\columnwidth]{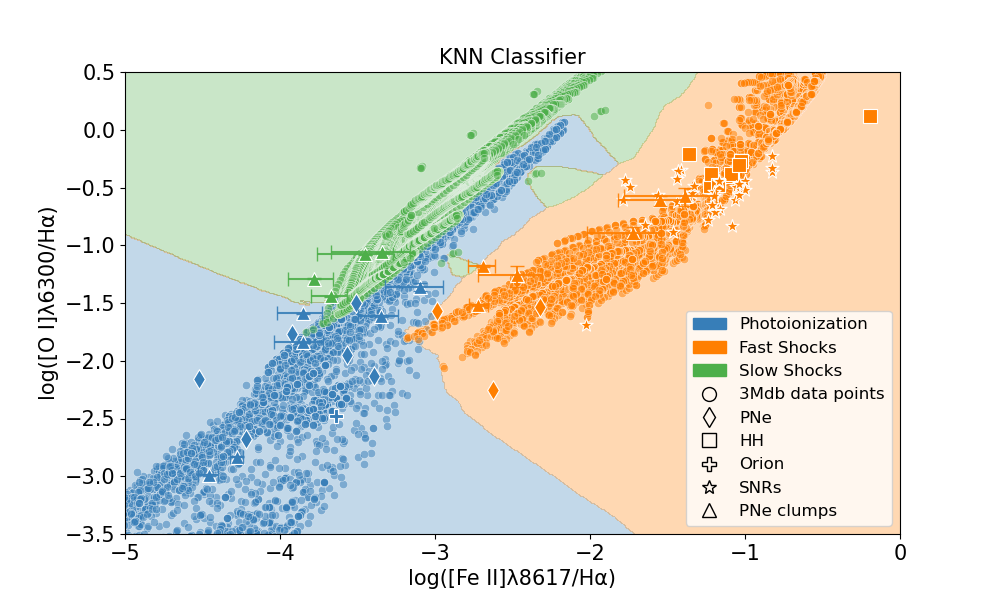}}
    \caption{KNN classifier algorithm results for log([O~{\sc i}] 6300~$\AA$/H$\upalpha$) versus log([Fe~{\sc ii}] 8617~$\AA$/H$\upalpha$) diagram. Different colors correspond to different excitation mechanisms, while different markers represent different kinds of objects.}
    \label{O_Fe_KNN}
\end{figure}

The supervised K-nearest neighbors \citep[KNN;][]{KNN} machine-learning algorithm was also used to train a model with the 3MdB data and five-fold cross-validation. Then, the model was employed in order to classify the data of our sample. The accuracy of the algorithm is 0.98, while other metrics of the algorithm such as precision, recall, and f1-score are nearly 1 for fast- and slow-shock classes and 0.96, 0.88, and 0.92 for the photoionization class, respectively (confusion matrix in Fig. \ref{knn_conf_matrix}), indicating that the sample is well separated into three different subgroups (Fig.~\ref{O_Fe_KNN}).

All the SNRs and HH objects are located in the fast-shock region. On the other hand, Orion, NGC~3132 SE, NGC~3242 clumps, NGC~6369 SW clump, Tc~1 clumps, He~2-86, M~1-25, IC~418, Pe~1-1, M~1-61, and NGC~7009 \citep[from][]{fang2011} are well placed in the photoionization region. Cn~1-5; M~1-32; M~1-91; the NGC~3132 NW and NE clumps; the NGC~6369 NE clump; and the IC~4406 clumps are located in the fast-shock region, while NGC~6153 clumps and NGC~7009 clumps lie in the slow-shock region. It is worth noting that most of the clumps in NGC~3242, NGC~6153, NGC~7009, and NGC~6369 match the intense [O~{\sc iii}]/H$\upalpha$ regions identified by \citet{guerrero2013}, further supporting the shock scenario. We therefore deduce that PNe clumps are more likely excited by shocks (either slow or fast). However, potential contamination of H$\upalpha$ emission by the host nebula may lead to lower line ratios, resulting in their misclassification as photo-dominated regions. Last but not least, in all clumps classified as slow shocks, [C~{\sc i}] 8727$\AA$ is detected.

\subsection{[Fe~{\sc ii}] emission lines as an $n_{\rm e}$ diagnostic}
\label{subsec:Dens_diag}

\begin{figure}[h!]     
    \centering{\includegraphics[width=0.48\textwidth]{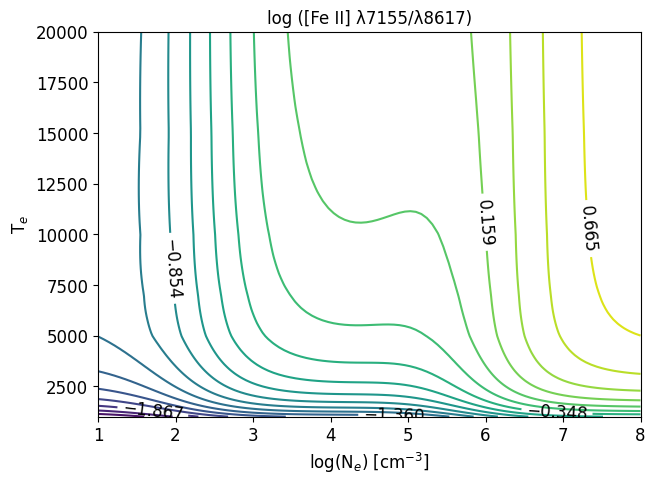}}
    \caption{log([Fe~{\sc ii}] 7155~$\AA$/8617~$\AA$) intensity ratio for different $T_{\rm e}$ and $n_{\rm e}$ using B15 atomic data. This diagram is similar for TZ18 atomic data as well.}
    \label{B15_pyneb}
\end{figure}

The [Fe~{\sc ii}] 7155~$\AA$/8617~$\AA$ intensity ratio has previously been used to derive electron density in the range of 10$^2$-10$^7$ cm$^{-3}$ \citep{bautista1998,bautista2015}. Thus, we reconstructed a grid of [Fe~{\sc ii}] line ratios for a wide range of $n_{\rm e}$ and $T_{\rm e}$ values employing {\sc PyNeb} (Fig.~\ref{B15_pyneb}). It is evident that the [Fe~{\sc ii}] ratio is $T_{\rm e}$ depended on the $n_{\rm e}$ range 10$^{3.5}$-10$^6$~cm$^{-3}$. Therefore, this line ratio should only be used for $n_{\rm e}$ determination below 10$^{3.5}$ cm$^{-3}$ or above 10$^6$ cm$^{-3}$.

\begin{figure}[h!]     
    \centering{\includegraphics[width=0.45\textwidth]{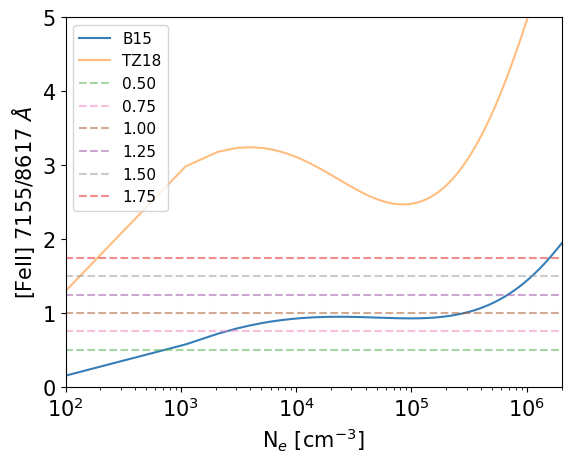}}
    \caption{$n_{\rm e}$ calculated from [Fe~{\sc ii}] 7155~$\AA$/8617~$\AA$ intensity ratio for $T_{\rm e}$=10.000 K using atomic data from B15 (blue lines) and TZ18 (orange line). Typical observed values of the ratio are shown by dashed colored lines.}
    \label{Fe2_den}
\end{figure}

The atomic data from \citet[][hereafter B15]{bautista2015} yields densities higher by a factor of 10$^2$-10$^4$ compared to those derived with the atomic data from \citet[][hereafter TZ18, recommended by \citet{mendoza2023}]{TZ2018} for typical observed ratio values (Fig.~\ref{Fe2_den}). Due to the complex nature of iron, it is really challenging to model it accurately, and small variations in the atomic data result in significant $n_{\rm e}$ differences \citep[see also][]{mendoza2023}. The high uncertainties in the [Fe~{\sc ii}] 7155~$\AA$/8617~$\AA$ ratio for our sample do not allow us to provide robust $n_{\rm e}$ estimates. As a reference, we provide the $n_{\rm e}$ estimated by the mean value of the [Fe~{\sc ii}] 7155~$\AA$/8617~$\AA$ ratio for the clumps in each PN of our sample: NGC~3242 $\sim$ 4$\cdot$10$^3$~cm$^{-3}$, NGC~6153 $\sim$ 3$\cdot$10$^5$~cm$^{-3}$, NGC~7009 $\sim$ 3$\cdot$10$^3$~cm$^{-3}$, NGC~3132 $\sim$ 10$^3$~cm$^{-3}$, NGC~6369 $\sim$ 6$\cdot$10$^5$~cm$^{-3}$, IC~4406 $\sim$ 5$\cdot$10$^2$~cm$^{-3}$, and Tc~1 $\sim$ 10$^6$~cm$^{-3}$.
Recent JWST observations of the Crab nebula have revealed filamentary structures rich in nickel, with an electron density of $n_{\rm e}$$\sim$3\,000 cm$^{-3}$, assuming $T_{\rm e}$=2\,400~K and infrared [Fe~{\sc ii}] diagnostic lines \citep{crab_jwst}.


\subsection{Nickel-and-iron abundance ratio}
\label{5.7}

In gaseous nebulae where Ni$^+$ and Fe$^+$ emission lines are prominent, the Ni/Fe abundance ratio is often found to deviate from the solar value. More specifically, super-solar Ni/Fe abundance ratios have been reported in previous studies \citep[e.g.,][]{henry1988,macalpine1989,hudgins1990,jerkstrand2015B,crab_jwst}. Super-solar Ni/Fe abundance ratios have also been estimated by \citet{bautista1996}, \citet{dennefeld1986}, \citet{maguire2018}, and \citet{liu2023} under the assumption that Ni/Fe=Ni$^+$/Fe$^+$, since singly ionized atoms are expected to dominate partially ionized regions. Higher ionization state ions were either absent (e.g., Ni$^{+2}$) or originated from different structures (e.g., Fe$^{+2}$).

Nonetheless, one must be careful when relying on the Ni/Fe=Ni$^+$/Fe$^+$ assumption for determining the gas-phase abundance ratios. In particular, one must consider the exposure of clumps to the hard UV radiation from hot central stars and/or shocks, both of which can result in non-negligible fractions of doubly ionized iron and nickel. For this reason, before employing the aforementioned assumption, we: 1) compared the spatial distribution of singly and doubly ionized iron emission lines (see Sect. \ref{fe2_fe3}); 2) thoroughly tested the validity of this assumption in a case study (see Sect. \ref{subsec:ngc7009}); and 3) searched for a potential detection of doubly ionized nickel emission lines in MUSE wavelength coverage.

\begin{figure}[h!]     
    \centering{\includegraphics[width=0.45\textwidth]{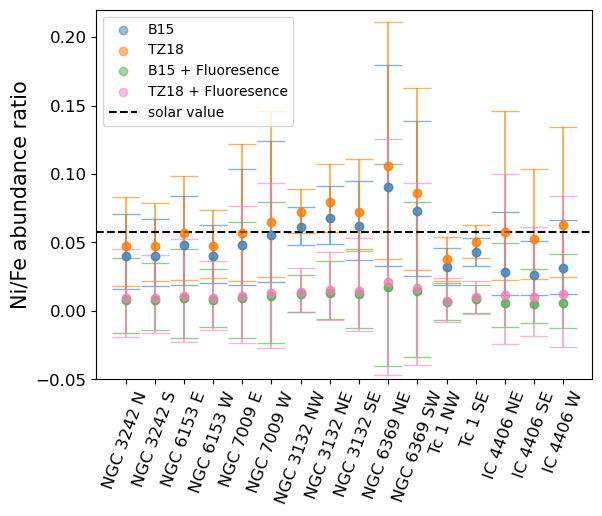}}
    \caption{Ni/Fe abundance ratio for each clump in our sample of PNe. All the abundances have been calculated for $T_{\rm e}$=10\,000~K and $n_{\rm e}$=1\,000 cm$^{-3}$ and Chianty v10.1 atomic data for nickel. The dashed black horizontal line corresponds to the iron solar abundance from \cite{lodders2010}. Green and blue dots were estimated utilizing B15 atomic data, and the former account for fluorescence in nickel emission. Similarly, the pink and orange dots represent TZ18 atomic data.}
    \label{Ni_Fe_abund}
\end{figure}

For our sample of PNe clumps, the Ni/Fe abundance ratios were derived from the optical lines and the {\sc PyNeb} package, considering iron atomic data from B15 and TZ18, and nickel atomic data from Chianty v10.1 \citet{chianti}. Assuming $T_{\rm e}$~=~10\,000~K and $n_{\rm e}$ =1\,000~cm$^{-3}$ \citep{akras2022,mari2023b, mari2023}, the overall Ni/Fe ratio is found to be close to the solar value \citep[0.058,][]{lodders2010} (orange and blue dots in Fig.~\ref{Ni_Fe_abund}).

\begin{figure}[h!]     
    \centering{\includegraphics[width=0.45\textwidth]{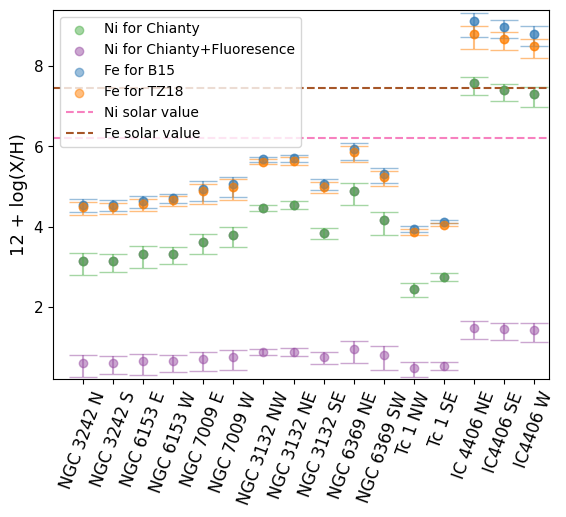}}
    \caption{Ni and Fe abundances, assuming Ni/H=Ni$^+$/H$^+$ and Fe/H=Fe$^+$/H$^+$, respectively. All the abundances were calculated for $T_{\rm e}$=10\,000 K and $n_{\rm e}$=1\,000 cm$^{-3}$ and Chianty v10.1 atomic data for nickel. The dashed pink and brown horizontal lines correspond to the solar abundance of nickel and iron, respectively. Green and purple dots correspond to nickel abundance, with the last ones accounting for fluorescence. Blue and orange dots corresponds to iron abundance utilizing B15 and TZ18 atomic data, respectively.}
    \label{Ni+Fe_abund}
\end{figure}

If fluorescence excitation of the [Ni~{\sc ii}] 7378~$\AA$ line is taken into account, the resulting abundances from our sample are significantly lower than the solar value (purple dot in Fig.~\ref{Ni+Fe_abund}). Similarly, Fig.~\ref{Ni_Fe_abund} shows that only in the no-fluorescence case (orange and blue dots), the Ni/Fe abundance ratio reaches its solar value, indicating that fluorescence excitation should be negligible, as expected in low-excitation regions \citep{zhang2006}. For the former calculation, we adopted $\sim$80$\%$ fluorescent contribution in [Ni~{\sc ii}] 7378~$\AA$ emission, based on models for the Orion nebula by \citet{lucy1995}. 

\citet{inglada2016} (hereafter DI16) investigated the Ni/Fe abundance ratio in PNe using the [Fe~{\sc iii}] and [Ni~{\sc iii}] optical lines. These authors argued that nickel atoms are more efficiently trapped in dust grains compared to iron atoms, which would explain their subsolar value. However, more recent studies have challenged this interpretation. \citet{crab_jwst} suggested that emission lines from doubly ionized nickel and iron underestimate the Ni/Fe abundance ratio, because Ni$^{+2}$ originates from lower temperature and lower density regions compared to Fe$^{+2}$. Additionally, \citet{mendez_delgado2021} derived unexpected Ni$^{+2}$ abundance estimates based on [Ni~{\sc iii}] emission lines and attributed them to inaccuracies in the available atomic data, which could further explain the subsolar Ni/Fe value reported by DI16. 

To ensure that the Ni/Fe abundance ratios of the clumps in our sample of PNe are reliable, we examined the extreme super-solar Ni/Fe abundance ratio ---60–75 times above the solar value--- of the Crab nebula reported by \citet{macalpine1989,macalpine2007}. For this exercise, we used {\sc PyNeb} along with the iron atomic data from B15/TZ18 and the nickel atomic data from Chianti v10.1. Assuming Ni/Fe=Ni$^+$/Fe$^+$, we found an Ni/Fe abundance ratio only 5.5-7 times above the solar value from \citet[0.058,][]{lodders2010} and 6-7.5 times from \citet[0.053,][]{scott2015}. Our values for the Crab nebula are consistent with those recently derived by \citet{crab_jwst}. 

All PNe in our sample, except for IC~4406, have Ni and Fe chemical abundances well below the solar value (Fig.~\ref{Ni+Fe_abund}). This suggests that Fe and Ni are either highly ionized  (e.g., Ni$^{+2}$ and Fe$^{+2}$) in the clumps, and the assumption Ni/Fe=Ni$^+$/Fe$^+$ leads to an underestimation of the abundances, or that both elements are depleted in dust grains. The latter scenario is the most probable, as PNe are not expected to have shocks capable of destroying large amounts of dust. Even in more extreme environments, such as HH objects, only $\sim$30-67$\%$ of dust is destroyed \citep{mesa2009,mendez-delgado2022}. Additionally, [Ni~{\sc iii}] and [Fe~{\sc iii}] are not detected in the clumps of our PNe sample.

As for IC~4406, iron and nickel abundances are significantly higher than the corresponding solar values. The faint emission lines measured in the nebula result in highly uncertain flux measurements, despite their indisputable detection. Therefore, we decided not to include IC~4406 in the following iron and nickel depletion calculations.

\begin{figure}[h!]     
    \centering{\includegraphics[width=0.45\textwidth]{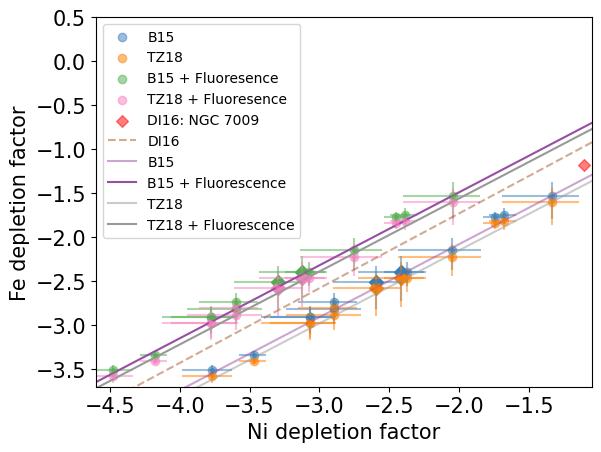}}
    \caption{Ni depletion factor versus Fe depletion factor in clump regions within PNe. Diamond shapes correspond to NGC~7009 values. All the clump abundances were calculated for $T_{\rm e}$=10\,000~K and $n_{\rm e}$=1\,000 cm$^{-3}$. Solid lines correspond to best-fit lines for different iron atomic data and fluorescent cases for nickel. The dashed line shows the best fit from DI16. Green and blue dots were estimated accounting for fluorescence (or lack thereof) in nickel using B15 atomic data. The pink and orange dots were derived with the TZ18 atomic data.}
    \label{Ni_Fe_depletion}
\end{figure}

The correlation between the Ni and Fe depletion factors\footnote{The depletion factor of an element X is defined as log(X/H)-log(X/H)$_{\odot}$.} is illustrated in Fig.~\ref{Ni_Fe_depletion} (with and without fluorescence). The slopes of the best-fit lines for our sample (0.83), obtained using the least-squares method ($R^2=0.98$ and Pearson's $r=0.99$), are in excellent agreement with previous studies (DI16:$y=0.85x-0.03$), using the same solar abundance data \citep{lodders2010}. However, our vertical intercept differs from the previous studies by -0.42 (B15) and -0.49 (TZ18) or by +0.17 (B15) and +0.1 (TZ18), depending on whether fluorescence excitation is taken into account or not, respectively. This discrepancy is probably caused by the different nickel and iron ions used in the two studies (singly ionized in our study, doubly ionized in DI16).

Overall, our sample exhibits a correlation between nickel and iron depletion factors, which vary over a wide range (spanning approximately 2.5 dex for Ni and 2 dex for Fe), which supports the validity of our abundances. We also concluded that a significant amount of Fe (and Ni) still remains locked in dust grains.

\subsection{The case study of NGC~7009}
\label{subsec:ngc7009}

NGC~7009 is an ideal source to examine Fe depletion using various techniques: (i) analyzing singly ionized lines under the assumption that Fe/H=Fe$^+$/H$^+$, (ii) combining singly and doubly ionized lines, and (iii) applying an ionization correction factor (ICF) to the doubly ionized lines. 

The outer pair of LISs \citep[K1 and K4,][]{goncalves2003} was only detected in the singly ionized lines, whereas both singly and doubly ionized lines are present in the inner pair (K2 and K3). The depletion factors for the outer LISs are $-$2.4 (K1) and $-$2.6 (K4), with a relative error of $\pm$0.3. For the inner LISs, we compute depletion factors of $-$2.3 (K2) and $-$2.6 (K3), assuming Fe=Fe$^+$+Fe$^{+2}$, $T_{\rm e}$=10\,000~K, and $n_{\rm e}$=5\,000~cm$^{-3}$ \citep{akras2022}. The agreement between the depletion factors of the inner and outer LISs supports our hypothesis that the latter are Fe$^+$ dominated, validating our assumption that Fe/H=Fe$^+$/H$^+$. If we only consider the [Fe~{\sc iii}] lines for the inner LISs and apply the ICF(Fe$^{+2}$) formulae from \citet{RR05}, the iron depletion factors are overestimated, ranging from $-$1.75 to $-$2.3.

It should be noted that our Fe depletion values are significantly lower than the value of $-$1.18 (red diamond in Fig.~\ref{Ni_Fe_depletion}) reported by DI16, which was derived using the [Fe~{\sc iii}] lines detected by \citet{fang2011}. To verify the reliability of our estimates, we recalculated the depletion using the same data \citep{fang2011} and the ICF formulae 2 and 3 from \citet{RR05}. The first ICF, based on photoionization models, yields a depletion factor of $-$1.18, consistent with the value reported by DI16. However, the second ICF, based on observational data, produces a value of $-$2.02, which is closer to our estimates.

We also computed the Fe depletion from the MUSE data by simulating the slit spectrum (position and orientation) from \citet{fang2011} using the {\sc satellite} code \citep{akras2022}. This analysis yielded Fe depletion factors of $-$2.11 and $-$2.57, based on the [Fe~{\sc iii}]~4881~$\AA$ line and the ICF formulae from \cite{RR05}, respectively. These values are consistent with our findings for both pairs of LISs. It should be noted that the observational ICF generally results in lower depletion factors compared to the theoretical ICF.

Regarding nickel, we computed a depletion factor of $-$2.3 for the outer LISs. We argue that our value is reliable due to the strong one-to-one relationship between the Fe and Ni depletion factors (Fig.~\ref{Ni_Fe_depletion}). For the inner LISs, we estimated depletion factors of $-$3.1 and $-$3.4 for K2 and K3, respectively, considering only the [Ni~{\sc ii}] 7378~$\AA$ line ([Ni~{\sc iii}] is not detected). Based on our previous analysis, we expect comparable Ni depletion factors for the inner and outer LISs, as  observed for Fe. Our Ni depletion factors are again lower than the value of $-$1.11 reported by DI16. Overall, the outer pair of LISs in NGC~7009 exhibits a high Ni$^+$/Ni$^{+2}$ ratio, making the assumption of Ni/H$\sim$Ni$^+$/H$^+, $ which is adequate for abundance estimations. 

Lastly, according to the KNN algorithms (Figs. \ref{ni_fe_k-means} and \ref{O_Fe_KNN}), the inner LISs (K2 and K3) are classified as photo-dominated, whereas the outer LISs (K1 and K4) are classified as slow-shock-dominated (Table \ref{ngc7009}). This distinction is expected, as K2 and K3, being closer to the central star, are exposed to more intense UV radiation. Therefore, the physical processes taking place in the two pairs of LISs in NGC~7009 are considerably different.

\begin{table}[h!]
\centering
\caption{Emission-line ratios including [O~{\sc i}] 6300~$\AA$, [Fe~{\sc ii}] 8617~$\AA,$ and [Ni~{\sc ii}] 7378~$\AA$ for the two pairs of LISs in NGC~7009.}
\resizebox{0.5\textwidth}{!}{%
\begin{tabular}{|c|c|c|c|c|}
\hline
   &log([O~{\sc i}]/H$\upalpha$)&log([Fe~{\sc ii}]/H$\upalpha$)&log([Ni~{\sc ii}]/H$\upalpha$)& Excitation \\ \hline
K1 & $-$1.08 & $-$3.45 & $-$3.67 & SS \\ \hline
K2 & $-$2.09 & $-$4.15 & $-$4.52 & UV \\ \hline
K3 & $-$1.72 & $-$3.71 & $-$4.20 & UV \\ \hline
K4 & $-$1.06 & $-$3.34 & $-$3.50 & SS \\ \hline
\end{tabular}
}
\tablefoot{Since H$\upalpha$ is saturated in MUSE data, we assumed a value of H$\upalpha$=2.85$\cdot$H$\upbeta$. The last column lists the excitation mechanism of each clump as UV (photoionization) and SS (slow shocks).}
\label{ngc7009}
\end{table}

\section{Conclusions}
\label{section6}

\subsection{Origin of Ni and Fe emissions in gaseous nebulae}
Heavy elements such as nickel and iron originate either from type Ia supernovae \citep[thermonuclear explosion of accreting white dwarfs; e.g., RCW 86 and Kepler's nebula;][]{dennefeld1986} or from core-collapse supernovae, where they are produced in the silicon-burning shell of massive stars \citep{jerkstrand2015A,jerkstrand2015B}. After being ejected into the interstellar medium and potentially interacting with stellar winds, they are finally condensed into dust grains \citep[e.g.,][]{barlow1978} in H~{\sc ii} regions and molecular interstellar clouds \citep[e.g.,][]{mesa2009}. 

The detection of Ni$^{+}$ and Fe$^{+}$ in gaseous nebulae can be explained by the fact that fast shocks (most commonly found in SNRs and HH objects) can efficiently destroy dust grains through thermal and nonthermal sputtering in the gas behind the shock front, as well as through grain–grain collisions \citep{hollenbach1989,jones1994,mouri2000}. Through these mechanisms, the elements are released back into the gas phase, making it possible to estimate their gas-phase abundances. Despite the fact that fast shocks effectively destroy dust, in slower shocks, dust destruction is incomplete until the recombination zone of the shock has been reached \citep{dopita2018}. For instance, \citet{mesa2009} estimated in HH202 (an HH object in the Orion nebula) that iron and nickel abundances increase by up to a factor of ten after the passage of a shockwave that destroys 30$\%$ to 50$\%$ of the dust, while \citet{mendez-delgado2022} studied HH514 and found $\sim$ 67$\%$ dust destruction.

Our analysis of 16 clumps in seven host PNe has shown that the clumps in NGC~3132 NW and NE, NGC~6369 NE, and IC~4406, where the highest Ni and Fe abundances are found, are also classified as fast-shock-dominated (Fig.~\ref{O_Fe_KNN}). These results corroborate that fast shocks are more efficient in destroying dust than slow shocks (Fig.~\ref{Ni+Fe_abund}). In the case of IC~4406, the clumps exhibit a high log([S~{\sc ii}]/H$\alpha$) ratio, ranging from $-$0.30 to $-$0.05, which further supports the shock hypothesis \citep{sab1977, leonidaki2013, kopsa2020, akras_2020_abel}. It is worth noting that ten out of 16 clumps in our sample of PNe, where Ni$^+$ and Fe$^+$ were detected, are located within confirmed or candidate binary systems (Table \ref{statistic_table}). This suggests that clump formation may be somehow associated with binary interactions.

\subsection{Clump-formation scenarios}

Two main mechanisms for clump formation have been proposed, as described by \citet{matsuura2009}. Clumps may form either during the AGB phase due to localized density enhancements or those in situ at the beginning of the PN phase.

Clumps can form by inhomogeneities in the stellar wind of the progenitor star \citep{bertoldi1989,reipurth1983,soker1998}. If these stellar winds were intrinsically knotty, probably due to binary interactions \citep{edgar2008}, the ejected clumps could evolve by merging with material previously or subsequently expelled from the central region \citep{meaburn2000}. Therefore, the outer clumps are older than the inner ones. In this scenario, the narrow [Ni~{\sc ii}] and [Fe~{\sc ii}] emission lines are expected to originate from the dense, mainly neutral knots, which are being swept up by the fast stellar wind. The broader [N~{\sc ii}] and [S~{\sc ii}] lines are associated from a bow shock surrounding the clumps, created by the clump-wind velocity difference. For example, a velocity difference of $\sim$30 km/s between [N~{\sc ii}] and [Ni~{\sc ii}] emissions has been observed in P-Cygni \citep{barlow1994}.

As a result, the broad [N~{\sc ii}] and [S~{\sc ii}] lines should appear less clumpy than the narrow [Ni~{\sc ii}] and [Fe~{\sc ii}] lines, since they arise from a larger volume \citep{meaburn2000}. Thus, observable velocity differences among various species in such clumps could help pinpoint their emission regions, whether within the molecular core or the bow shock. Furthermore, detecting distinct velocity components within clumps would support the hypothesis that clumps form through stellar-wind inhomogeneities, which in turn could indicate the presence of a binary system in the nebula. To test this hypothesis, high-spectral-resolution observations of clumps embedded in PNe are necessary to gain insights into their formation history. 

Another possible scenario for clump formation suggests that Ni$^{+}$ and Fe$^{+}$ are formed in situ, in small, dense regions that are illuminated by high-energy photons. The intense UV radiation from the central star ionizes the nebula, and the ionization front propagates through the gas. However, the ionization front is not smooth or in pressure equilibrium with the ambient gas, as hydrodynamical models have shown \citep{bedijn1981}, and this could lead to the formation of turbulent instabilities (Reyleigh-Taylor). Turbulent instabilities could also form from the interaction between the slow and fast progenitor winds \citep{garcia2006}. In this scenario, clumps situated closer to the central star are expected to be older than those located further away. 

The H atoms are ionized on the side of the gas facing the central star and emit photoelectrons toward the molecular gas (photoevaporation flow). At the same time, the temperature in this region increases along with the pressure, which forces the gas to expand. Therefore, the hot ionized gas compresses the rest of the molecular gas on the outer side, while it expands freely toward the central star. Due to conservation of momentum, the remaining cold molecular gas starts moving outwards, as in a rocket-jet system. Thus, this “rocket effect” \citep{Oort1955,mellema1998} can create highly condensed and compressed clump regions.

\section{Summary}
\label{section7}
This study focused on the presence of low-ionization Ni and Fe emission lines in planetary nebulae. Using MUSE IFU archival data, we detected nickel and iron emission lines in seven PNe. Specifically, we identified 16 clump structures rich in nickel and iron. The most prominent lines of these heavy elements were [Ni~{\sc ii}]~7378~$\AA$ and [Fe~{\sc ii}] 8617~$\AA$.

New clumps were discovered in NGC~3132 and IC~4406. [Ni~{\sc ii}] and [Fe~{\sc ii}] emission lines were found to emanate directly from the LISs in NGC~3242, NGC~6153, and NGC~7009. A spatial offset of 0.5-1\arcsec~between the [Fe~{\sc ii}] 8617~$\AA$ and [C~{\sc i}] 8727~$\AA$ was measured for the LISs of NGC~3242 and NGC~7009. The peak of the [Fe~{\sc ii}] 8617~$\AA$ line was found to be further away from the central star in the NGC~3242 clumps and the K4 clump in NGC~7009. 

From our analysis, these small, discrete, and isolated structures show a possible link with binarity, X-ray emission, and molecule-rich PNe. In some cases, they appear to be co-spatial with the LIS already detected in characteristic emission lines (e.g.,~[N~{\sc ii}]). However, we kept our sample small in order to provide significant statistical results. 

The hypothesis that nickel and iron co-exist in such clumps in a wide variety of gaseous nebulae was tested through the log([Ni~{\sc ii}]~7378~$\AA$/H$\upalpha$) versus log([Fe~{\sc ii}]~8617~$\AA$/H$\upalpha$) diagram, and a strong correlation was found. Log([Ni~{\sc ii}] 7378~$\AA$/H$\upalpha$) and log([Fe~{\sc ii}] 8617~$\AA$/H$\upalpha$) < $-$2.20 very likely define the boundary zone between shock-dominated (HH objects and SNRs) and photo-dominated nebulae (PNe and Orion). 

From the log([O~{\sc i}]~6300~$\AA$/H$\upalpha$) versus log([Fe~{\sc ii}]~8617~$\AA$/H$\upalpha$) diagram and a machine-learning-algorithm approach, distinct regimes were found to distinguish shock-dominated and photo-dominated gases. Ten out of 16 clumps in our sample were classified as shock-excited. High-velocity shocks exhibit extensive dust destruction and are characterized by higher depletion factors. 

Moreover, the electron density was examined using the [Fe~{\sc ii}] 7155~$\AA$/8617~$\AA$ diagnostic-line ratio. This ratio only seems to be sensitive to density below 10$^{3.5}$ cm$^{-3}$ or above 10$^{6}$ cm$^{-3}$. In addition, different atomic data resulted in significant variations in the results.

The Ni/Fe abundance ratio derived for the clumps in our PNe sample was calculated using different atomic data, assuming Ni/Fe$\sim$Ni$^+$/Fe$^+$ and considering the possible contribution of fluorescence in the [Ni~{\sc ii}] 7378~$\AA$. Our Ni/Fe values are systematically below or close to the solar value. This result supports the idea that nickel atoms are stuck more efficiently on dust grains than iron atoms. The fact that both Ni and Fe abundances are below their solar value suggests that either higher ionization state ions are present or that a significant fraction of nickel and iron atoms remain depleted in dust grains. Given that shocks can partially destroy dust grains depending on their velocity, the depletion scenario seems more probable. Furthermore, the absence of detectable [Fe~{\sc iii}] and [Ni~{\sc iii}] emission in PNe clumps ---at least within the MUSE optical wavelength range--- further supports the depletion scenario.

In the particular case of NGC~7009, both the inner (K2, K3) and the outer (K1, K4) pairs of LISs exhibit comparable iron and nickel depletion, as expected. The detection of doubly ionized iron in the inner LISs, while being absent in the outer LISs is likely a key factor to improving our understanding of the formation, evolution, and origin of LISs in PNe.

\begin{acknowledgements}
We are grateful to the anonymous referee for their constructive comments, which helped improve the content and the clarity of this work. We also thank Jeremy Walsh for kindly providing the reduced data cube of Tc 1. We wish to thank the "Summer School for Astrostatistics in Crete" for providing training on the statistical methods adopted in this work. The research project is implemented in the framework of H.F.R.I call “Basic research financing (Horizontal support of all Sciences)” under the National Recovery and Resilience Plan “Greece 2.0” funded by the European Union – NextGenerationEU (H.F.R.I. Project Number: 15665). JGR acknowledges financial support from the Agencia Estatal de Investigaci\'{o}n of the Ministerio de Ciencia e Innovaci\'{o}n (AEI- MCINN) under Severo Ochoa Centres of Excellence Programme 2020-2023 (CEX2019-000920-S), and from grant PID-2022136653NA-I00 (DOI:10.13039/501100011033) funded by the Ministerio de Ciencia, Innovaci\'{o}n y Universidades (MCIU/AEI) and by ERDF "A way of making Europe" of the European Union. DRG acknowledges FAPERJ (E-26/200.527/2023) and CNPq (403011/2022-1; 315307/2023-4) grants. 

The results from NGC~7009, NGC~3132, IC~418, NGC~6369, NGC~6563 and IC~4406 are based on public data released from the MUSE commissioning observations at the VLT Yepun (UT4) telescope under Programs ID 60.A-9347(A), 60.A-9100(A), 60.A-9100(H) and 60.A-9100(G), respectively. Lastly, NGC~6369 and IC~4406  analysis is based on data obtained from the ESO Science Archive Facility with DOI: https://doi.org/10.18727/archive/41.
\end{acknowledgements}

\bibliographystyle{aa} 
\bibliography{biblio} 

\begin{appendix}
\onecolumn
\section{Supplementary emission-line maps}
\label{supplementary_maps}
\begin{figure*}[h!]     
    \centering{\includegraphics[width=0.9\textwidth]{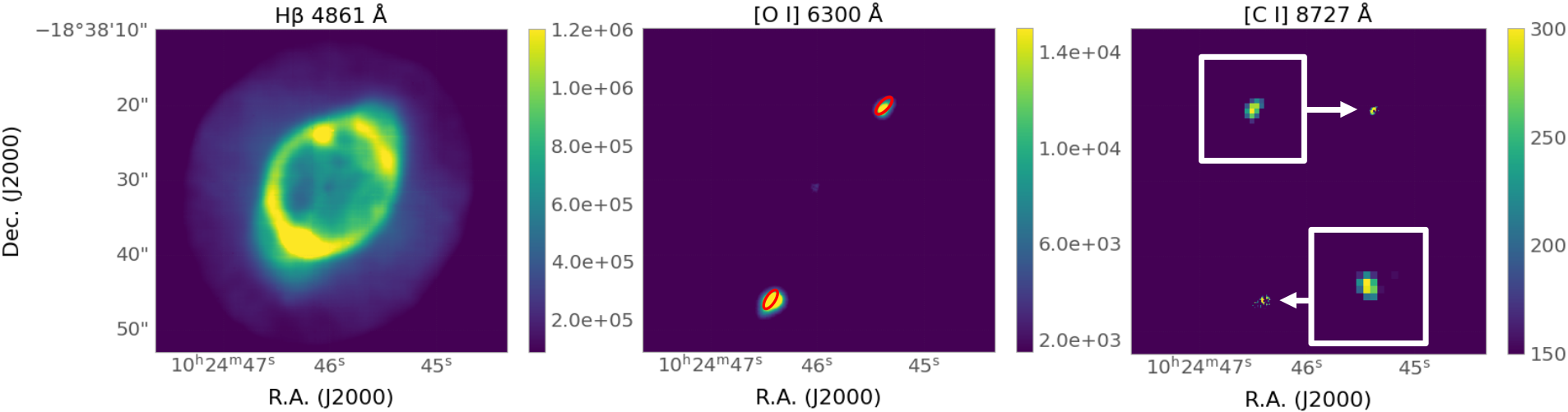}}
    \caption{NGC~3242 emission-line maps of H$\upbeta$, [O~{\sc i}] 6300~$\AA$ and [C~{\sc i}] 8727~$\AA$. The regions where fluxes were measured are displayed on the [O~{\sc i}] 6300~$\AA$ map. Color-bar values are in units of 10$^{-20}$ erg s$^{-1}$ cm$^{-2}$.}
    \label{ngc3242_append}
\end{figure*}

\begin{figure*}[h!]     
    \centering{\includegraphics[width=0.9\textwidth]{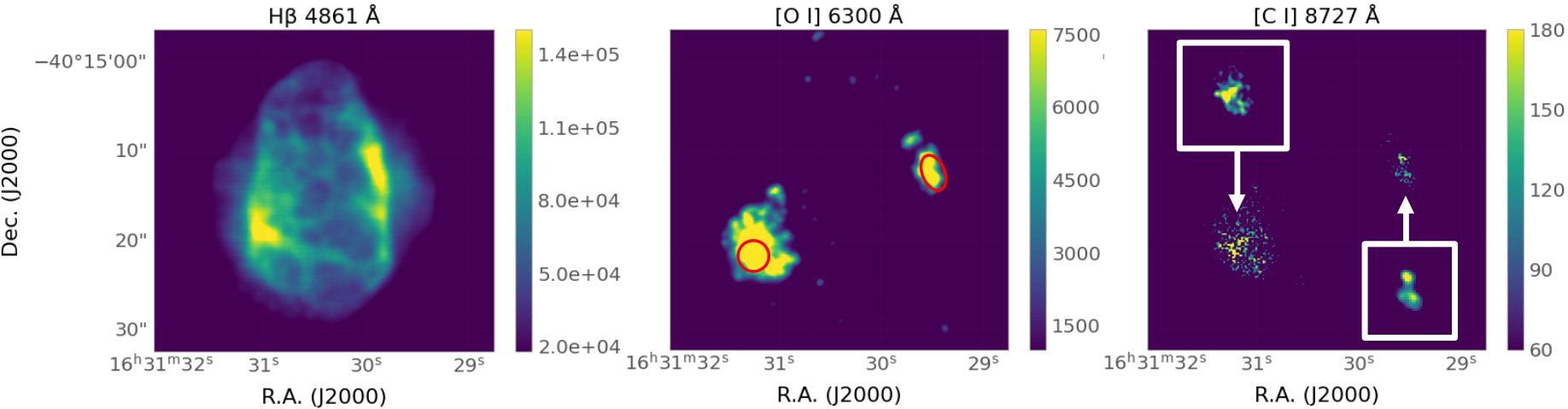}}
    \caption{Same as Fig.~\ref{ngc3242_append} for NGC~6153.}
    \label{ngc6153_append}
\end{figure*}

\begin{figure*}[h!]     
    \centering{\includegraphics[width=0.6\textwidth]{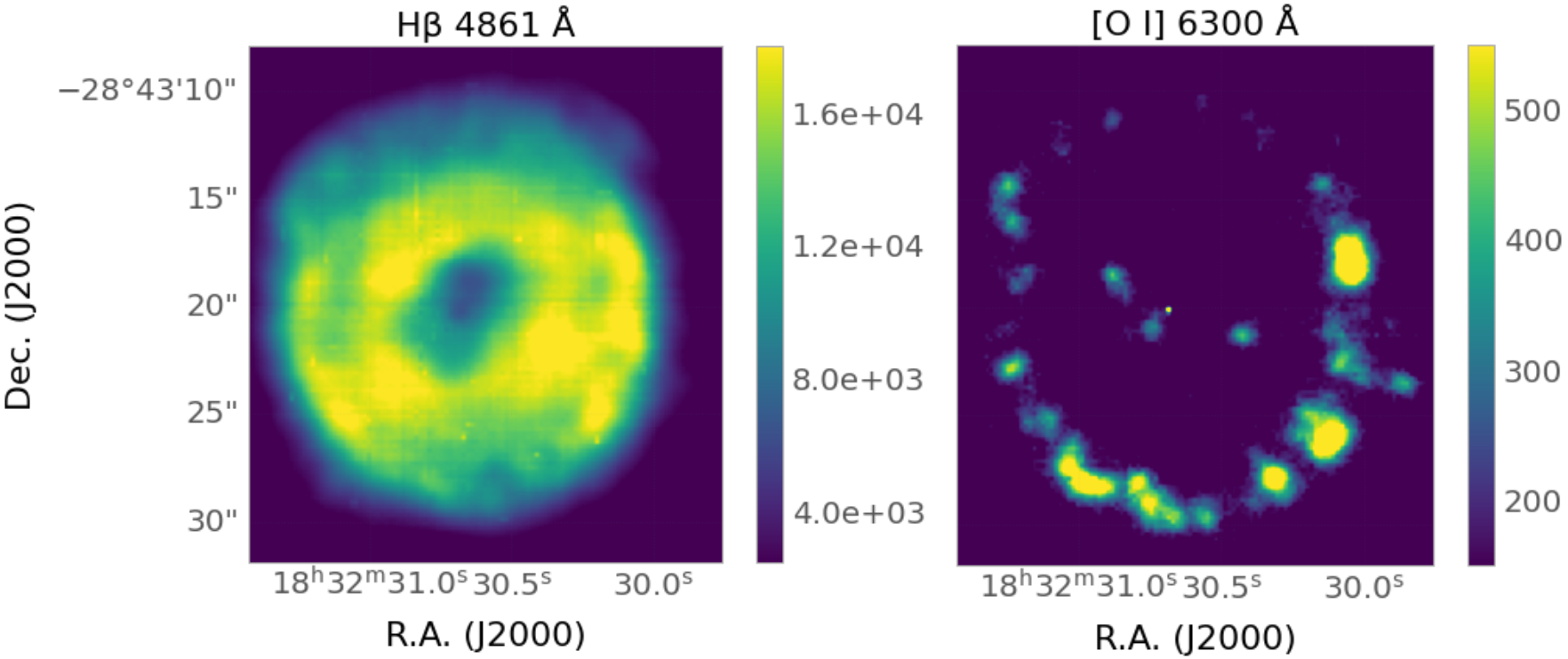}}
    \caption{Same as Fig.~\ref{ngc3242_append} for Hf~2-2, without [C~{\sc i}] 8727~$\AA$.}
    \label{hf2-2_append}
\end{figure*}

\begin{figure*}[h!]     
    \centering{\includegraphics[width=0.8\textwidth]{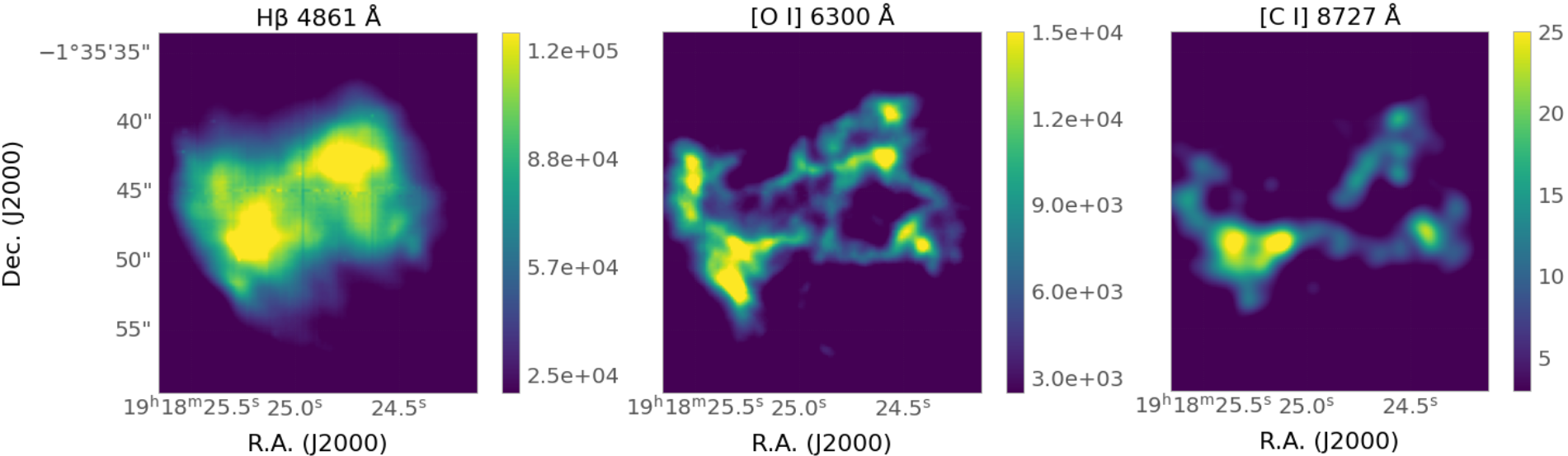}}
    \caption{Same as Fig.~\ref{ngc3242_append} for NGC~6778.}
    \label{ngc6778_append}
\end{figure*}

\begin{figure*}[h!]     
    \centering{\includegraphics[width=0.9\textwidth]{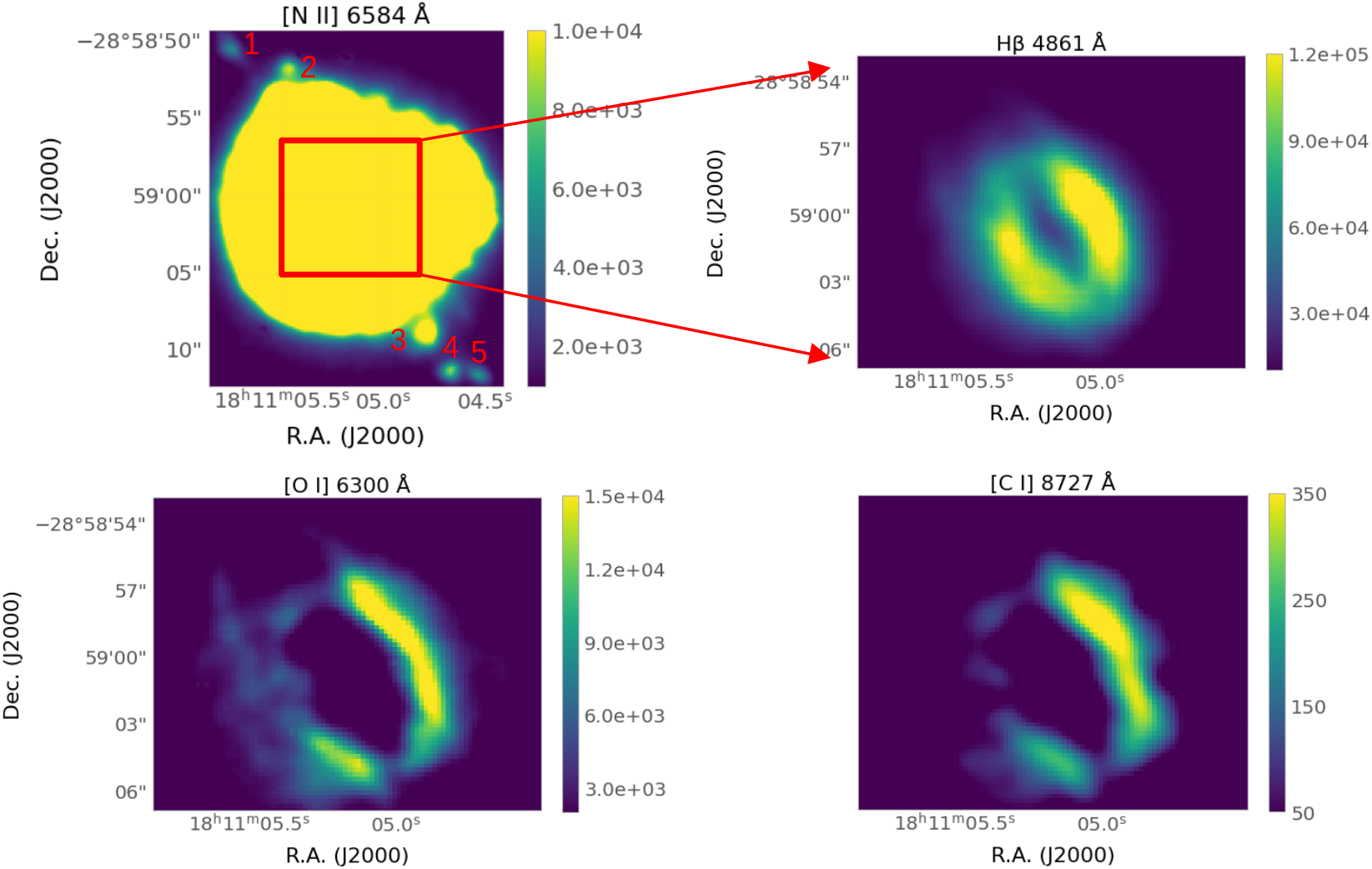}}
    \caption{M~1-42 emission-line maps of [N~{\sc ii}], H$\upbeta$, [O~{\sc i}] 6300~$\AA$ and [C~{\sc i}] 8727~$\AA$. Numbers 1 to 5 denote the different LISs aligned along the major axis of the PN.}
    \label{m1-42_append}
\end{figure*}

\begin{figure*}[h!]     
    \centering{\includegraphics[width=0.5\textwidth]{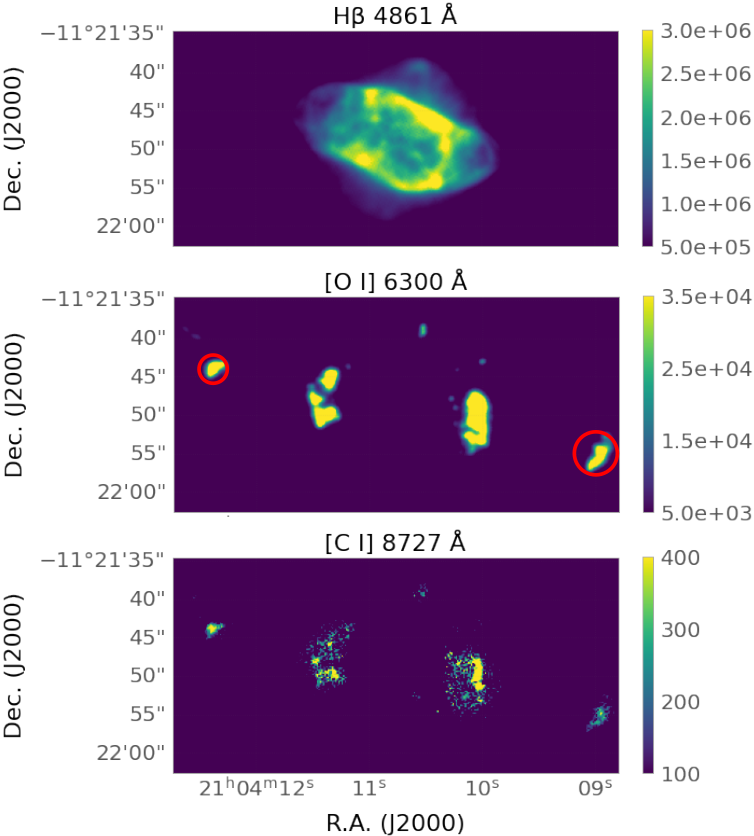}}
    \caption{Same as Fig.~\ref{ngc3242_append} for NGC~7009.}
    \label{ngc7009_append}
\end{figure*}

\begin{figure*}[h!]     
    \centering{\includegraphics[width=1\textwidth]{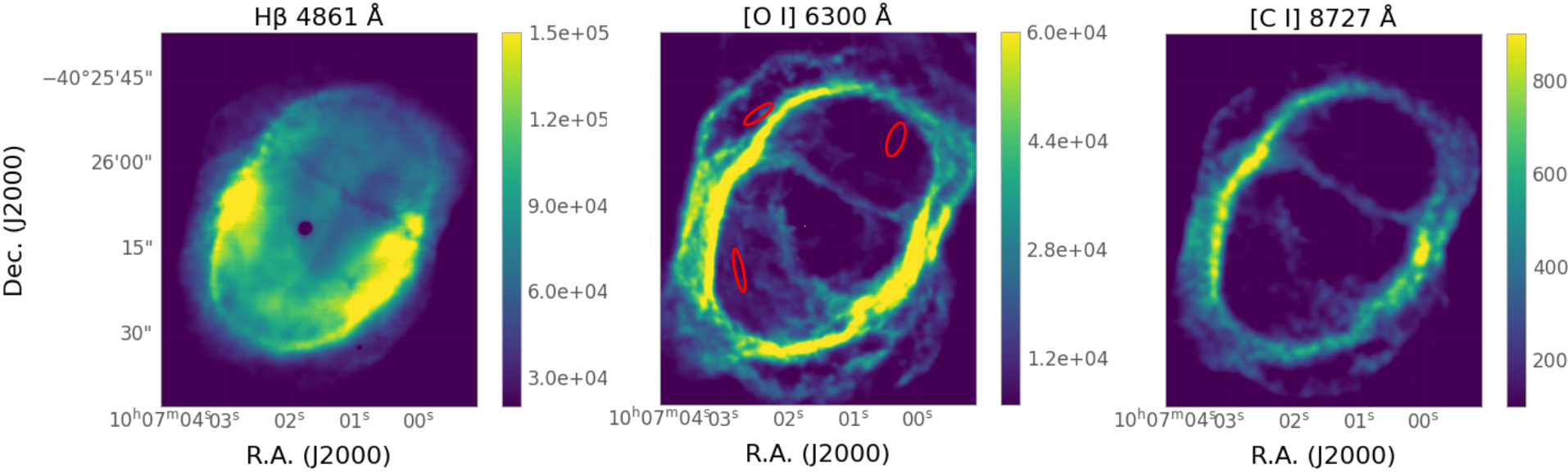}}
    \caption{Same as Fig.~\ref{ngc3242_append} for NGC~3132.}
    \label{ngc3132_append}
\end{figure*}

\begin{figure*}[h!]     
    \centering{\includegraphics[width=1\textwidth]{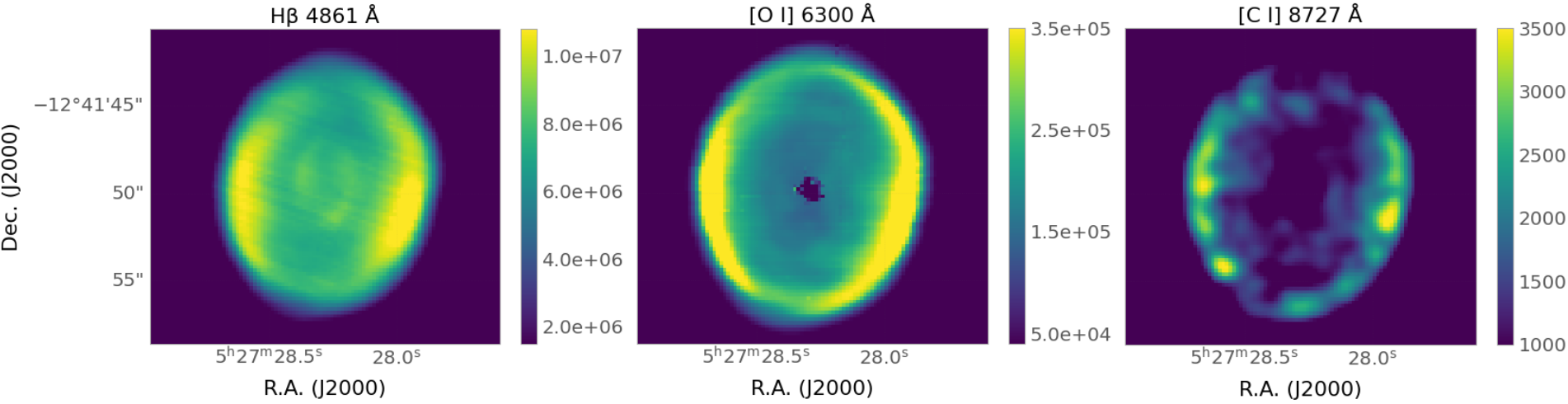}}
    \caption{Same as Fig.~\ref{ngc3242_append} for IC~418.}
    \label{ic418_append}
\end{figure*}

\begin{figure*}[h!]     
    \centering{\includegraphics[width=1\textwidth]{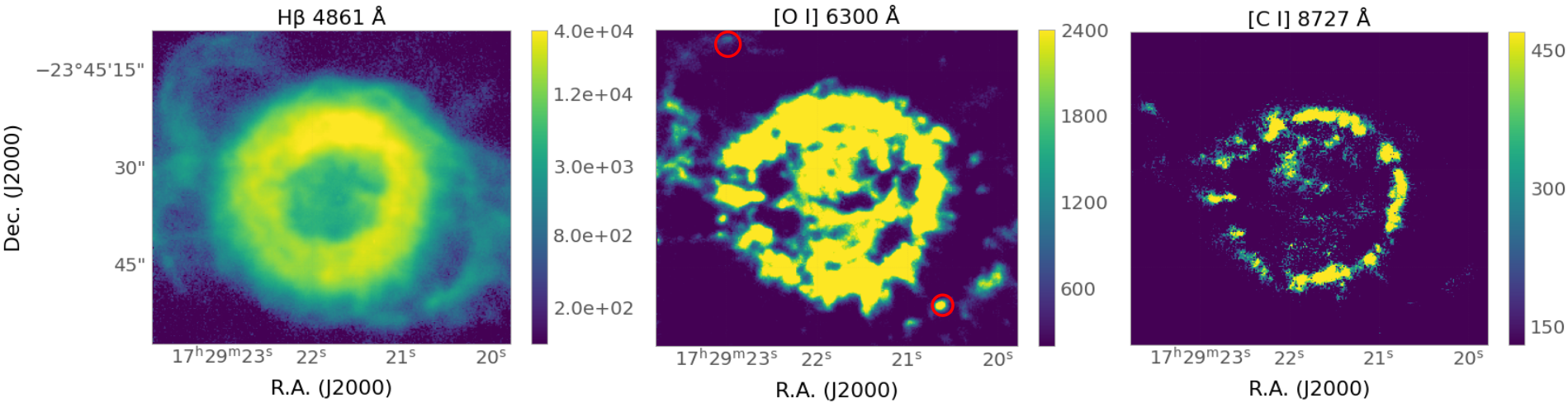}}
    \caption{Same as Fig.~\ref{ngc3242_append} for NGC~6369.}
    \label{ngc6369_append}
\end{figure*}

\begin{figure*}[h!]     
    \centering{\includegraphics[width=1\textwidth]{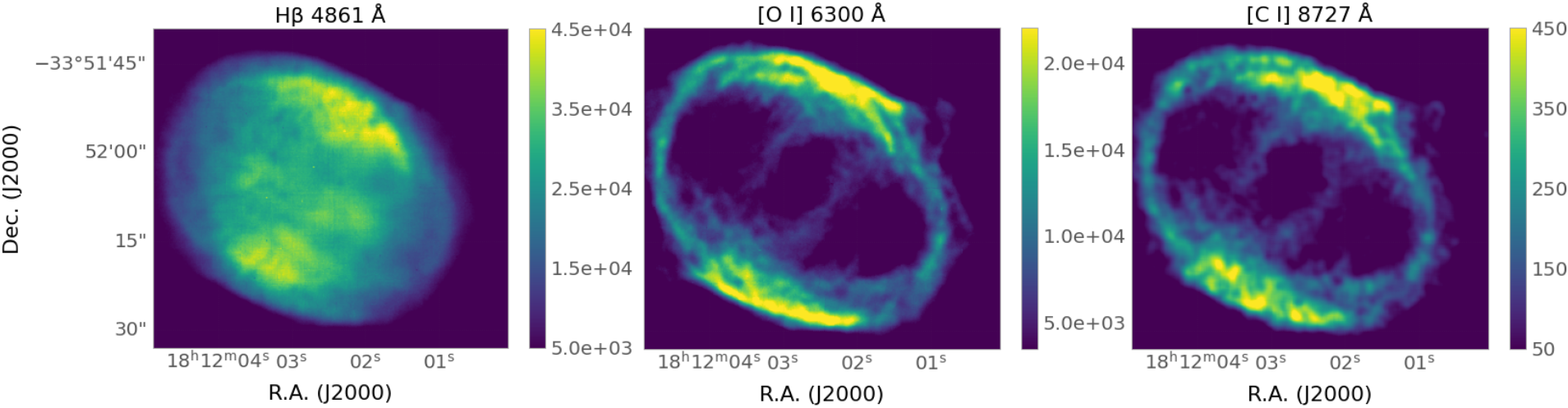}}
    \caption{Same as Fig.~\ref{ngc3242_append} for NGC~6563.}
    \label{ngc6563_append}
\end{figure*}

\begin{figure*}[h!]     
    \centering{\includegraphics[width=0.5\textwidth]{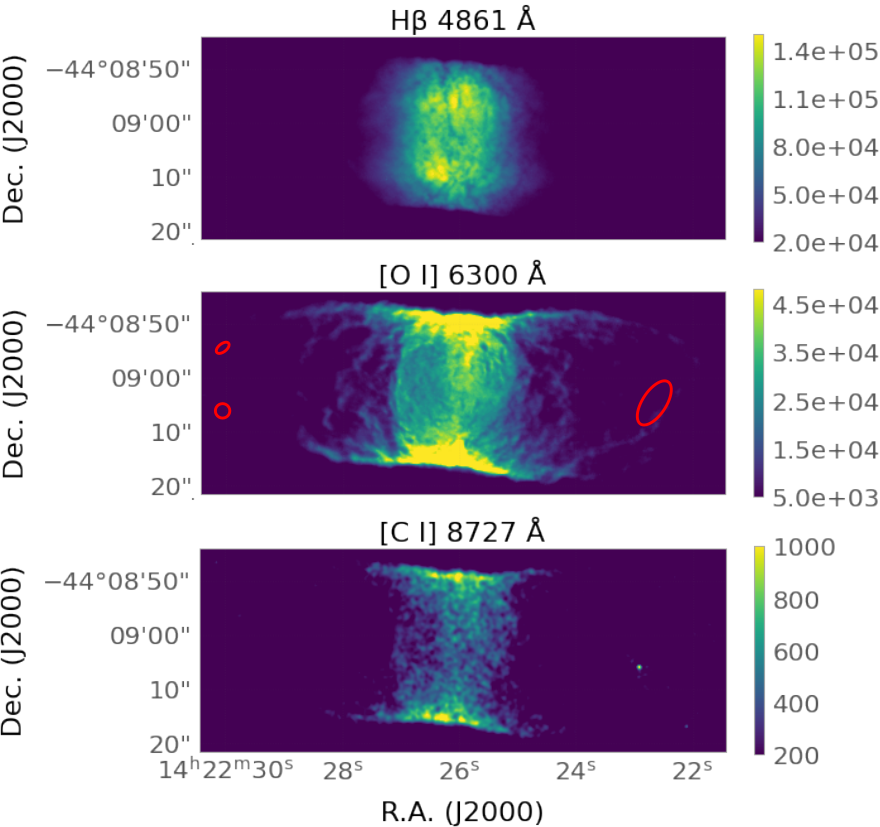}}
    \caption{Same as Fig.~\ref{ngc3242_append} for IC~4406.}
    \label{ic4406_append}
\end{figure*}

\begin{figure*}[h!]     
    \centering{\includegraphics[width=1\textwidth]{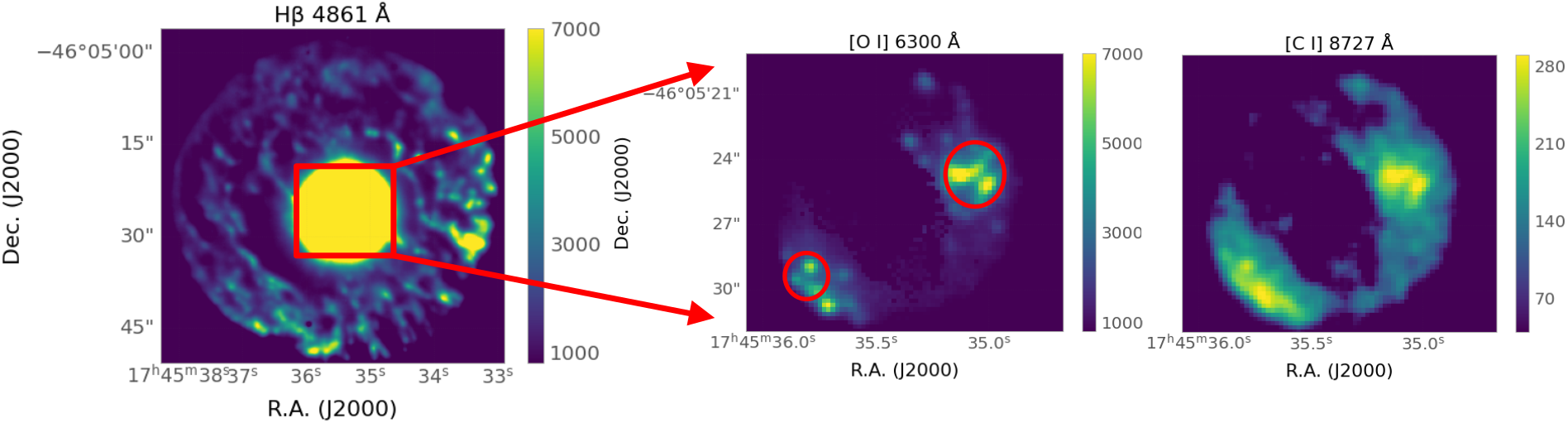}}
    \caption{Same as Fig.~\ref{ngc3242_append} for Tc~1.}
    \label{tc1_append}
\end{figure*}

\FloatBarrier 
\twocolumn

\section{Integrated spectrum}
\label{int_spec}

\begin{figure}[h!]     
    \centering{\includegraphics[width=0.39\textwidth]{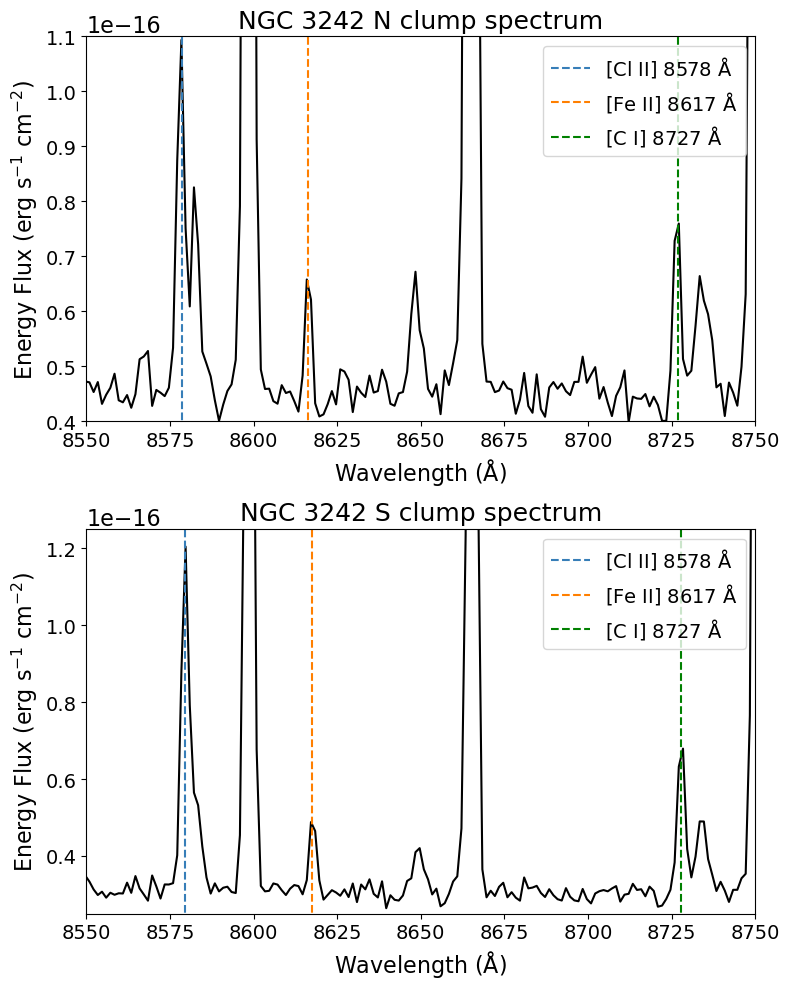}}
    \caption{\footnotesize{The spectrum of two nebular structures (N and S clumps) in NGC~3242. The [Cl~{\sc ii}] 8578~$\AA$, [Fe~{\sc ii}] 8617~$\AA$ and [C~{\sc i}] 8727~$\AA$ emissions lines are indicated with blue, orange and green vertical dot lines, respectively. H {\sc i} 8598~$\AA$ and 8665 $\AA$, He {\sc i} 8649 $\AA$ and 8733~$\AA$ emissions are also present in the plot. The number in the top left corner indicates the scaling factor applied in the y-axis values.}}
    \label{ngc3242_spec1}
\end{figure}
\begin{figure}[h!]     
    \centering{\includegraphics[width=0.39\textwidth]{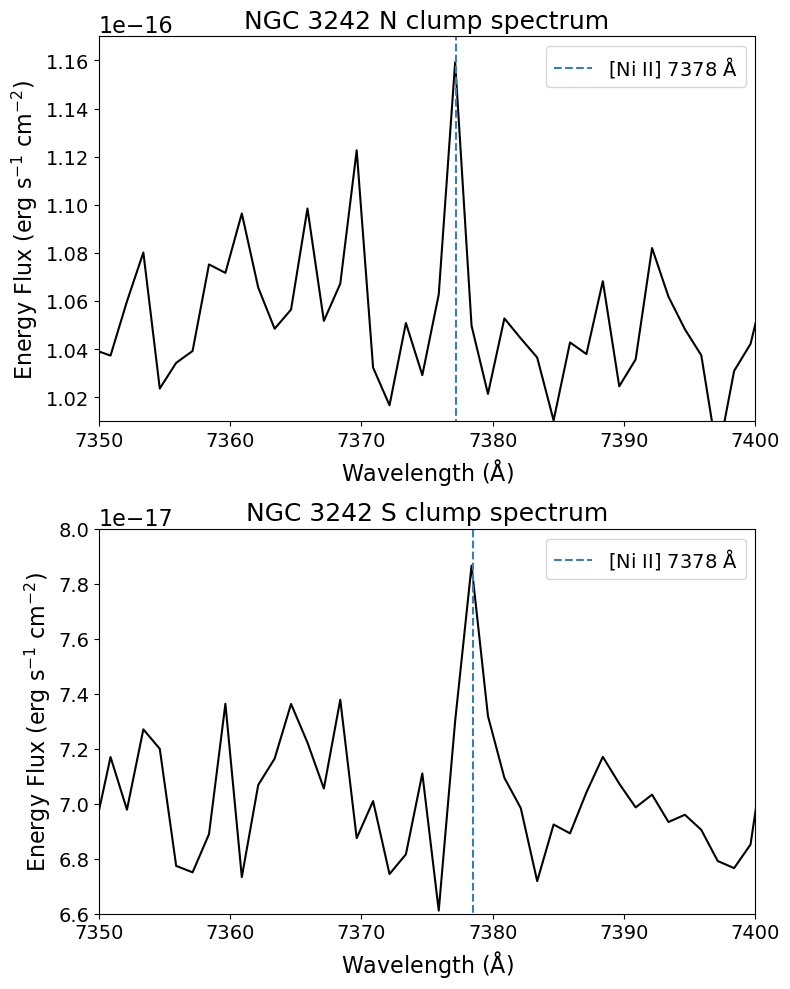}}
    \caption{\footnotesize{The spectrum of two nebular structures (N and S clumps) in NGC~3242 covering the wavelength range from 7350 to 7400~$\AA$. The [Ni~{\sc ii}] 7378~$\AA$ emission line is indicated with blue vertical dot line. The number in the top left corner indicates the scaling factor applied in the y-axis values.}}
    \label{ngc3242_spec2}
\end{figure}

\begin{figure}[h!]     
    \centering{\includegraphics[width=0.45\textwidth]{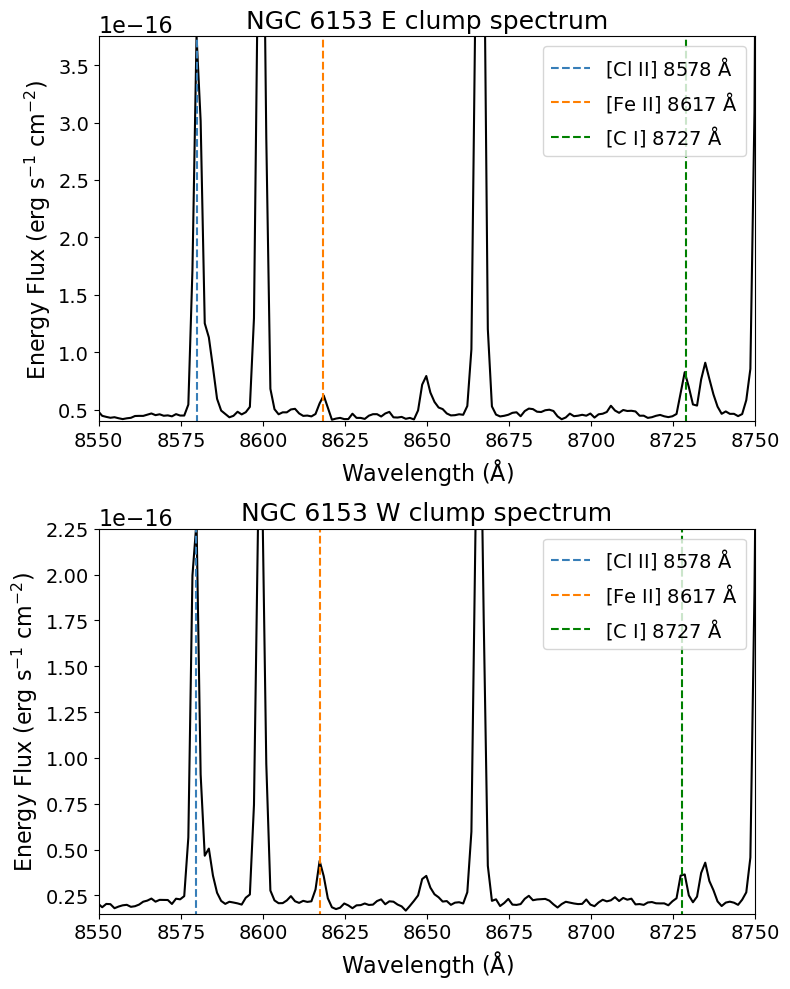}}
    \caption{Same as Fig.~\ref{ngc3242_spec1} for NGC~6153.}
    \label{ngc6153_spec1}
\end{figure}
\begin{figure}[h!]     
    \centering{\includegraphics[width=0.45\textwidth]{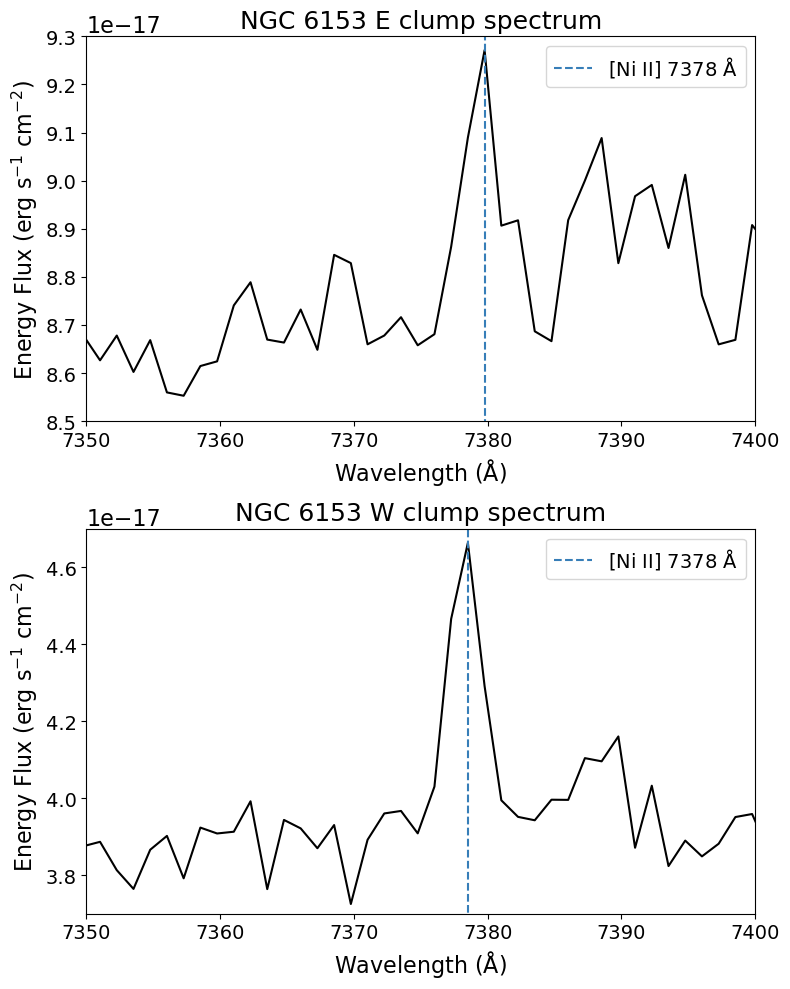}}
    \caption{Same as Fig.~\ref{ngc3242_spec2} for NGC~6153.}
    \label{ngc6153_spec2}
\end{figure}

\begin{figure}[h!]     
    \centering{\includegraphics[width=0.48\textwidth]{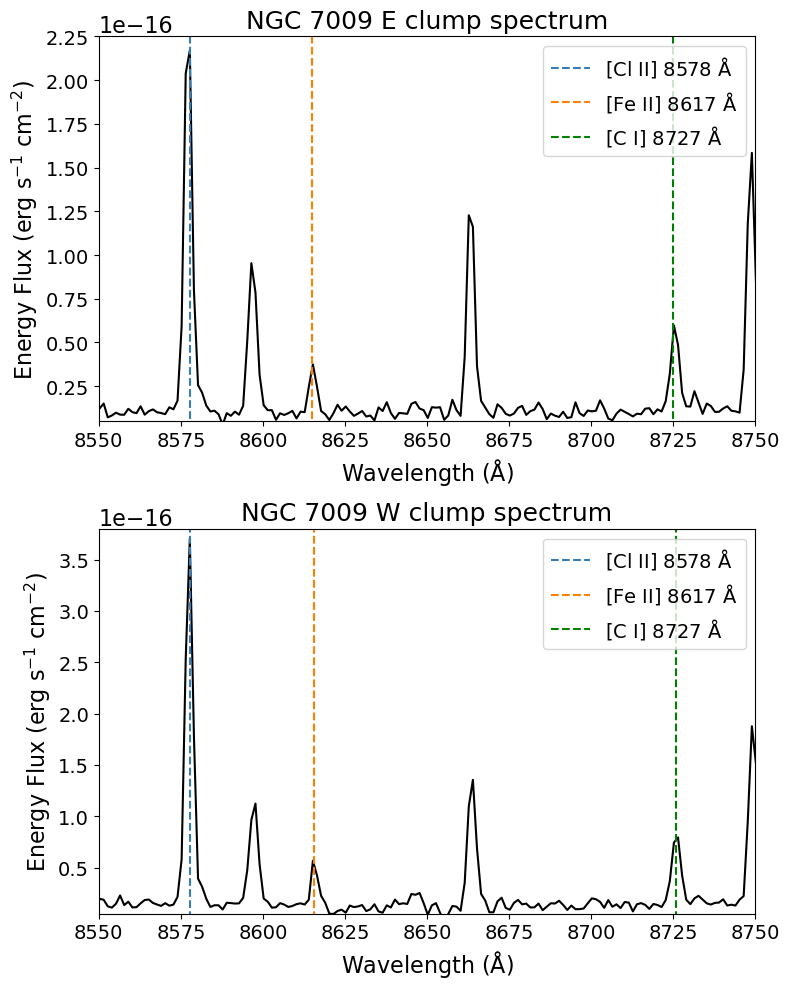}}
    \caption{Same as Fig.~\ref{ngc3242_spec1} for NGC~7009.}
    \label{ngc7009_spec1}
\end{figure}
\begin{figure}[h!]     
    \centering{\includegraphics[width=0.48\textwidth]{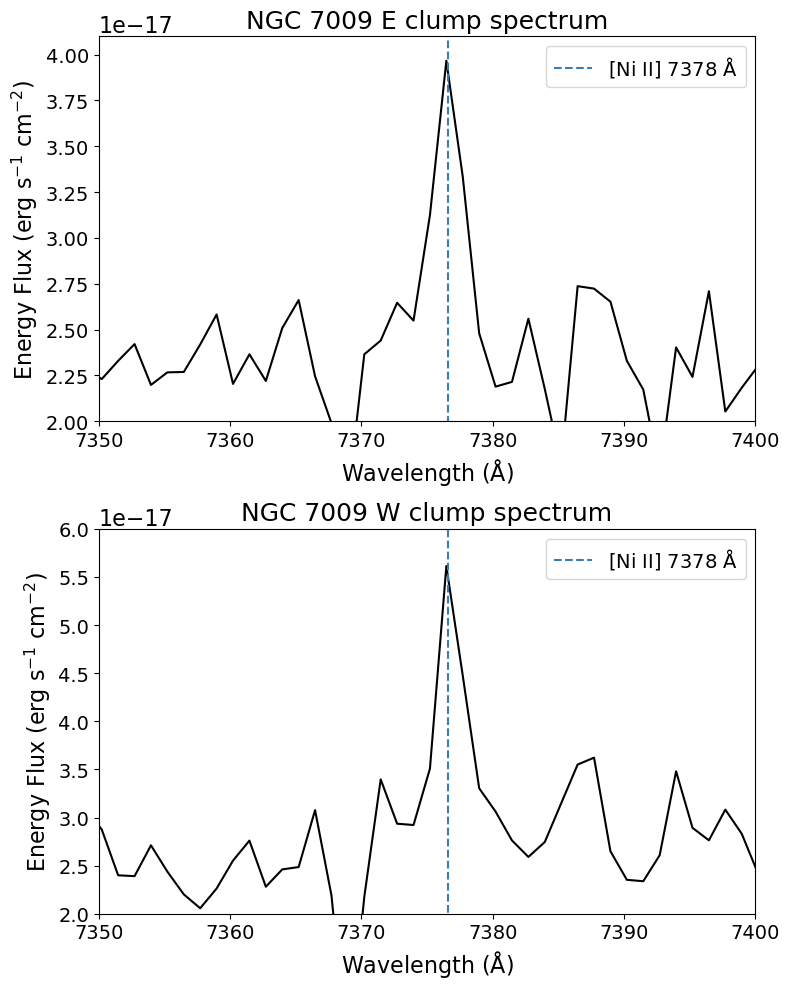}}
    \caption{Same as Fig.~\ref{ngc3242_spec2} for NGC~7009.}
    \label{ngc7009_spec2}
\end{figure}

\begin{figure}[h!]     
    \centering{\includegraphics[width=0.48\textwidth]{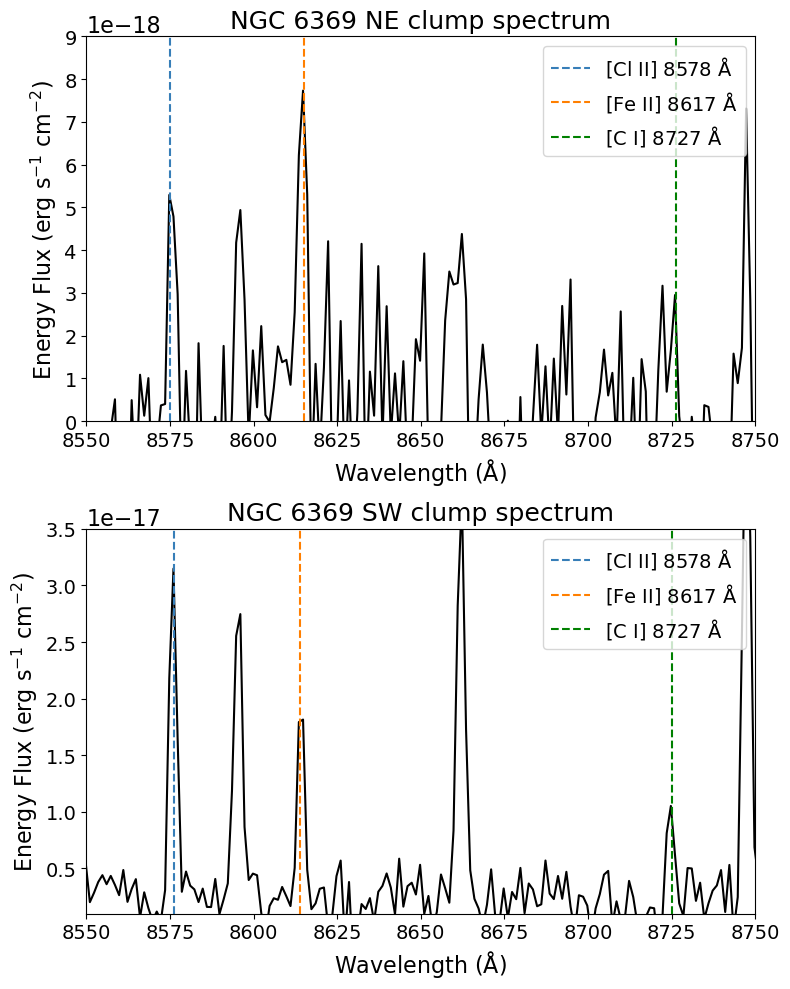}}
    \caption{Same as Fig.~\ref{ngc3242_spec1} for NGC~6369.}
    \label{ngc6369_spec1}
\end{figure}
\begin{figure}[h!]     
    \centering{\includegraphics[width=0.48\textwidth]{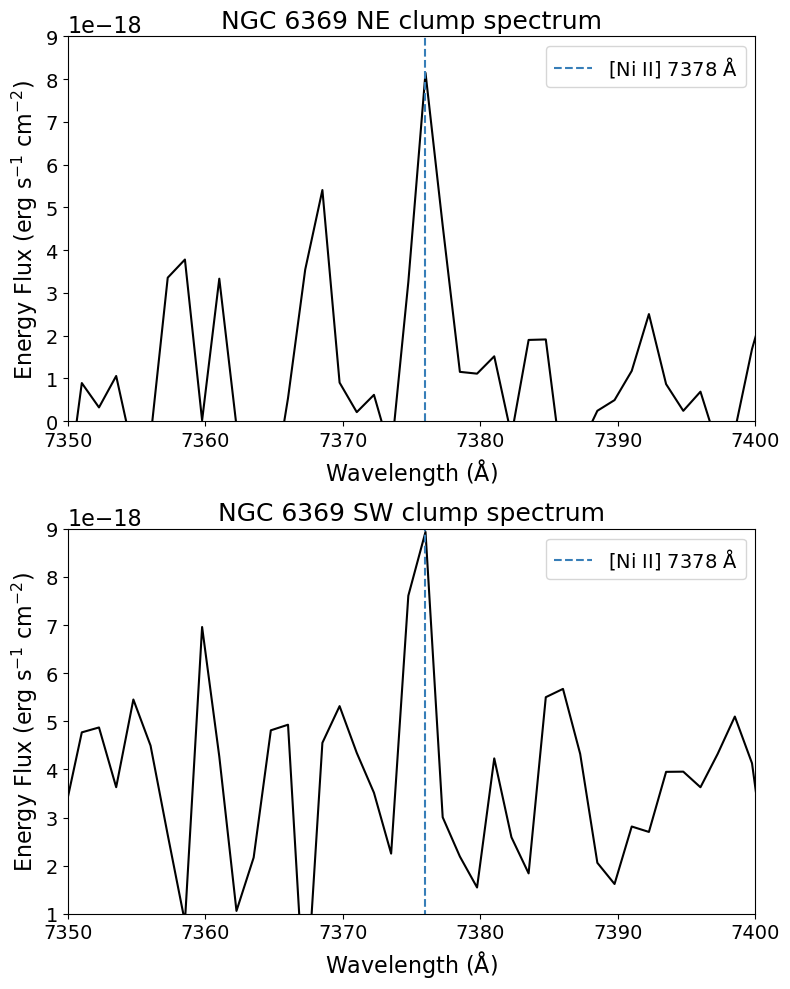}}
    \caption{Same as Fig.~\ref{ngc3242_spec2} for NGC~6369.}
    \label{ngc6369_spec2}
\end{figure}

\begin{figure}[h!]     
    \centering{\includegraphics[width=0.5\textwidth]{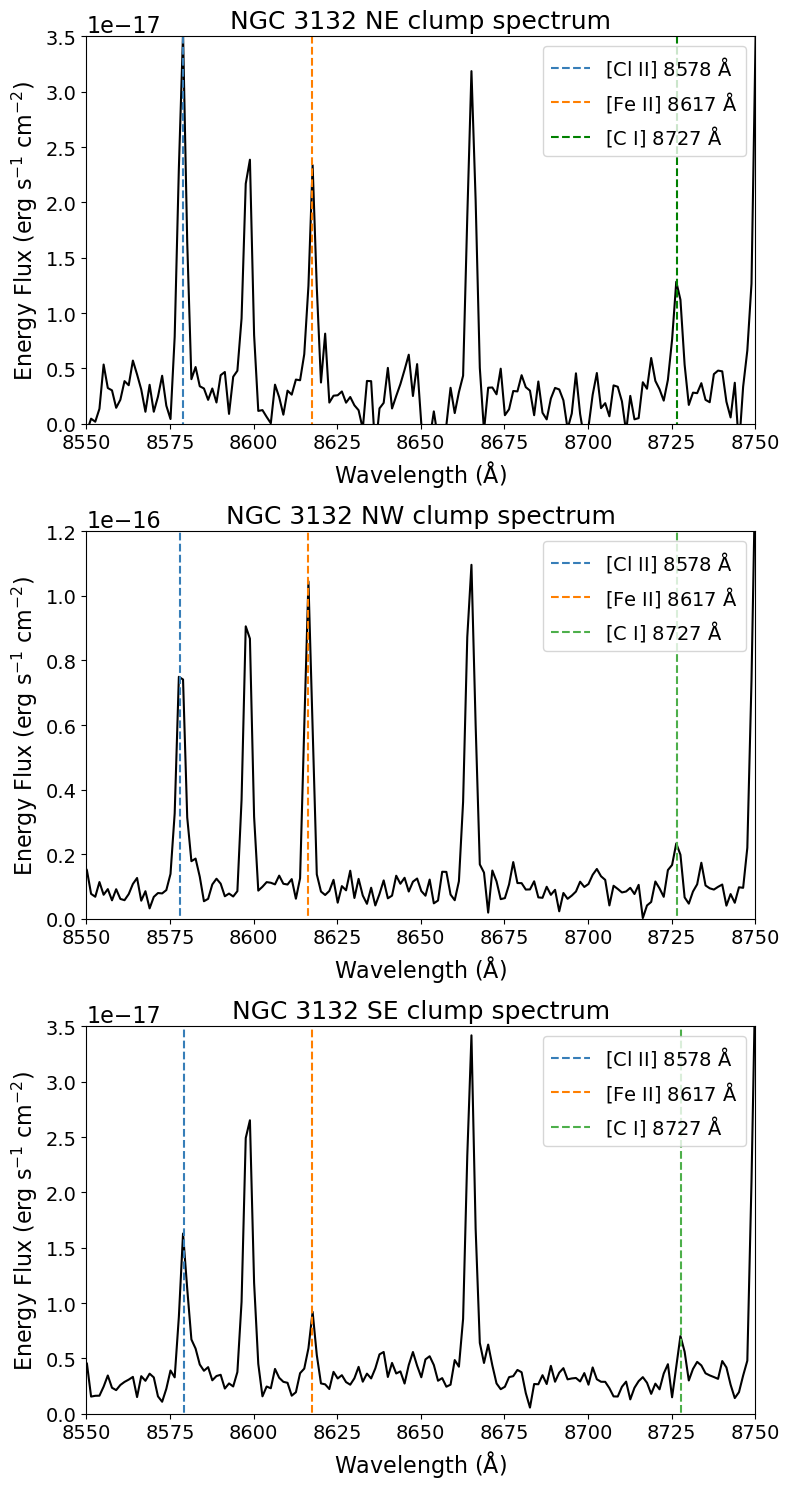}}
    \caption{Same as Fig.~\ref{ngc3242_spec1} for NGC~3132.}
    \label{ngc3132_spec1}
\end{figure}
\begin{figure}[h!]     
    \centering{\includegraphics[width=0.5\textwidth]{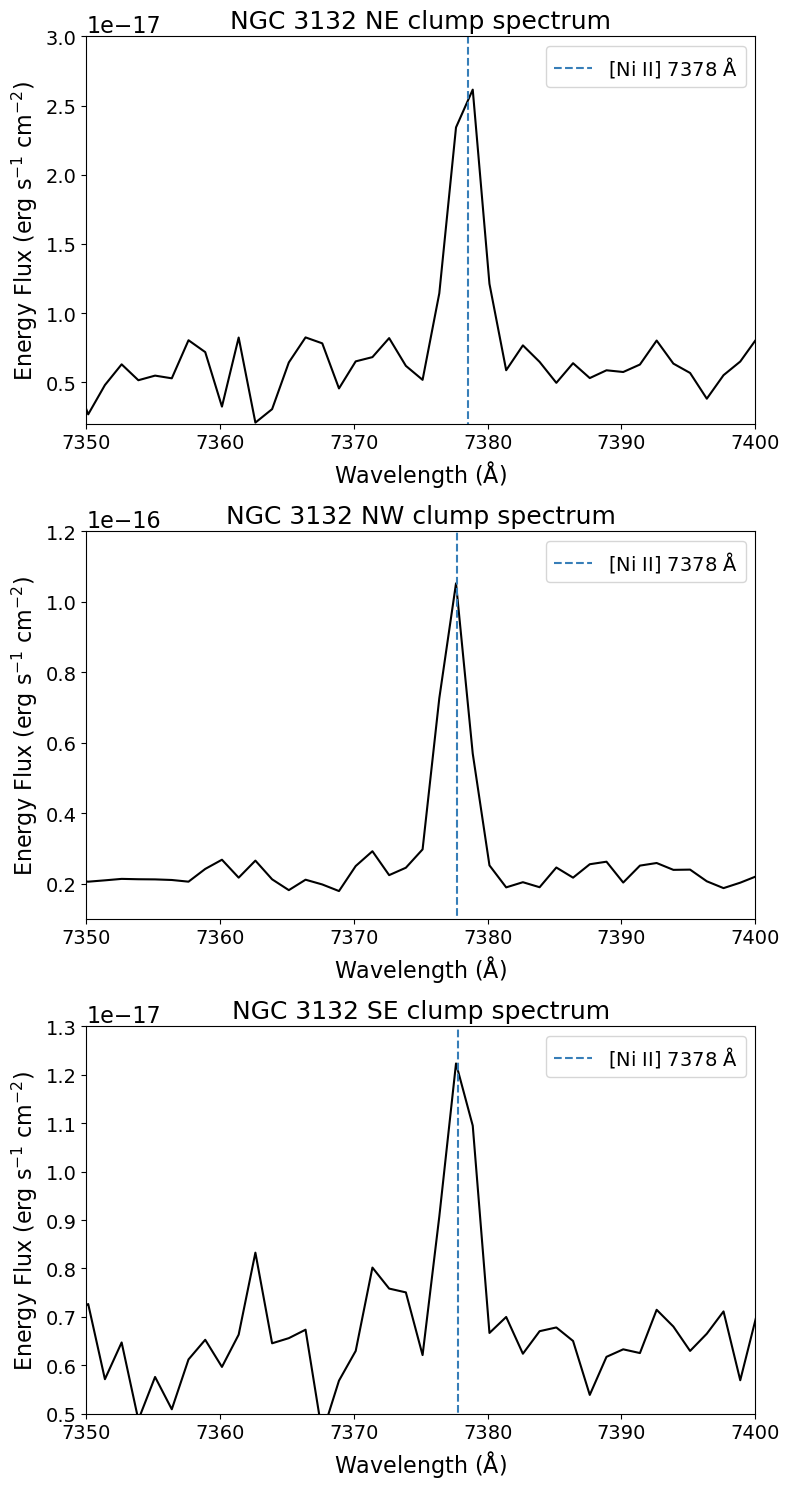}}
    \caption{Same as Fig.~\ref{ngc3242_spec2} for NGC~3132.}
    \label{ngc3132_spec2}
\end{figure}

\begin{figure}[h!]     
    \centering{\includegraphics[width=0.5\textwidth]{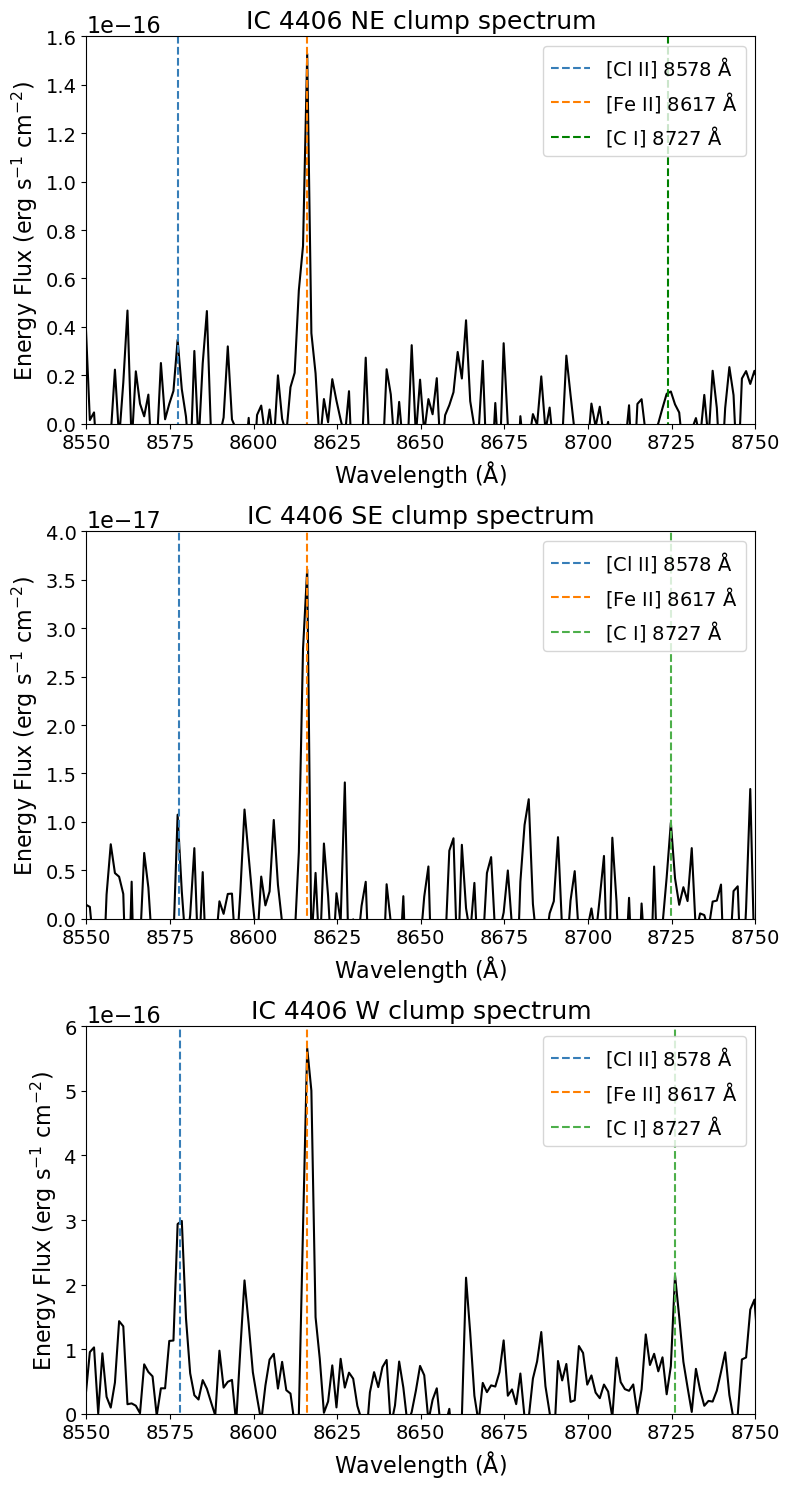}}
    \caption{Same as Fig.~\ref{ngc3242_spec1} for IC~4406.}
    \label{ic4406_spec1}
\end{figure}
\begin{figure}[h!]     
    \centering{\includegraphics[width=0.5\textwidth]{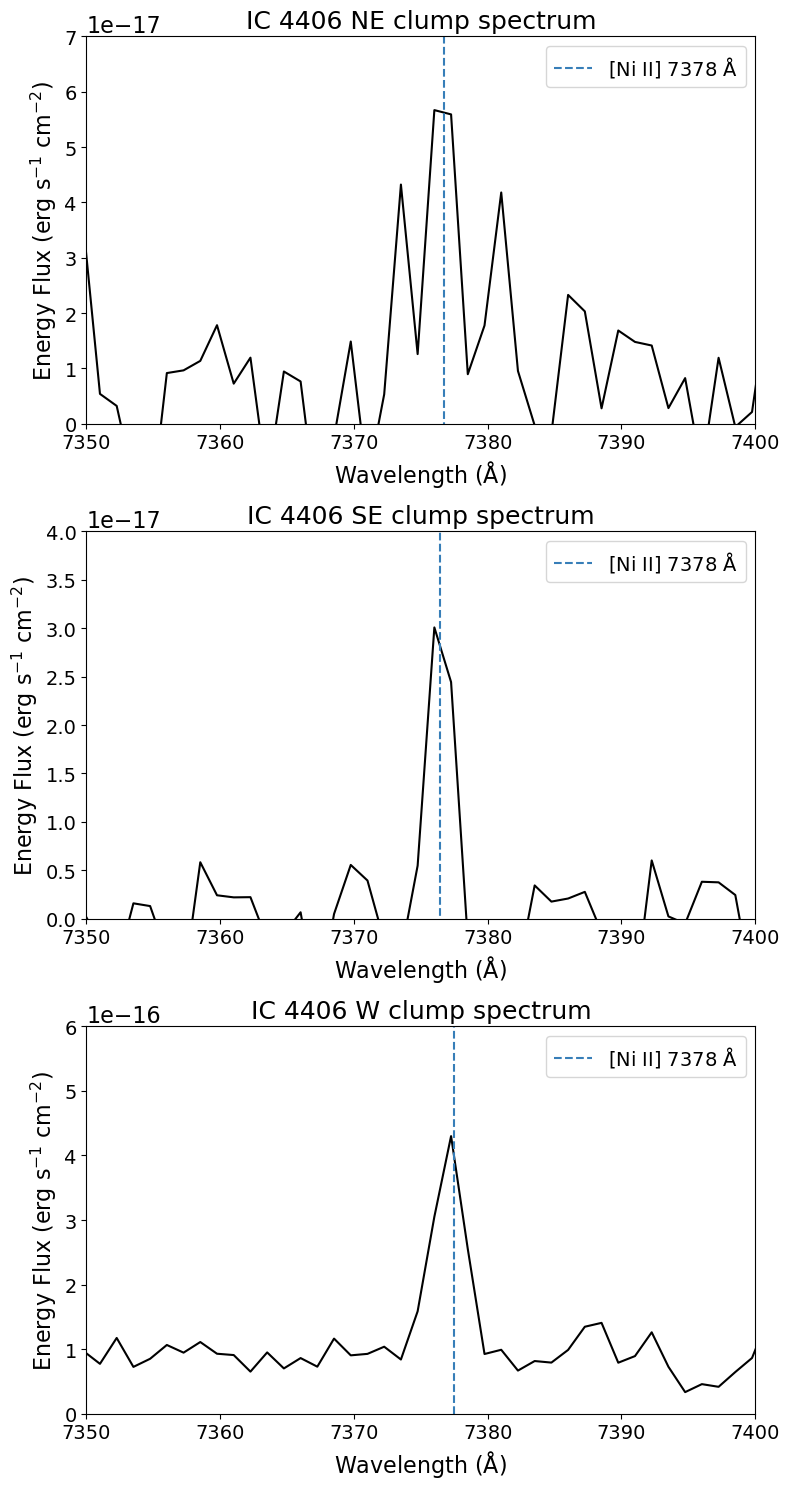}}
    \caption{Same as Fig.~\ref{ngc3242_spec2} for IC~4406.}
    \label{ic4406_spec2}
\end{figure}

\begin{figure}[h!]     
    \centering{\includegraphics[width=0.49\textwidth]{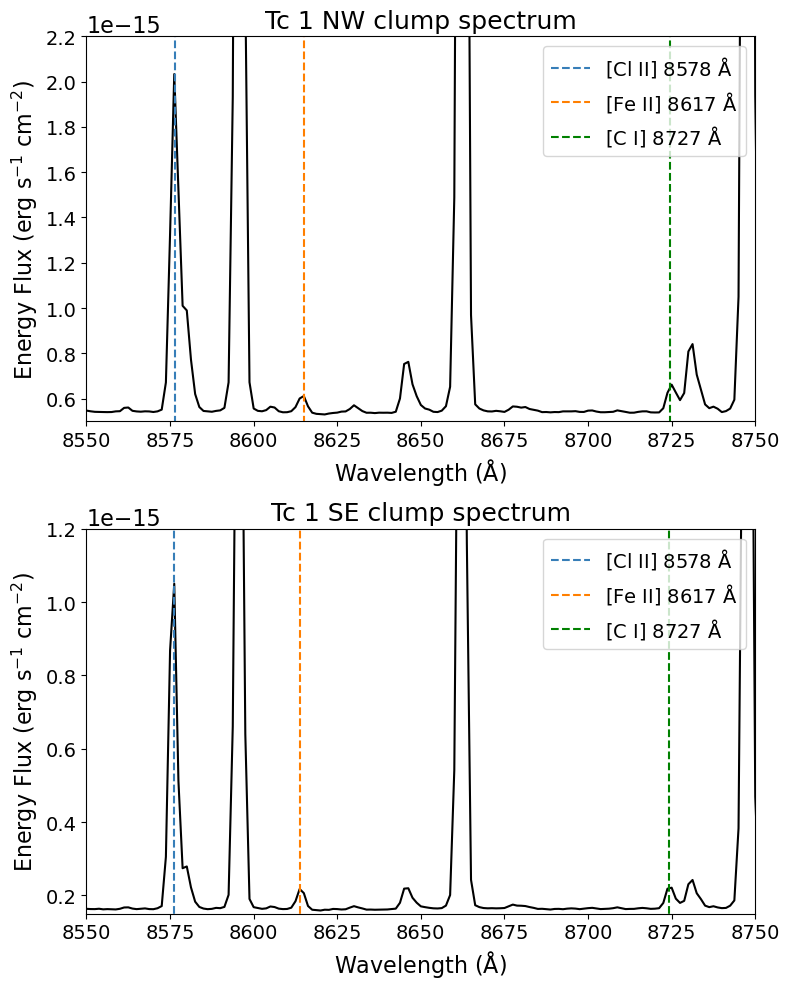}}
    \caption{Same as Fig.~\ref{ngc3242_spec1} for Tc~1.}
    \label{tc1_spec1}
\end{figure}
\begin{figure}[h!]     
    \centering{\includegraphics[width=0.49\textwidth]{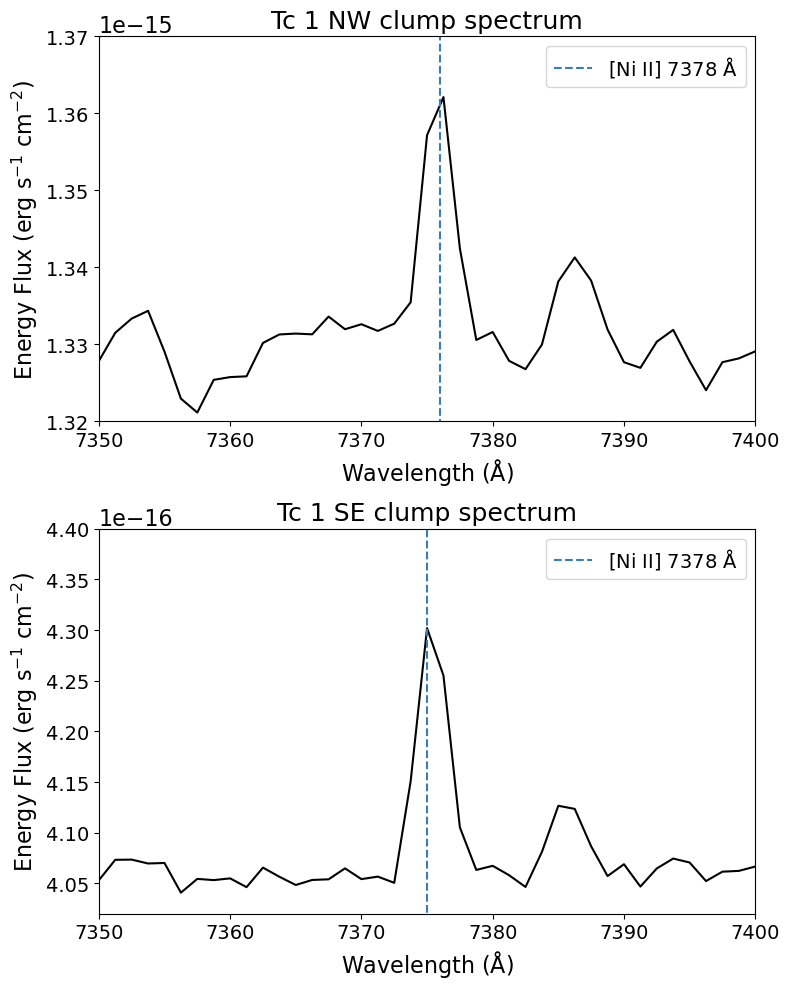}}
    \caption{Same as Fig.~\ref{ngc3242_spec2} for Tc~1.}
    \label{tc1_spec2}
\end{figure}

\FloatBarrier
\onecolumn
\section{Extra Figures}

\begin{center}
\begin{minipage}{0.5\textwidth}
  \centering
  \includegraphics[width=0.8\linewidth]{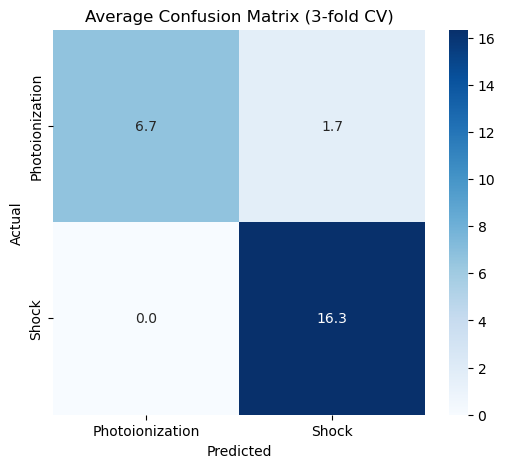}
  \captionof{figure}{Mean confusion matrix for K-Means classifier.}
  \label{kmeans_conf_matrix}
\end{minipage}%
\begin{minipage}{0.5\textwidth}
  \centering
  \includegraphics[width=1\linewidth]{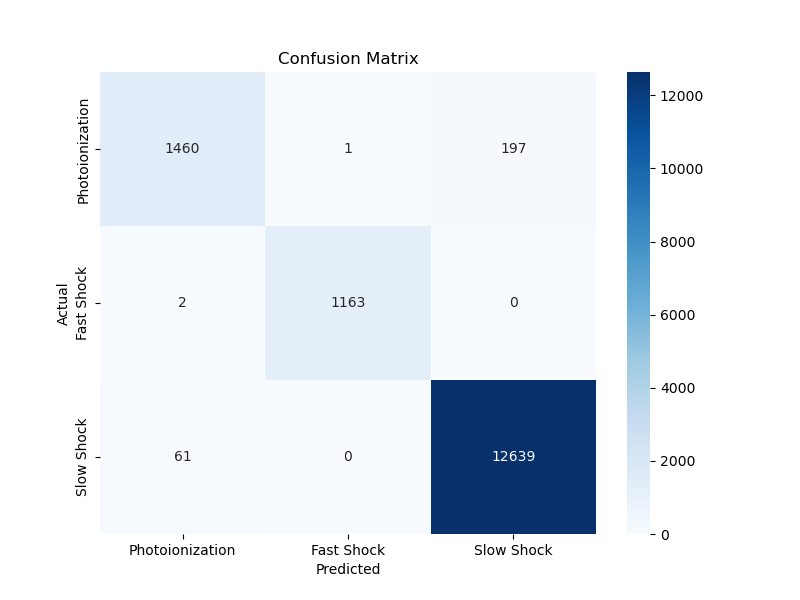}
  \captionof{figure}{Confusion matrix for KNN classifier.}
  \label{knn_conf_matrix}
\end{minipage}
\end{center}

\FloatBarrier 

\section{Extra Tables}

\begin{table*}[h!]
\centering
\caption{Parameters selected for the photoinization models from 3Mdb database.}
\begin{tabular}{|c|}
\hline
Abundances: \citet{delgado_inglada_2014} \\ \hline
Central Star SED: Blackbody                                                                                              \\ \hline
Dust: No                                                                                                                  \\ \hline
Density Law: Constant                                                                                                     \\ \hline
Mean Ionization Parameter (U): 10$^{-6}$ --  1                                                                                                      \\ \hline
(log(L$_{\rm bol}$) \textless 4.2) \& (log(L$_{\rm bol}$) \textgreater 3.4 | T$_{\rm eff}$ \textgreater 10$^5$) \& (log(L$_{\rm bol}$) \textgreater 1.5$^{-5}$ $\cdot$ T$_{\rm  eff}$ - 0.25) \\ \hline
H$_{\rm mass}$ \textless 1 M$_{\odot}$                                                                                           \\ \hline
n$_{\rm H}$ $\cdot$ r$_{\rm out}$ $^3$: 2 $\cdot$ 10$^{53}$ -- 3 $\cdot$ 10$^{56}$                                                                     \\ \hline
H$\upbeta$ / (4$\pi$ $\cdot$ r$_{\rm out}$ $\cdot$ 206265)$^2$: 10$^{-15}$ -- 10$^{-11}$                                                                                      \\ \hline
\end{tabular}
\tablefoot{L$_{\rm bol}$ is the bolometric luminosity in solar units, T$_{\rm eff}$ is the effective temperature in K, n$_{\rm H}$ is the mean hydrogen density, r$_{\rm out}$ is the outer radius in cm and H$\upbeta$ is the H$\upbeta$ luminosity in erg s$^{-1}$.}
\label{3mdb_uv_params}
\end{table*}

\begin{table*}[h!]
\centering
\caption{Parameters selected for the shock models from 3Mdb database.}
\begin{tabular}{|c|c|c|c|}
\hline
                          & Fast Shocks      & Fast Shocks (Incomplete)& Slow Shocks      \\ \hline
Abundances                & \citet{allen2008} & \citet{allen2008}     & \citet{delgado_inglada_2014}  \\ \hline
Velocity (km s$^{-1}$)    & 100 -- 1\,000       & 100 -- 1\,000              & 10-- 100          \\ \hline
Pre-Shock Density (cm$^{-3}$)& 10 -- 10\,000    & 10 -- 10\,000              & 10 -- 10\,000      \\ \hline
Pre-Shock Temperature (K) & \textless 15\,000& \textless 15\,000       & \textless 15\,000\\ \hline
Magnetic Field ($\upmu$G) & \textless 10     & \textless 10            & 1 \\ \hline
Cutoff Temperature (K)    & -                & \textless 15\,000       & \textless 15\,000\\ \hline
\end{tabular}
\label{3mdb_shock_params}
\end{table*}

\end{appendix}

\end{document}